\documentclass[10pt]{article}

\def\nn{\nonumber}

\def\be{\beta}

\def\la{\lambda}

\def\const{{\rm const}}

\def\be{\begin{equation}}
\def\ee{\end{equation}}
\def\bea{\begin{eqnarray}}
\def\eea{\end{eqnarray}}
\newcommand{\de}{{\partial}}
\newcommand{\td}{{\tilde{d}}}
\newcommand{\tD}{{\tilde{D}}}
\newcommand{\akapit}{\hskip 15pt}
\newcommand{\Arth}{{\rm Arth}}
\newcommand{\artanh}{{\rm ar \, tanh }}
\newcommand{\arcoth}{{\rm ar \, coth }}

\newcommand{\sgn}{{\rm sgn }}
\newcommand{\calD}{{\cal D}}
\newcommand{\calE}{{\cal E}}
\newcommand{\calI}{{\cal I}}
\newcommand{\calK}{{\cal K}}

\newcommand{\calM}{{\cal M}}

\newcommand{\Btheta}{{B_{\vartheta}}}

\newcommand{\unin}{\in \!\!\! / \,}

\begin{document}

\begin{center}

\begin{flushright}
  hep-th/0410237
\end{flushright}

\vskip 8em
{\LARGE \bf \sc Nonsupersymmetric intersecting}

\vskip 1em
{\LARGE \sc branes in supergravity}

\vskip 6em
{\large \bf Marcin P. Flak}

\vskip 1em
{\small \tt Marcin.Flak@fuw.edu.pl}

\vskip 1em
{\it Institute of Theoretical Physics, Department of Physics, Warsaw University\\ Ho\.za 69, 00-681 Warsaw, Poland}

\vskip 6em
{Ph.D. thesis written under supervision of \\ prof. Krzysztof A. Meissner \\ presented to the faculty of 
Warsaw University Physics Department \\ and recommended for acceptance}

\end{center}

\vskip 6em

%%%%%%%%%%%%%%%%%%%%%%%%%%%%%%%%%%%%%%%%%%%%%%%%%%%%%%%%%%%%%%%%

\begin{abstract}

In this doctoral thesis a model of many orthogonally commonly intersecting 
delocalized branes with neither harmonic gauge nor any other 
extra conditions is discussed. Further a method of
solving equations of motion of the model is given. It is proved
that the model reduces to the so called Toda-like system which
is solvable at least in
several cases relevant for realistic brane configurations. The
solutions generally can break supersymmetry. 
Examples of the solutions are given and some their
properties are considered in more detail. Especially the presence
and interpretation of singularities is discussed and the relation
between energy and charge density of the solution. A certain
duality in the space of solutions is described connecting two
seemingly different elements of the space. It is shown that the
solution dual to the supersymmetric one breaks supersymmetry, but
it still possesses some features usually attributed only to
solutions preserving supersymmetry. In particular for the dual
solution equality between energy and charge density holds.

\end{abstract}

%%%%%%%%%%%%%%%%%%%%%%%%%%%%%%%%%%%%%%%%%%%%%%%%%%%%%%%%%%%%%%%%%
%%%%%%%%%%%%%%%%%%%%%%%%%%%%%%%%%%%%%%%%%%%%%%%%%%%%%%%%%%%%%%%%%
%%%%%%%%%%%%%%%%%%%%%%%%%%%%%%%%%%%%%%%%%%%%%%%%%%%%%%%%%%%%%%%%%
\newpage
. 
\vskip 10em
{\LARGE Acknowledgments}

\vskip 6em

{\large I would like to thank my supervisor, Professor Krzysztof 
Meissner, for providing me a guidance during the long journey through 
the amazing world of elementary particle physics. Not only did he make me
familiar with secrets of this world but also supported me in 
overcoming difficulties which I often encountered in the course of my 
studies. I am very grateful for having him for a mentor.}

%%%%%%%%%%%%%%%%%%%%%%%%%%%%%%%%%%%%%%%%%%%%%%%%%%%%%%%%%%%%%%%%%
%%%%%%%%%%%%%%%%%%%%%%%%%%%%%%%%%%%%%%%%%%%%%%%%%%%%%%%%%%%%%%%%%
%%%%%%%%%%%%%%%%%%%%%%%%%%%%%%%%%%%%%%%%%%%%%%%%%%%%%%%%%%%%%%%%%
\newpage
\tableofcontents

%%%%%%%%%%%%%%%%%%%%%%%%%%%%%%%%%%%%%%%%%%%%%%%%%%%%%%%%%%%%%%%%%
%%%%%%%%%%%%%%%%%%%%%%%%%%%%%%%%%%%%%%%%%%%%%%%%%%%%%%%%%%%%%%%%%
%%%%%%%%%%%%%%%%%%%%%%%%%%%%%%%%%%%%%%%%%%%%%%%%%%%%%%%%%%%%%%%%%
\newpage
\section{Introduction}

\akapit
Branes are one of the most interesting topics that
appeared in the theoretical physics of elementary particles in the
last years. Treated at the beginning as just some special
solutions of supergravity models they slowly gathered importance
to the point that now they are treated as hypothetically more
fundamental than strings. In superstring theories branes are
identified with sources of Ramond-Ramond charges. These branes, so
called D-branes, are equivalently described as hypersurfaces where
ends of open strings are attached. Consequently, the standard type
I string theory with freely propagating open strings can be
understood as a theory on a ten dimensional brane worldvolume. But
there is also another sort of branes -- NS-branes and a special
representative of the category is a fundamental string serving as
an elementary object of the whole string theory. This fact allows
to make a conjecture that there can be a theory -- a
generalization of string theory -- where not strings but branes
are the fundamental objects. The hypothesis suggests that this
tentative M-theory could be realized as a quantum theory of
membranes but the explicit realization of this idea has not yet
been found.

But even restricted to string theory the role of branes is
extraordinarily important. Branes belong to the spectrum of the
theory, so the knowledge of their properties is necessary for the
better understanding of string theory. Discovery of branes
significantly enlarged the number of known states in the theory,
and since they don't have a string interpretation their presence
in the spectrum sheds new light on the whole theory. Several
classes of states can be distinguished depending on their
features. For example BPS states, which saturate so called
Bogomolny bound and preserve at least part of supersymmetry. The
remaining non--BPS states fall into two main categories: stable
and unstable where the last one contains tachyons. Non--BPS stable
states are especially interesting because the physics we see
(which is obviously nonsupersymmetric) is probably the low energy
limit of such a state with (softly) broken supersymmetry.

A fact worth to note at this moment is that branes are in general
objects of certain dimensionality living in a higher dimensional
theory. So it is tempting to interpret the four dimensional Universe
we observe as the brane immersed in ten dimensional spacetime of
string theory.

Since branes can be treated as an extension of some already
well know (at least theoretically) objects as magnetic monopoles
or fundamental strings, they allow to look at some previously
discovered facts from a wider perspective and find connections
among them. For example theorems telling that classical
superstrings are permitted only for $D=3,4,6,10$ and the maximal
dimension for supergravity is eleven are consequences of a rule
determining how many dimensional superbranes are possible to
introduce in a given spacetime. The idea of branes provides also a
method for studying Yang-Mills theories in diverse dimensions.
This comes from a fact that a theory induced on the brane
worldvolume by the surrounding string theory is in general gauge
invariant. So one can understand properties of a given Yang-Mills
theory in connection with properties of the branes. In particular
BPS branes give a description of super-Yang-Mills theories, while
non-BPS branes -- gauge theories with broken supersymmetry.

Branes can also be regarded from the purely classical point of
view as solutions of field equations in supergravity or more
generally in systems with gravity coupled to antisymmetric tensors
and other fields. This method of description of branes is very
important because it gives some exact information which cannot be
obtained with the use of perturbative methods peculiar for string
theory. The main subject of this work is dedicated to this part of
the theory of branes, especially to such brane solutions of
supergravity equations of motion which can break supersymmetry. We
aim in this work to find and analyze new brane solutions that are
in general nonsupersymmetric and which describe many intersecting
branes supported either by the same type of field or by different
fields.

Before we go over to the main part of the work we present a short
explanation what branes are and why they are interesting. In
chapter \ref{From_SM_GR_to_branes} a path is presented which leads
from the Standard Model and General Relativity -- the theories
extremely well describing physics from the smallest to the biggest
scales experimentally testable at this moment -- through Grand
Unification Theories, models with extra dimensions, supersymmetry
and supergravity to bosonic strings, superstrings and finally
M-theory, currently thought to be the best candidate for the
unified theory of all interactions.

Various kinds of branes are introduced afterwards, in chapter
\ref{branes_quantum}. First of all a possibility of constructing
quantum theories based on branes as its fundamental objects
(instead of point particles or strings) is examined. A result
of so called brane scan is presented, which tells for what
combinations of a spacetime and the brane worldvolume dimensions
it is possible to introduce a supersymmetric brane. The discussion
however ends with the conclusion that such a theory as
we understand it now cannot be consistently quantized. Therefore
we gather information about branes by treating them as sources of
fields of antisymmetric forms and we obtain generalization of the
famous Dirac quantization rule for magnetic monopoles. Next we
discuss branes appearing in M- and string theories in particular
D-branes, NS-branes and M-branes and give arguments that branes are
the necessary part of the theories. We present the classification
of the brane states with respect to the saturation of the Bogomolny
bound and show some examples of non-BPS unstable and stable
states. We check also how the various kinds of branes behave
under string dualities or dimensional reductions. The observations
allow us to conjecture that all the branes can be various
manifestations of only one class of more fundamental objects.

Finally in chapter \ref{sugra_branes} supergravity description of
branes is given starting with detailed discussion of a simple but
very instructive example -- single component brane solution in the
harmonic gauge. Some generalizations of the example including
black branes and systems with many intersecting branes are also
presented.

The last chapter consists of a construction of a model of many
orthogonally commonly intersecting delocalized branes with neither
harmonic gauge nor any other extra conditions. Further a method of
solving equations of motion of the model is given. It is proved
that the model reduces to the so called Toda-like system (after
adequate redefinition of radial coordinate: $r \rightarrow
\vartheta(r)$, where $\vartheta(r)$ is in general not a harmonic
function in flat space). The system is solvable at least in
several cases relevant for realistic brane configurations. The
solutions generally can break supersymmetry and the supersymmetric
solutions are generally distinguished by their very specific
properties. Examples of the solutions are given and some their
properties are considered in more detail. Especially the presence
and interpretation of singularities is discussed and the relation
between energy and charge density of the solution. A certain
duality in the space of solutions is described connecting two
seemingly different elements of the space. It is shown that the
solution dual to the supersymmetric one breaks supersymmetry, but
it still possesses some features usually attributed only to
solutions preserving supersymmetry. In particular for the dual
solution equality between energy and charge density holds.

%%%%%%%%%%%%%%%%%%%%%%%%%%%%%%%%%%%%%%%%%%%%%%%%%%%%%%%%%%%%%%%%%
%%%%%%%%%%%%%%%%%%%%%%%%%%%%%%%%%%%%%%%%%%%%%%%%%%%%%%%%%%%%%%%%%
%%%%%%%%%%%%%%%%%%%%%%%%%%%%%%%%%%%%%%%%%%%%%%%%%%%%%%%%%%%%%%%%%
\newpage
\section{From the Standard Model and General Relativity to
superstrings.}
\label{From_SM_GR_to_branes}

\akapit
Nowadays physicists, especially those interested in
theories of elementary particles sometimes seem to be modern
incarnations of the legendary King Arthur's knights. Similarly to
the ancient warriors, they spend a lot of time and devote most of
their efforts to achieve one goal for which there is even no known
proof that it really exists. The goal however is not the Holy
Graal. The physicists are searching for something even more
miraculous -- the Theory of Everything, how is often called the
hypothetical unified theory of all interactions. The road expected
to lead to the goal is very difficult, winding and with many
branches ending as blind alleys. Fortunately there are no dragons
or other beasts lurking for inadvertent travellers. But striding
the way one can find several other creatures lying at the shoulder
of the road as a milestones of scientific progress. Creatures,
which in many cases are still not fully domesticated and which at
any time can make a surprise for the explorer. The list of the
creatures is long and contains grand unification theories,
supersymmetries and supergravities, extra dimensions, strings and
superstrings and one of the latest discovery for which this work
is dedicated -- branes. Let us look at some of their features
more closely.

%%%%%%%%%%%%%%%%%%%%%%%%%%%%%%%%%%%%%%%%%%%%%%%%%%%%%%%%%%%%%%%%%
%%%%%%%%%%%%%%%%%%%%%%%%%%%%%%%%%%%%%%%%%%%%%%%%%%%%%%%%%%%%%%%%%
\subsection{General Relativity and the Standard Model.}

\akapit
In modern physics we know four fundamental interactions
described by two completely distinct theories. One of them is
Einstein's General Relativity \cite{Weinberg1, MTW, HawkingI, Held} 
describing classically gravitational forces. The other is the 
Standard Model \cite{Glashov, SW, Weinberg2, Salam}, 
quantum field theory of electromagnetic, weak and strong interactions.

The General Relativity Theory is based on Equivalence Principle
i.e. a postulate of an invariance under general coordinate
transformations (in the usual formulation) or the local $ISO(3,1)$
or Poincar\'{e} symmetry in a tangent space (in so called
Einstein--Cartan formulation). The theory describes particles of
spin 2 (gravitons) and has a very elegant and simple structure
because its lagrangian is given just by
$\frac{1}{2\kappa}R[g]+\lambda$. One of the most significant
achievements of the theory is a direct link between physics and
geometry because the kinetic term for gravitons $R$ is
simultaneously the Ricci curvature of space-time. Therefore it
allows to interpret the gravitational forces as effects of a
non-flat space-time geometry. But the theory has a very important
defect: it is consistent only classically.

On the contrary, the Standard Model is a {\it bona fide} quantum
theory. It is of Yang-Mills \cite{YM} type based on local non-abelian gauge
symmetry group $SU(3) \times SU(2) \times U(1)_Y$ where the group
$SU(3)$ is responsible for strong interactions while electroweak
forces are related to $SU(2) \times U(1)_Y$. The particle spectrum
contains spin 1/2 fermions in different representations and spin 1
gauge bosons constituting $(8,1)$, $(1,3)$ and $(1,1)$
representations of the gauge group. The gauge bosons which are
directly responsible for a mediation of the interactions are
described as coefficients of a connection of the gauge symmetry
group. All the fermions are divided into three generations
characterized by different mass scale but filling up the identical
pattern of representations: $(3,2)_{1/6}$ for left-handed quarks,
$(1,2)_{-1/2}$ for left-handed leptons, $(3,1)_{2/3}$ and
$(3,1)_{-1/3}$ for right-handed quarks and $(1,1)_{-1}$ for
right-handed leptons (plus $(1,1)_{0}$ if we include right-handed
neutrinos). The theory is chiral i.e. a transformation
interchanging left-handed fermions with right-handed ones is not a
symmetry of the theory. The chirality forbids fermions to attain a
mass from the usual Dirac mass term: $\overline{\phi_R} m \phi_L +
\overline{\phi_L} m \phi_R$ because such a term would not be gauge
invariant. A necessary part of the model is the Higgs mechanism \cite{Higgs, Higgs2} of
a spontaneous breaking of the $SU(2) \times U(1)_Y$ symmetry into
the electromagnetic $U(1)_{em}$. This mechanism makes fermions and
three vector bosons carrying weak interactions massive but to make
the mechanism working there should necessarily exist additional,
still experimentally not observed fields -- scalar Higgs bosons.
With the exception of this point requiring confirmation, the
Standard Model is a consistent quantum theory, renormalizable,
anomaly free and last but not least confirmed in all existing
experiments with a fantastic precision.

%%%%%%%%%%%%%%%%%%%%%%%%%%%%%%%%%%%%%%%%%%%%%%%%%%%%%%%%%%%%%%%%%
%%%%%%%%%%%%%%%%%%%%%%%%%%%%%%%%%%%%%%%%%%%%%%%%%%%%%%%%%%%%%%%%%
\subsection{The need for unification}

\akapit The Standard Model and General Relativity are surprisingly
successful in prediction and description of all of the
experimentally observed physical phenomena at the fundamental
level. They have however significant theoretical disadvantages. A
validity of both of the theories is limited and both of them are
constructed under assumptions that seem arbitrary from a purely
theoretical point of view. What is more, some aspects of one of
the theories are in contradiction with the other, the most
profound example being quantum character of the Standard Model and
classical one of General Relativity. Therefore one cannot obtain a
consistent physical theory just by joining them together. The most
popular hypothesis says that the theories are only extreme limits
of some yet unknown unified theory. This theory is expected to be
unified not only in a sense of its completeness in a
phenomenological description of the Universe, but it also should
unify all the four interactions reducing them to low energy limit
symptoms of a single fundamental force. The history of physics
notes similar facts in the case of electricity and magnetism in
nineteenth century and in the case of electromagnetism and weak
interactions thirty years ago. We believe that there are no
reasons to forbid occurring it again. But before we start to
follow the way leading to the unified theory let us list the main
problems arising in the Standard Model or the General Relativity
Theory and questions which cannot be answered by these theories.

\begin{itemize}

\item
The main disadvantage of the Standard Model is its arbitrariness.
The agreement with experiments even in the simplified version with
no right--handed neutrinos requires determination of 18 seemingly
arbitrary and uncorrelated parameters (3 coupling constants, 6
masses of quarks, 3 masses of leptons, 3 quark mixing angles, one
phase and two parameters of the Higgs potential).

\item
Similarly arbitrary is a choice of the gauge group and a choice of
the fields multiplets appearing in the Model (constrained only by
vanishing of anomalies) .

\item
A next mysterious thing is a number of the fermionic generations.
We do not know why exactly three generations are observed, whether
we should expect next generations at higher energy level and why
fields in different generations have different masses.

\item
We also do not know why all known generations are chiral and
whether the chirality is a physical rule still true for the higher
generations if they exist.

\item
There is unexplained reason for which electric charge is
quantized. For the anomaly cancellation it is enough if a certain
combination of charges vanish, but the condition does not state
that, for example, the up quark charge is exactly $2/3$ of the
positron charge.

\item
In classical General Relativity problems are of different nature.
To find any solution one has to provide the sources for the
space-time curvature i.e. the energy--momentum tensor. To describe
the evolution of the Universe we have to assume that the
energy-momentum tensor is extremely unnatural - besides the well
understood matter and radiation content it has to contain the so
called cosmological constant which in comparison with any
theoretical estimates is too small by tens orders of magnitude.

\item
The only intrinsic parameter of General Relativity is $\kappa$ --
the gravitational constant. It creates even bigger problems. Since
it is dimensional it describes not only a relative strength (or
rather weakness) of the gravity compared to other interactions but
gives a specific length and energy scale called the Planck's
scale:
 \be
  l_{Planck} \simeq 10^{-35} m \simeq
  \left( 10^{-19} GeV \right)^{-1} \simeq M_{Planck}^{-1}.
 \ee
The theory with such a scale cannot be quantized with usual
methods, because when an energy of an interaction exceeds the the
Planck's mass the theory becomes strongly coupled and
nonrenormalizable. However following the example how the Fermi
theory of weak interactions was replaced by the Weinberg-Salam model
one should expect an appearance of a new quantum physics at the
Planck scale and a new theory of gravity being a generalization of
General Relativity.

\item
Possible existence of the new physics at the Planck scale
automatically causes so called hierarchy problem. Calculating
quantum corrections to masses of the low energy (electroweak
scale) particles we find that the corrections should be very
large. Barring a possibility that these masses are just fine-tuned
with extreme precision there should exist a mechanism restricting
the interaction between low energy and high energy particles and
then preventing a mixing of the both scales.

\item
At the classical level solutions of General Relativity equations
of motion contain singularities (like the famous Big-Bang
singularity). Therefore the theory predicts existence of objects
that cannot be satisfactorily described by the theory itself and
these objects may be different or even disappear in the larger
theory

\item
Although General Relativity directly connects energy and geometry
of a space-time, it does not answer a question why the observed
Universe is flat and has exactly three space and one time
direction. And indeed it is possible to define the theory in a
space-time with arbitrary dimensionality, signature and geometry.

\end{itemize}

One could idealistically think, that to define the Unified Theory
of All Interactions it should be enough to write an adequate set
of assumptions determining a consistent theory. On one hand the
number of assumptions should be as small as possible to answer a
request for "naturalness" of the theory. But on the other the
assumptions should be strong enough to allow a derivation of only
a single theory which obviously has to reduce to the Standard
Model and General Relativity in the low energy limit.
Unfortunately till today nobody has found any nontrivial and fully
consistent quantum theory let alone to write down such a set of
axioms. Some physicists even doubt the possibility of formulation
of the Unified Theory that way (if of course it could be
formulated anyhow).

But there is another way perhaps more efficient. We should extract
the principles constituting the Standard Model and General
Relativity and slightly generalize or relax one or few of them. If
after such a redefining it is not possible to construct a theory
more unified, less arbitrary and still compatible in some limit
with the starting ones, then we should step back and try with some
other set of principles. If it is possible then we should suppose
that it was a correct step towards the Unified Theory.

Let us describe some of the steps that presumably lead
towards the Unified Theory \cite{Mohapatra}.

%%%%%%%%%%%%%%%%%%%%%%%%%%%%%%%%%%%%%%%%%%%%%%%%%%%%%%%%%%%%%%%%%
%%%%%%%%%%%%%%%%%%%%%%%%%%%%%%%%%%%%%%%%%%%%%%%%%%%%%%%%%%%%%%%%%
\subsection{Grand Unification Theories. }

\akapit Investigating by the renormalization group methods a behavior of
the three coupling constants of the Standard Model one finds that
their values almost converge at some large ($\sim 10^{16}$ GeV)
energy scale. The natural hypothesis is that perhaps the three
coupling constants are replaced at this scale by a single constant
and the group product $SU(3) \otimes SU(2) \otimes U(1)_Y$ by a
single simple Lie group of larger gauge symmetry. If it really
happens a real unification of strong and electroweak interactions
is attained in the similar way as the electromagnetic and weak
interactions are unified at $\sim$ 100 GeV. The idea of a Grand
Unified Theory (GUT) \cite{PS, GG} has several attractive theoretical features
for example it can explain why the $U(1)$ charges are quantized.
Let us discuss the most attractive GUT -- $SO(10)$ theory \cite{FMin}. 
A fundamental representation of $SO(10)$ is ${\bf 16}$ and under
symmetry breaking $SO(10) \rightarrow SU(3) \otimes SU(2) \otimes
U(1)_Y$ it decomposes as \cite{Slansky}:
 \be
  {\bf 16} \rightarrow
   ({\bf 1} \otimes {\bf 1})_2 \oplus
   (\overline{\bf{3}} \otimes {\bf 1})_{-\frac43} \oplus
   ({\bf{3}} \otimes {\bf 2})_{\frac13} \oplus
   ({\bf{1}} \otimes \overline{\bf 2})_1 \oplus
   (\overline{\bf{3}} \otimes {\bf 1})_{-\frac23} \oplus
   ({\bf 1} \otimes {\bf 1})_0,
  \label{SO(10)_rozklad}
 \ee
where the numbers in subscripts denote values of the $U(1)_Y$
electroweak hypercharge. As we see the representations appearing
in the above formula precisely agree with the representations
containing spinor fields in the Standard Model (both quarks and
leptons) including the right-handed neutrino.

However GUTs still suffer most of the problems peculiar for the
Standard Model. All these theories are full of arbitrary
parameters, do not answer the questions of the number of
generations or chirality and do not propose any way to incorporate
gravity. Since a new energy scale (the grand unification scale)
appears in the theories which is many orders of magnitude higher
than the electroweak scale, the hierarchy problem is even more
pronounced. Additionally there is a lot of Lie groups containing
the Standard Model symmetry group as their subgroup and therefore
being possible bases for construction of different GUTs and we
have no clear indication how the correct one should be chosen.

%%%%%%%%%%%%%%%%%%%%%%%%%%%%%%%%%%%%%%%%%%%%%%%%%%%%%%%%%%%%%%%%%
%%%%%%%%%%%%%%%%%%%%%%%%%%%%%%%%%%%%%%%%%%%%%%%%%%%%%%%%%%%%%%%%%
\subsection{Extra dimensions.}

\akapit A proposal to unify gravity with other interactions was 
first put forward by Kaluza and Klein in 1920s 
\cite{Kaluza, Klein1, Klein2}. It consist in introducing
additional dimensions of a space-time \cite{MKKT} and interpreting
the additional geometrical symmetries of higher--dimensional
gravity as the gauge symmetries of the four--dimensional world.
This is not a new idea. If we look how electricity and magnetism
is unified in the Maxwell theory of electromagnetism we can see
that the key of the unification is to replace a three dimensional
space with separated time by a four dimensional space-time. So a
conjecture that a similar mechanism can be used in case of gravity
seems to be very plausible.

Let us start with a $D$ dimensional theory where $d$ dimensions
are infinite (or very large) and $D-d$ dimensions are small (i.e.
the inverse radius is larger than presently available energies and
therefore impossible for direct detection). In other words we
assume that at a low energy level a $D$-dimensional space-time
$\calM_D$ splits into a product $\calM_d \times \calK$, where
$\calK$ is relatively small. A natural assumption is that a length
scale of the small directions is of the order of the Planck's
scale.

With such a splitting $G_D$ -- the original symmetry group of
the general coordinate transformations defined on $\calM$ --
breaks into a product $G_d \times G_{D-d}$ of symmetries defined
on $\calM_d$ and $\calK$ respectively. Any field in an arbitrary
representation of $G_D$ decomposes into a sum of products of
representations of $G_d \times G_{D-d}$. But for a $d$-dimensional
observer who cannot excite momentum along $\calK$, group $G_{D-d}$
is seen as a gauge symmetry group and with different choices of
topology of $\calK$ we can in principle obtain any gauge symmetry
group.

From the $d$-dimensional point of view each of the resulting
fields has an additional dependence on $D-d$ continuous parameters
-- coordinates on $\calK$. If $\calK$ is a compact manifold it
leads to very interesting conclusions. As an example let us
consider a field $f(X_D)$ satisfying on $\calM$ an equation
$\calD_D f = 0$, where $\calD_D$ is an adequate wave operator. For
wave operators on products of manifolds it is possible to write
$\calD_D= \calD_d + \calD_{D-d}$ where $\calD_d$ commutes with
$\calD_{D-d}$. Therefore:
 \be
  f(x_D)=\sum_g f^{(g)} (x_d) v^{(g)} (y_{D-d}),
 \ee
where $g$ runs over some countable set. The $f^{(g)}$, $v^{(g)}$
are eigenvectors of the operators $\calD_d$ and $\calD_{D-d}$
respectively and constitute a complete, orthogonal systems. Then:
 \be
  \calD_D f (X_D) = \sum_g v^{g} (y_{D-d})
  \left( \calD_d + m^{(g)}_{D-d} \right) f^{(g)} (x_d),
 \ee
where $m^{(g)}_{D-d}$ are eigenvalues of $\calD_{D-d}$. So, in $d$
dimensions we see a tower of the so called Kaluza--Klein modes
$f^{(g)}(x_d)$ each characterized by a mass (for fermions) or mass
squared (for bosons) $m^{(g)}_{D-d}$. It is plausible to identify
stages of the tower with the subsequent generations of the fields.

It would be extremely appealing if one could justify the Standard
Model gauge group in such a purely geometrical way -- it turns out
however that it is impossible with the main obstacle being the
chirality of fermions in the Standard Model.

%But the above mechanism in its simplest version fails when it
%is applied to the problem of chirally asymmetric fermions.
%If $\calK$ is compact and there is no gauge symmetry group in $D$
%dimensions then it is impossible to obtain
% a chiral spectrum of fermion generations via the compactification method.
%So we need to relax at least one of the assumptions.
%But if gauge symmetry is present in the high dimensional theory then
%all the questions about specific
% choice of the group and its representations are back.
%On the other hand if the small space is not necessary compact then
%other additional restrictions are needed to
% preserve discrete character of the Kaluza-Klein modes.
%The eventual compactness or non-compactness of $\calK$ is of course
%only a part of a wider problem.
%If it is allowed to discuss many possible space-times $\calM$ and
%$\calK$ of various dimensionality, topology and
% geometry, why the only one choice is realized by the Nature?
%What distinguishes the one among an infinite set of possibilities?

%%%%%%%%%%%%%%%%%%%%%%%%%%%%%%%%%%%%%%%%%%%%%%%%%%%%%%%%%%%%%%%%%
%%%%%%%%%%%%%%%%%%%%%%%%%%%%%%%%%%%%%%%%%%%%%%%%%%%%%%%%%%%%%%%%%
\subsection{Supersymmetry and supergravity.}

\akapit Looking at the Standard Model one can notice a strange asymmetry
between bosons and fermions. Each kind of fields appears in the
Model in different representations of the gauge group and because
of that they play different roles. Spin 1 bosons are in the
adjoint representation of the gauge group and can be interpreted
as carriers of interactions, while spin $1/2$ fermions are in
fundamental (or trivial) representations and act only as pure
matter and there is no the slightest hint that bosons and fermions
could be connected.

In the middle of 1970s however it was realized that in string
theory the seemingly fundamental difference between bosons and
fermions is alleviated and there exists there a symmetry
connecting bosons and fermions -- it was called supersymmetry 
\cite{GolfLik, WessZ, GGRS, Nilles, Ferrara}. An
immense theoretical effort over the last 20 years was devoted to
the application of this idea for different theories and most of the
nearest future experimental efforts in particle physics is aimed
at one goal -- a discovery of the supersymmetric partners of the
Standard Model particles.

The starting point for a construction of supersymmetry group is to
assume that generators of the group are fermions, so a
supersymmetry transformation of a bosonic field is gives a fermion 
field and vice versa. In a consequence, any
supersymmetry representation contains equal number of bosonic and
fermionic degrees of freedom and all fields belonging to a single
representation have to have the same mass. The last statement
stays in contradiction with experiments because supersymmetric
partners of known particles are not observed. Therefore, if
supersymmetry is realized in nature it cannot be exact but must
be (spontaneously) broken. But even broken supersymmetry has many
interesting features \cite{FY, OR}.

The assumption of supersymmetry puts rigorous constraints on a
theory highly determining a number and a kind of terms which can
appear in its lagrangian. For example the cosmological term is
forbidden so the cosmological constant problem is less severe (by
60 orders of magnitude) bringing down its scale to the
supersymmetry breaking scale.

Thanks to the fermion-boson symmetry supersymmetric theories are
usually less divergent than their nonsupersymmetric counterparts
and can help to solve the hierarchy problem. The idea comes from
an observation that in the perturbation expansion corrections from
bosonic excitations have to be accompanied by fermionic ones and
the latter contribute with an opposite sign. As a result many of
the corrections cancel each other reducing the degree of
divergence (for scalars from the quadratic to the logarithmic).

Supersymmetry constitutes also a doorway by which gravity can be
introduced to join the other interactions. A first observation is
that the supersymmetry generators belong to an algebra that
necessarily contains the Poincar\'e algebra as its subalgebra. So
we can conjecture that there should exist a supersymmetric theory
being an extension of the general relativity. Indeed, if we want
to construct a local supersymmetry we need a spin $3/2$
vector-spinor field. They play a role of a connection of the
supersymmetry group analogously like the gauge vectors are
connections of the gauge symmetries groups. But the vector-spinors
have to be in one supermultiplet with spin 2 fields, which
naturally can be interpreted as fields carrying gravitational
interactions. This is the reason why the locally supersymmetric
theories are usually called supergravities \cite{FNF, DZumino}.

%One of the  features of supergravities is that they are
%nonrenormalizable -- if they were fundamental theories this fact
%would be disturbing but now they are always treated as just
%effective theories (which are always nonrenormalizable).

Supersymmetry has also another feature especially interesting in
an association with extra dimensional theories 
\cite{SalamS, Tanii}. Denoting by $N$ a
number of irreducible spinorial supersymmetry generators one can
talk about $N$ supersymmetries. From the four--dimensional point
of view acting with a supersymmetry generator changes a spin
projection of the field by 1/2. It is natural not to introduce
fields with spin higher than 2, because there are not known
consistent interacting quantum theories describing them so the
maximal number of four dimensional supersymmetries is $N=8$. This
leads to the conclusion that the highest number of dimensions with
supersymmetry is generally twelve and with one time direction it
is eleven (since $N=1$ in $D=11$ corresponds to $N=8$ in $D=4$).

In this way we come to the eleven dimensional supergravity
\cite{CJS}. The theory is constructed only from one supermultiplet
$(g_{MN}, \Psi_M, C_{MNR})$ where the $\Psi_M$ is a Majorana
spinor and its action is:
 \bea
  S_{D=11} &=& \int_\calM d^{11}X
   \left\{
     \sqrt{|g|}
     \left[
       R
       + \frac12 \overline{\Psi}_M
       \Gamma^{MNR} D_N \left( \frac{\omega+\hat{\omega}}{2} \right)
       \Psi_R-\frac{1}{48} H^{MNRS} H_{MNRS}
     \right.
   \right. \nn \\
   &-&
   \left.
     \left.
       \frac{1}{384}
       \left(
         \overline{\Psi}_M \Gamma^{MNRSTU} \Psi_N +
         12 \overline{\Psi}^R \Gamma^{ST} \Psi^U
       \right)
       \left(H_{RSTU}+\hat{H}_{RSTU}\right)
     \right]
   \right. \nn \\
   &+&
   \left.
     \frac{1}{144^2}\epsilon^{M_1 \ldots M_{11}} H_{M_1 \ldots M_4}
     H_{M_5 \ldots M_8} C_{M_9 \ldots M_{11}}
   \right\},
  \label{D11sugra}
 \eea
where $H=dC$ is a strength of the potential $C$, $D_M$ is a
covariant derivative, $\omega$ a spin connection and:
 \bea
  \hat{H}_{MNRS} &=& H_{MNRS}+
  \frac32 \overline{\Psi}_{[M}\Gamma_{NR} \Psi_{S]}, \\
  \hat{\omega}_{M \bar{N}\bar{R}} &=& \omega_{M \bar{N}\bar{R}}
  +\frac{1}{16} \overline{\Psi}^{S}\Gamma_{M\bar{N}\bar{R} ST} \Psi^{T}.
 \eea
The supergravity transformations laws are given by:
 \bea
  \delta_\eta \Psi_M &=&
    \left[
      D_M (\hat{\omega}) - \frac{1}{288}
      \left(
        \Gamma^{NRST} {}_M - 8 g_M^N \Gamma^{RST}
      \right)
      \hat{H}_{NRST}
    \right]
    \eta, \label{D11sugra_psi_eta} \\
  \delta_\eta e_M^{\bar{N}} &=& - \frac14 \overline{\eta}
  \Gamma^{\bar{N}} \Psi_M, \label{D11sugra_g_eta}\\
  \delta_\eta C_{MNR} &=& - \frac34 \overline{\eta}
  \Gamma_{[MN} \Psi_{R]},
 \label{D11sugra_A_eta}
 \eea
where $e_M^{\bar{N}}$ is an elfbein and $\eta$ is a spinorial
parameter of supergravity. In the above we replaced the coefficient 
$1/2\kappa$ which should accompany the curvature term $R$ with
the number $1$. The operation is allowed if we assume to work with 
such units system where $\kappa=1/2$.

It is instructive to check the number of physical degrees of
freedom of the fields. In arbitrary dimension they are given by
(the number for a Dirac spinor should be multiplied by 2 and for a
Majorana-Weyl spinor divided by 2) :
\begin{itemize}
 \item $\# g_{MN} = \frac{1}{2}(D-2)(D-1) - 1$,
 \item $\# C_{M_1 \ldots M_n} = \frac{1}{n!}(D-2)\ldots(D-n-1)$,
 \item $\# \Psi_M = (D-3)2^{[D/2]-1}$.
\end{itemize}
So, if $D=11$, a simple calculation gives $44$ degrees for the
metric tensor and $84$ for the antisymmetric rank $3$ potential.
Their sum is $128$ and is precisely equal to the number of
on-shell degrees of freedom of the gravitino.

The eleven dimensional supergravity was thought in the past as an
excellent candidate for the Unified Theory. Firstly, it is simple
and elegant. Secondly, quite natural, because it is naturally
distinguished in a set of all supersymmetry theories as its
maximal element. Moreover a compactification from eleven to four
dimensions on $\calK={\rm CP}(2) \times S^2 \times S^1$ could even
produce the gauge group of the Standard Model, but the mechanism
cannot give a chiral four dimensional theory because of lack of
gauge symmetry in the $D=11$ supergravity.

Let us describe also some lower dimensional supergravities. In ten
dimensions there are three possible theories described as: $(N_L,
N_R) = (1,1), \; (2,0), \; (1,0) $ where $N_L$ and $N_R$ count
numbers of left-handed and right-handed generators respectively.

The $(1,1)$ theory \cite{GianiP, CampbellW, HuqN} 
can be derived by a dimensional reduction form
the eleven dimensional one, since the $D=11$ Majorana spinor
splits into two Majorana-Weyl spinors of opposite chirality in
$D=10$. An action of the theory truncated to only bosonic part
reads:
 \bea
  S_{(1,1)} &=& S_{NS}+S_{R}+S_{CS}, \label{sugra_IIA_ac} \\
  S_{NS} &=& \int_\calM d^{10}X \sqrt{|g|}
  \left( R - \frac12 \de_M \phi \de^M \phi
          - \frac12 e^{-\phi} \left| H_{[3]} \right|^2 \right),
          \label{sugra_IIA_ac_NS} \\
  S_{R} &=& - \frac12 \int_\calM d^{10}X \sqrt{|g|}
  \left( e^{3\phi/2} \left| F_{[2]} \right|^2 +
              e^{\phi/2} \left| \hat{F}_{[4]} \right|^2 \right),
              \label{sugra_IIA_ac_R} \\
  S_{CS} &=& - \frac12 \int_\calM F_{[4]} \wedge F_{[4]} \wedge C_{[2]},
  \label{sugra_IIA_ac_CS}
 \eea
where $\phi$ is a scalar field and:
 \bea
  F_{[k]} &=& d A_{[k-1]} \quad \mbox{for} \quad k=2,4, \\
  H_{[3]} &=& d C_{[2]}, \\
  \hat{F}_{[4]} &=& F_{[4]} + A_{[1]} \wedge H_{[3]}.
 \eea
The terms $S_{NS}$ and $S_{R}$ describe respectively so called
Neveu-Schwarz-Neveu-Schwarz and Ramond-Ramond sectors of the
theory (the meaning of this terminology will be explained later)
and $S_{CS}$ is Chern-Simons term.

The $(2,0)$ supergravity \cite{Schwartz, HoweWest} 
is at first sight quite odd. It cannot be
derived from dimensional reduction of eleven dimensional
supergravity. Moreover it is impossible to write down a lagrangian
because the theory contains an antisymmetric field of rank 5,
which is selfdual i.e.:
 \be
   \hat{F}_{[5]} = \ast \hat{F}_{[5]}.
   \label{samodual_1}
 \ee
However, the equations of motion derived from an action with
bosonic part given by:
 \bea
  S_{(2,0)} &=& S_{NS}+S_{R}+S_{CS},
  \label{sugra_IIB_ac} \\
  S_{NS} &=& \int_\calM d^{10}X \sqrt{|g|}
  \left( R - \frac12 \de_M \phi \de^M \phi
     - \frac12 e^{-\phi} \left| H_{[3]} \right|^2 \right),
     \label{sugra_IIB_ac_NS} \\
  S_{R} &=& - \frac12 \int_\calM d^{10}X \sqrt{|g|}
  \left( e^{2\phi} \left| F_{[1]} \right|^2 +
     e^{\phi} \left| \hat{F}_{[3]} \right|^2 +
     \frac12 \left| \hat{F}_{[5]} \right|^2 \right),
     \label{sugra_IIB_ac_R} \\
  S_{CS} &=& - \frac12 \int_\calM A_{[4]} \wedge H_{[3]} \wedge F_{[3]},
  \label{sugra_IIB_ac_CS}
 \eea
 where:
 \bea
  F_{[k]} &=& d A_{[k-1]} \quad \mbox{for} \quad k=1,3,5, \\
  H_{[3]} &=& d C_{[2]}, \\
  \hat{F}_{[3]} &=& F_{[3]} - A_{[0]} \wedge H_{[3]}, \\
  \hat{F}_{[5]} &=& F_{[5]} - \frac12 A_{[2]} \wedge H_{[3]} +
  \frac12 C_{[2]} \wedge F_{[3]} 
 \eea
with (\ref{samodual_1}) as an extra condition are just the
equations of motion of the $(2,0)$ supergravity.

Setting equal to zero one of the generators of the $(1,1)$ or $(2,0)$
theories one reduces them to the $(1,0)$ theory \cite{BRWN, ChaplineM}. 
The bosonic part of its action is:
 \be
  S_{(1,0)} = \int_\calM d^{10}X \sqrt{|g|}
    \left(R - \frac12 \de_M \phi \de^M \phi -
    \frac12 e^{-\phi} \left| F_{[3]} \right|^2 \right).
    \label{sugra_I_ac} \\
 \ee

There one more supergravity known in ten dimensions being an
extension of $(1,1)$ theory, called massive supergravity or Romans \cite{Romans}
theory. An action of the theory can be obtained from the action
$S_{(1,1)}$ given in (\ref{sugra_IIA_ac}) by adding a scalar field
$M$ and a $10$-form $F_{[10]}$ in the following way:
 \be
  S_{(1,1)massive}= \tilde{S}_{(1,1)}-
  \int_\calM \frac12 d^{10}X \sqrt{|g|} e^{5/2 \phi} M^2 +
  \int_\calM M F_{[10]},
   \label{sugra_IIAmass_ac}
 \ee
where $\tilde{S}_{(1,1)}$ is the same as (\ref{sugra_IIA_ac})
after the substitution:
 \bea
  F_{[2]} &\rightarrow& F_{[2]} + M C_{[2]}, \\
  F_{[4]} &\rightarrow& F_{[4]} + \frac12 M C_{[2]} \wedge C_{[2]}.
 \eea
Field $M$ is as an auxiliary one and can be integrated over in the
action giving quite complicated combination of other fields.

As we see in 11 dimensions a supersymmetric theory is necessarily
a supergravity but in 10 dimensions a supersymmetry generator can
be a Majorana--Weyl spinor leading to a gauge supermultiplet (with
fields of spin not higher than 1). There exists a rule that chiral
Yang-Mills and simultaneously supersymmetric theories are allowed
only if $N=1$. In such a case nothing prevents the $(1,0)$
supergravity for coupling to matter which is supersymmetric, gauge
symmetric but chirally asymmetric \cite{ChaplineM}. Such theories can be
compactified to a chiral, gauge symmetric lower dimensional
theory, if the supersymmetry preserved in the target space is not
higher then $N=1$. In the case of the $(1,0)$ supergravity the
required compactification scheme is obtained if $\calK$ is
Calabi-Yau manifold (i.e. a complex manifold of complex dimension
$3$ and with $SU(3)$ holonomy group \cite{Hubsch}). The reasoning is as follows:
we have to compactify 6 dimensions; the holonomy group in 6
dimensions is $SO(6)$ which is locally isomorphic to $SU(4)$. If
the holonomy of the manifold is $SU(3)$ (i.e. it is a Calabi-Yau
manifold) then the holonomy breaks three out of four
supersymmetries which can emerge after compactification from $10$
to $4$ dimensions. Simultaneously we can identify a subgroup of
the gauge group with the holonomy group so it gives a mechanism of
gauge symmetry breaking. For example starting with $E(8)$ gauge
symmetry, which is a natural candidate because of reasons
described later, one obtains:
 \be
  E(8) \rightarrow E(6) \otimes SU(3) \rightarrow E(6)
 \ee
 and afterwards:
 \be
  E(6) \rightarrow SO(10) \otimes U(1).
 \ee
But $SO(10)$ is one of the most suitable GUT theory thanks to the
behavior of its multiplets under reduction to $SU(3) \otimes SU(2)
\otimes U(1)$, (see \ref{SO(10)_rozklad}).

In spite of their attractive features supergravities have a major
drawback -- they are nonrenormalizable. They are consistent only
at a classical level but not at the quantum one. But even if they
are not the final goal one is tempted to think that they indicate
the correct way toward the Unified Theory.

%%%%%%%%%%%%%%%%%%%%%%%%%%%%%%%%%%%%%%%%%%%%%%%%%%%%%%%%%%%%%%%%%
%%%%%%%%%%%%%%%%%%%%%%%%%%%%%%%%%%%%%%%%%%%%%%%%%%%%%%%%%%%%%%%%%
\subsection{Bosonic strings.}

\akapit The result of search for the unification of all interactions
summarized in the previous section can seem quite
disappointing. Even if we recovered an interesting idea that could
lead to unification of gravity with other fundamental forces it
turned out that a theory incorporating the idea was still
nonrenormalizable. But for now we were considering only theories
where the fundamental objects were defined as point particles.
What if the point-like structure is only a simplification, the
particles can be extended objects and at some scale (for example
the Planck's scale) it is necessary to take into account this
fact? So, let us describe a theory with points replaced by strings
\cite{GSW, Kaku, Pol1}. Then the Feynman diagrams, which in the
case of point particles were networks of crossing worldlines now
become homomorphic to two dimensional manifolds. But at any order
of the perturbation expansion a number of topologically inequivalent
two-folds is significantly smaller than a number of
topologically inequivalent line-like diagrams, so the theories of
strings at first sight seem to be more convergent than the point
particles' ones. And really they are. They are only known examples
of consistent interacting quantum theories which include both
gauge interactions and gravity, they are expected to be finite at every level of
perturbation expansion and they are not anomalous. But this is not
the only advantage of the string theories. The other include:

\begin{itemize}
\item
They contain only two dimensionful parameters -- speed of light
$c$ and the string scale $\la_s$.
\item
They can be constructed only in a specific dimension of a spacetime.
\item
Fields of various spins and masses are quantum excitations of a
single string.
\item
At the low energy limit string theories reduce to supergravities
possibly coupled to super Yang Mills.
\item
There is only a small number of inequivalent string theories and
connections among them (dualities) suggest that they are all
only special cases of one theory (M--theory).
\end{itemize}

Let us describe in more detail a construction of the string theories.
It is natural to start with an analog
of a relativistic point particle action $S=-m\int ds$ (so called
Nambu-Goto action \cite{Nambu,Goto}):
 \be
  S_{NG} = -\frac{1}{2\pi\alpha'} \int d^2 \xi
   \sqrt{ \left|\det \left( \de_a X^M \de_b X^N \eta_{MN} \right)\right|},
   \label{NG_str_action}
 \ee
where $X^M(\xi^a)$ are space-time coordinates giving a
localization of the string worldsheet with respect to two
parameters: timelike $\xi^1$ and spacelike $\xi^2$, indices $a,b$
run over values $1$ and $2$. The $\alpha'$ is a string coupling
constant and it is related to a string tension by
$T=1/(2\pi\alpha')=1/\la_s^2$. Because equations of motion derived
from the action (\ref{NG_str_action}) could be equivalently
obtained from another action which is free of roots of $X^\mu$, it
is more convenient to construct string theories on a base of so
called Polyakov action \cite{Polyakov}:
 \be
  S_P = -\frac{1}{4\pi\alpha'} \int d \xi^2 \sqrt{|\gamma|}
  \gamma^{ab} \de_a X^M \de_b X^N \eta_{MN}.
  \label{P_str_action}
 \ee
In the above we introduced a worldsheet metric $\gamma_{ab}(\xi)$.
The action is invariant under general transformations of
coordinates $\xi$ and global Poincar\'{e} transformations in
space-time. Additionally it exhibits local Weyl symmetry given by:
 \bea
  {X'}^M(\xi) &=& X^M(\xi), \label{Weyl_inv_1} \\
  {\gamma'}_{ab}(\xi) &=& e^{2\omega(\xi)}\gamma_{ab}(\xi).
  \label{Weyl_inv_2}
 \eea
This is an important fact, because three parameters of the local
symmetries exactly agree with a number of $\gamma_{ab}$ degrees of
freedom. Therefore the field describing internal structure of a
string can be completely gauged out from the physical theory and
even in the quantum theory we have well defined distances on the
world-sheet. Thanks to (\ref{Weyl_inv_1} - \ref{Weyl_inv_2}) a worldsheet of a
propagating string can be described not only as a general two
dimensional real manifold but also as a Riemann surface, it means
as a complex one-fold. This is a starting point for developing so
called conformal field theory (CFT) which gives an apparatus
allowing to evaluate an amplitude for any scattering process in
the string theory. In such a description there are deep subtleties
in the Wick rotation from the physical Minkowski to the Euclidean
signature on the world sheet - it is a beautiful mathematical
result that such a rotation can be done and that the amplitudes
coincide. This result justifies the common approach to string
theory by the powerful formalism of conformal field theory.

One distinguishes several categories of strings characterized as
open or closed and oriented or unoriented. The open ones are
homeomorphic to an open interval and have to satisfy the following
boundary conditions:
 \be
  \frac{\de}{\de \xi^2} X^M (\xi^1,0) = 0 =
  \frac{\de}{\de \xi^2} X^M (\xi^1,l),
  \label{b_c_open}
 \ee
where $\xi^2=0$ and $\xi^2=l$ describe free ends of the string.
Analogously the closed ones are homeomorphic to a circle, so the
points $\xi^2=0$ and $\xi^2=l$ should be identified in this case
and the boundary conditions are:
 \bea
  X^M (\xi^1,\xi^2) &=&  X^M (\xi^1,\xi^2+l),
  \label{b_c_closed1} \\
  \frac{\de}{\de \xi^2} X^M (\xi^1,\xi^2) &=&
  \frac{\de}{\de \xi^2} X^M (\xi^1,\xi^2+l).
  \label{b_c_closed2}
 \eea
The open string conditions are more restrictive. While on the
closed strings there can live two infinite series of quantum
excitations related to left and right moving waves
 \be
  X^M(\xi^1,\xi^2) = X^M_L(\xi^2 - \xi^1) + X^{\mu}_R(\xi^2 + \xi^1)
 \ee
where $X^M_L$ and $X^M_R$ are independent,  on the open strings
they are combined into one series of stable waves.

The oriented and the unoriented strings are defined as strings
with worldsheets of respectively oriented or unoriented manifolds.
The orientability is a global feature of manifolds and in the
interacting theory all the string worldsheets are joined and form
a single connected manifold. So, it leads to a conclusion that the
oriented and the unoriented strings can interact only within their
classes. By use of a purely topological arguments it can be
checked that an interaction among open strings always can produce
a closed one, while for the closed strings it is possible to
impose consistent restrictions forbidding them to interact with
the open ones. Therefore there are four possible kinds of
interacting string theories:
\begin{itemize}
 \item theories with only closed oriented strings,
 \item theories with only closed unoriented strings,
 \item theories with open and closed strings, all oriented,
 \item theories with open and closed strings, all unoriented.
\end{itemize}

Quantizing the theory derived from (\ref{P_str_action}) with an
assumption that it should preserve Lorentz invariance one obtains
a spectrum of quantum states with masses given by:
 \be
   m^2 = \frac{s^2}{\alpha'} \left(N+\frac{2-D}{24} \right),
   \label{s_spectrum}
 \ee
where $D$ is a dimension of space-time, $N$ enumerates levels of
excitations, and $s$ is a number equal to $1$ in the case of open
strings and $2$ in the case of closed strings. Counting number of
states at each level and checking their behavior under Lorentz
transformations one finds that only $N=1$ can form massless
representations in $D$ dimensional space-time. Simultaneously,
from the condition $m^2=0$ for $N=1$ one obtains $D=26$ as a
critical dimension i.e. the number of dimensions where strings can
live. For the oriented open strings the massless representation is
a vector and for the oriented closed strings it is a
multiplication of two vectors which decomposes into irreducible
rank $2$ tensors: a traceless symmetric tensor, an antisymmetric
tensor and a scalar. The condition of unorientability reduces a
number of possible massless fields disallowing vectors and
antisymmetric tensors.

Consider an interaction among several external open strings.
Because ends of the external strings are distinguished points on a
worldsheet, each possible interaction can be characterized by a
specific order of the points and this order is invariant under the
worldsheet reparametrization. This means that the open strings in
a distinction to the closed ones have additional degrees of freedom
attached to their endpoints. They are known as Chan-Patton degrees
of freedom and can be described in terms of gauge symmetry \cite{ChanPatton}. More
detailed analysis shows that possible gauge groups are $U(N)$ in
the case of oriented string and $SO(N)$ or $Sp(N)$ in the case of
unoriented strings.

Until now we have discussed strings in flat space-time but it is
possible to consider strings propagating in other backgrounds. A
first step of such an extension is to replace the constant flat
metric $\eta_{MN}$ from (\ref{P_str_action}) by a general
coordinate dependent metric $g_{MN}(X)$ and add couplings to other
massless states of the oriented closed string: the antisymmetric
tensor $C_{MN}(X)$ and scalar $\phi(X)$. Then one obtains a
nonlinear sigma model action which takes the form:
 \be
  S_P = -\frac{1}{4\pi\alpha'} \int d^2 \xi \sqrt{|\gamma|} \left[
   \left( \gamma^{ab} g_{MN}(X)\! +\! i \epsilon^{ab} C_{MN} (X) \right)
   \de_a X^M \de_b X^N
   \!\!+\! \alpha' R[\gamma] \phi(X) \right].
 \ee
The above theory is a renormalizable theory of fields $X^M (\xi)$.
It is consistent only if Weyl invariance is preserved on a quantum
level, it means when the following conditions leading to a
cancellation of Weyl anomaly are satisfied:
 \bea
  0 &=& \alpha' \left( R[g]_{MN} + 2 \nabla_M \nabla_N \phi -
  \frac{1}{4} H_{MNR}H^{MNR} \right)
       + O({\alpha'}^2), \\
  0 &=& \alpha' \left( -\frac{1}{2} \nabla^M H_{MNR} +
  \nabla_M \phi H^{MNR} \right) + O({\alpha'}^2), \\
  0 &=& \alpha' \left( -\frac{1}{2} \nabla^2 \phi +
  \nabla^M \phi \nabla_M \phi
        - \frac{1}{24} H_{MNR} H^{MNR} \right) + O({\alpha'}^2),
 \eea
where $H=dC$. These conditions look like equations of motion of
some theory and indeed it is possible to find an action from which
they can be derived:
 \be
  S_{string} =  \int d^{26}x \sqrt{|g|} \; e^{-2\phi}
  \left[ R[g] - \frac{1}{12} H_{MNR}H^{MNR}
   + 4 \de_M \phi \de^M \phi + O(\alpha') \right].
   \label{lea_boson_str_frame}
 \ee
In this way one arrives at an effective theory describing massless
modes of the closed oriented string. The above action is written
in a formalism known as a string frame, in which the curvature
term is given by $\sqrt{|g|} \; e^{-2\phi} R[g]$.
By a redefinition of fields:
 \bea
  g_{MN} &\rightarrow& \exp\left(-\frac{\tilde{\phi}}{\sqrt{2(D-2)}} \right)
  \tilde{g}_{MN}, \\
  \phi &\rightarrow& \sqrt{\frac{D-2}{8}} \; \tilde{\phi}
 \eea
it can be transformed to so called Einstein frame where the
curvature term is $\sqrt{|\tilde{g}|} \; R[\tilde{g}]$ and the
whole action:
 \be
  S_{Einstein} =  \int d^{26}x \sqrt{|\tilde{g}|} \;
  \left[ R[\tilde{g}]
 - \frac{1}{12} e^{-\tilde{\phi}/\sqrt{3}} \tilde{H}_{MNR}\tilde{H}^{MNR}
 + \frac{1}{2} \de_M \tilde{\phi} \de^M \tilde{\phi} + O(\alpha') \right].
   \label{lea_boson}
 \ee
However the theory (\ref{lea_boson}) is obviously not a good
candidate for an effective string theory being simultaneously a
generalization of the Standard Model and General Relativity,
because it does not contain fermions. Another important
disadvantage is that the fields at the lowest level in the
spectrum given by (\ref{s_spectrum}) are not massless but
tachyonic with a negative mass square described by $N=0$. So, the
interacting theory cannot be stable since any excited state
including the massless ones should decay into tachyons. It was
conjectured that the presence of the tachyons is a consequence of
a wrong vacuum choice and there should be a mechanism shifting the
theory to the correct vacuum similarly as the Higgs mechanism
makes it with the Standard Model. But this idea did not give a
satisfactory result.

%%%%%%%%%%%%%%%%%%%%%%%%%%%%%%%%%%%%%%%%%%%%%%%%%%%%%%%%%%%%%%%%%
%%%%%%%%%%%%%%%%%%%%%%%%%%%%%%%%%%%%%%%%%%%%%%%%%%%%%%%%%%%%%%%%%
\subsection{Superstrings and M-theory.}

\akapit Fortunately there is another method avoiding the
shortcomings of the bosonic string theory described above but
saving its virtues. The key idea is to apply supersymmetry to
strings and therefore introduce so called superstrings. There are
two variants of the theory, each more convenient in different but
complementary aspects. Fortunately both lead to at least partially
equivalent results.

The first one is known as spacetime supersymmetry or
Green-Schwarz theory \cite{GreenS1, GreenS2, GreenS3}. 
In this case derivatives $\de_a X^M$ in
(\ref{P_str_action}) are replaced by:
 \be
  \Pi^M_a = \de_a X^M - i \overline{\theta}^\alpha \gamma^M
  \de_a \theta^\alpha
  \label{super_co}
 \ee
 and the resulting action is:
 \be
  S_{GS} = -\frac{1}{4\pi\alpha'} \int d \xi^2 \sqrt{|\gamma|}
  \gamma^{ab} \Pi_a^M \Pi_b^N \eta_{MN}.
  \label{GS_str_action}
 \ee
where $\theta$ is a spinor in $D$ dimensions, $\alpha=1, \ldots,
N$ with $N$ giving a number of global supersymmetries. The
expression (\ref{super_co}) is invariant under transformations of
the supersymmetries:
 \be
  \delta_\epsilon \theta^\alpha = \epsilon^\alpha,
  \qquad
  \delta_\epsilon \overline{\theta}^\alpha = \overline{\epsilon}^\alpha,
  \qquad
  \delta_\epsilon X^M = i\overline{\epsilon}^\alpha\Gamma^M \theta^\alpha.
 \ee
However it can be checked, that the $\theta^\alpha$ fields have
twice too many components than can be determined by solving the
equations of motion of the theory. So, the additional symmetry is
needed to gauge away the undesired degrees of freedom and possibly
redefining (\ref{GS_str_action}) to incorporate new terms allowing
a closure of the action under the new symmetry.

To write the appropriate action it is convenient to introduce
a superspace formalism with supercoordinates: $Z^{\bf M} = (X^M,
\theta^\alpha)$ and a supervielbein $E_{\bf M}^{\overline{\bf
M}}$ where the indices $\overline{\bf M} = (\overline{M}, \alpha)$
label tangent space coordinates. Define $E_a^{\overline{\bf
M}}=\de_a Z^{\bf M} E_{\bf M}^{\overline{\bf M}}$ and then the
action reads:
 \bea
  S_{GS} = \frac{1}{4\pi\alpha'} \int d^d \xi \left(
   - \frac{1}{2} \sqrt{|\gamma|}
   \gamma^{ab} E_a^{\overline{M}} E_b^{\overline{N}}
   \eta_{\overline{M}\overline{N}} +
   \frac{1}{d!} \epsilon^{ab} E_a^{\overline{\bf M}}
   E_b^{\overline{\bf N}} B_{\overline{\bf N} \overline{\bf M}}
   \right).
  \label{GS+WZ_str_action}
 \eea
The action is invariant under local fermionic transformations
called $\kappa$ symmetry:
 \be
  \delta_\kappa Z^{\bf M} E_{\bf M}^{\overline{M}} = 0
  \qquad
  \delta_\kappa Z^{\bf M} E_{\bf M}^{\alpha} =
  (1+\Gamma)^\alpha_\beta \kappa^\beta(\xi),
 \ee
where
 \be
  \Gamma^\alpha_\beta
   = \frac{-i}{2!\sqrt{|\gamma}|} \epsilon^{ab}
   E_a^{\overline{M}}E_b^{\overline{N}}
    (\Gamma_{\overline{M}\overline{N}})^\alpha_\beta.
 \ee
It is important that the action (\ref{GS+WZ_str_action}) has to
incorporate the Wess-Zumino term with antisymmetric tensor to
achieve $\kappa$ invariance. But on the other hand the action is
not supersymmetric for arbitrary $N$ and $D$ as 
(\ref{GS_str_action}) is. It can be checked that such a situation
occurs only for $N\leq 2$ and $D=3,4,6,10$.

A quantization program for the Green-Schwarz theory encounters
serious difficulties and it was carried out only in the light-cone
gauge. But there is another supersymmetrisation method for
strings, more convenient for the quantization but with the
manifest space-time supersymmetry lost. The method is known as
world-sheet supersymmetry or Ramond-Neveu-Schwarz theory 
\cite{Ramond, NeveuS, GSO1, GSO2}. The idea
in this case is to add to (\ref{P_str_action}) fields $\psi^M$
being Majorana spinors on the worldsheet but vectors in the
spacetime:
 \be
  S_{world \; sheet \; SUSY} =
   -\frac{1}{4\pi\alpha'} \int d \xi^2 \sqrt{|\gamma|}
   \gamma^{ab} \left(
   \de_a X^M \de_b X^N -
   i \overline{\psi}^M \gamma_a \de_b \psi^N \right) \eta_{MN}.
 \ee
Then, for any given $M$ the pairs $(X^M, \psi^M)$ are scalar
supermultiplets of $N=1$ supersymmetry. Quantizing the theory one
obtains a critical dimension $D=10$. It is also possible to
consider extended $N>1$ supersymmetries. But it occurs that for
$N=2$ the critical dimension is $D=2$ what is unreasonable for a
theory with eventual applications but can be interesting as a
playground for theoretical experiments. For $N>2$ the critical
dimension is negative what is obviously unacceptable.

The spinors living on a string worldsheet have to obey one of two
possible boundary conditions, known as Ramond (R) and
Neveu-Schwarz (NS) ones:
 \be
  \begin{array}{rcll}
   \psi(\xi^1+l,\xi^2) &=& + \psi(\xi^1,\xi^2) & \quad \mbox{R}, \\
   \psi(\xi^1+l,\xi^2) &=& - \psi(\xi^1,\xi^2) & \quad \mbox{NS}, \\
  \end{array}
 \label{RNS_cond}
 \ee
what gives two kinds of quantum states. Additionally all quantum
states can be divided with respect to the worldsheet fermion
number operator $(-1)^F$ being an extension of the chirality
operator with two eigenvalues $+1$ and $-1$. The projection from
the space of all states onto states of given fermion number is
called the GSO (Gliozzi-Scherk-Olive) projection.

Identically as for the bosonic strings, on the open superstrings
only stable waves can exist, so the whole quantum spectrum can be
classified by four sectors labelled as R+, R-, NS+ and
NS-. The lowest state in NS- is a tachyon, but in NS+, R-,
R+ it is a massless vector and two massless Majorana-Weyl
spinors with opposite chirality respectively. On the closed
superstrings left and right moving waves are independent of each
other, so the spectrum in this case is given by sectors described
as pairs of the open superstrings sectors. Total number of the
closed superstring sectors is $10$ and not $16$ because
combinations of NS(-) with the other possibilities are forbidden
by the quantum level matching rule. The rule says that tachyons
are only in (NS-,NS-) sector and the lowest states in the
remaining sectors are massless. Massless fermionic states are
contained in the sectors labelled by NS+ accompanied by R+ or
R-, while the sectors with massless bosonic states are described
by (NS+,NS+) and all possible parings among R+'s and R-'s.
In short all massless bosons can be classified as NS-NS or R-R.

Each choice of several sectors can lead to a different superstring
theory. A number of such choices is possibly very huge. For
example in a case of closed superstrings it is equal to $2^{10}$.
Fortunately only a few choices give a consistent interacting
theory with tachyon free and supersymmetric spectrum.

\vskip 2 em

{\bf Type IIA and IIB superstrings.}

\vskip 1em

Those are closed oriented superstrings theories with two supersymmetries.
They are constructed of the following sectors:
 \bea
  IIA \qquad (NS+,NS+) \quad (R+,NS+) \quad (NS+,R-) \quad (R+,R-), \\
  IIB \qquad (NS+,NS+) \quad (R+,NS+) \quad (NS+,R+) \quad (R+,R+).
 \eea
In the above $R+$ can be replaced by $R-$, what changes chirality
of all fermions, but leads to equivalent physical theories. What
is important, in the type IIA theory left and right moving
massless fermion states have opposite chirality but in the type
IIB theory all fermions have the same chirality. In the language
of the GSO projection for the IIB theory the same GSO projection is
chosen for both left and right moving states, while for the IIA
the opposite ones.

By the low energy limiting procedure analogous to that described
for bosonic string one can derive effective theories for massless
states of type IIA and IIB superstrings and find that they
coincides respectively with the $(1,1)$ and $(2,0)$ supergravities
in ten dimensions. Because massless bosonic states in the
superstring theory belong to two disjoint sectors R-R or NS-NS
the corresponding fields in the supergravity theory can be
described in the same way. This is an explanation of the
classification given in (\ref{sugra_IIA_ac}) and (\ref{sugra_IIB_ac}).

\vskip 2 em

{\bf Type I $SO(32)$ superstrings.}

\vskip 1 em

This is a theory of unoriented closed and open strings and
possesses only one supersymmetry. It can be obtained form the IIB
theory by the requirement of unorientability and removing states
not satisfying this condition. Particularly from the two sectors
containing massless fermions only one linear combination survives:
(NS,R)+(R,NS). But the theory is inconsistent unless open string
states NS+ and R+ are also included. The additional states
have Chan-Patton degrees of freedom, so the theory is gauge
symmetric with the gauge group $SO(N)$ or $Sp(N)$. The requirement
of vanishing anomaly further constrains the gauge group to only
$SO(32)$. In the low energy limit the theory converges into the
$(1,0)$ supersymmetry coupled with $SO(32)$ super-Yang-Mills.

\vskip 2 em

{\bf Heterotic $E(8) \otimes E(8)$ and $SO(32)$ superstrings.}

\vskip 1 em

Another possibility to construct a superstring theory is to
combine the superstring constraints on a half of the closed
strings states (say: on the right-moving) and the bosonic string
constraints on the second half (left-moving) in a way which leads
to removing tachyons from the physical spectrum. Because
superstring lives in ten dimensions and the bosonic string in 26,
the additional 16 dimensions of bosonic string should be
compactified. This process leads to a gauge symmetry. Only two
choices of the gauge group are acceptable to achieve a consistent
theory with tachyon free and supersymmetric spectrum: $E(8)
\otimes E(8)$ and $SO(32)$. In the low energy limit these theories
become the $(1,0)$ supersymmetry coupled with $E(8) \otimes E(8)$
or $SO(32)$ super-Yang-Mills respectively. The massless spectrum
of type I and heterotic $SO(32)$ theories coincide, however they
are different if the massive states are taken into account.

\vskip 2 em

In this way one finally obtains exactly five consistent string
theories which satisfy conditions requiring supersymmetric and
tachyon free spectrum. If the conditions are relaxed, other string
theories can be introduced as type $0$ theory or a few heterotic
theories with gauge groups other than the described before. In
most cases they are simultaneously not supersymmetric and
tachyonic, but there is one exception: the heterotic $SO(16) \otimes
SO(16)$ superstring which is not supersymmetric but still
tachyon-free.

An extensive research on string theories has recently shown, that
these five consistent string theories are all connected by a net
of dualities. There are two kinds of such dualities. One is
T-duality \cite{GPR}, which is perturbative, what means that it
works precisely at every level of perturbative expansion.
T-duality connects theories compactified on n-torus and is
described in general by $O(n,n,{\bf \rm Z})$ group. In the
simplest case it connects a theory compactified on a circle of
radius $R$ with another theory compactified on a circle of radius
$R'=\alpha'/R$. The second kind of duality: S-duality \cite{FILQ, Rey}
is nonperturbative and it is described by $SL(2,{\rm \bf Z})$
group. In particular it establishes relations between weakly and
strongly coupled limits of two theories. Let us list examples of
the dualities:
\begin{itemize}
\item
Heterotic $E(8)\otimes E(8)$ theory is T-dual to heterotic
$SO(32)$ theory.
\item
Type IIA and type IIB theory compactified on odd dimensional tori
are T-dual.
\item
Type IIA and type IIB theories compactified on even dimensional
torus are T-selfdual.
\item
Type I $SO(32)$ theory compactified on odd dimensional torus is
T-dual to type II A theory.
\item
Type I $SO(32)$ theory compactified on even dimensional torus is
T-dual to type II B theory.
\item
Type IIB theory is S-selfdual,
\item
Type I $SO(32)$ and heterotic $SO(32)$ theories are S-dual.
\end{itemize}
In the above no S-dual partners for type IIA and heterotic
$E(8)\otimes E(8)$ superstrings are shown since these cases need a
special treatment. Applying S-duality to these theories one finds
that their duals do not coincide with any known superstring
theory. Moreover the dual theories seem to live in 11 rather than
10 dimensions. Thanks to duality between type IIA and heterotic
$E(8)\otimes E(8)$ superstrings one deduces that their S-duals
have to be the same theory for which the name was coined:
M--theory. However, besides a name the theory is generally
unknown. We can describe only several features of it based on the
S-duality relation with superstrings and conjecture that its low
energy limit should be the only allowed 11 dimensional
supergravity (\ref{D11sugra}). So we can add to the previous list:
\begin{itemize}
\item
Type IIA theory is S-dual to M-theory compactified on a circle
$S^1$.
\item
Heterotic $E(8)\otimes E(8)$ theory is S-dual to M-theory
compactified on an orbifold $S^1/{\bf \rm Z_2}$.
\end{itemize}

Existence of the net of dualities among all known superstring
theories leads naturally to the conclusion that all these theories
are only specific sectors of some really unified, probably unique
theory. The theory obviously should be at least 11 dimensional and
is usually called M-theory \cite{HullT, Witten, Schwarz, Duff1} 
identically as the previously
introduced dual partner for type IIA and heterotic $E(8)\otimes
E(8)$ superstrings. In this picture S and T dualities should be
understood as symmetries of M-theory transforming one of its
sectors into another. There is a conjecture that S and T dualities
are only subgroups of more general symmetry group of the whole
theory called U-duality.

%%%%%%%%%%%%%%%%%%%%%%%%%%%%%%%%%%%%%%%%%%%%%%%%%%%%%%%%%%%%%%%%%
%%%%%%%%%%%%%%%%%%%%%%%%%%%%%%%%%%%%%%%%%%%%%%%%%%%%%%%%%%%%%%%%%
%%%%%%%%%%%%%%%%%%%%%%%%%%%%%%%%%%%%%%%%%%%%%%%%%%%%%%%%%%%%%%%%%
\newpage
\section{Branes in quantum theories.}
\label{branes_quantum}

\akapit If it is possible to construct string theories one could
wonder why do not introduce a theory of higher dimensional objects
like membranes? Such an idea is not a new one. In 1962 Dirac
proposed a model where the elementary particles were described in
terms of modes of a vibrating membrane \cite{Dirac}. The membrane
Dirac theory was based on a action which was a straight analog of
the Nambu-Goto string action (\ref{NG_str_action}). Curiously,
that Nambu and Goto have written their actions a few years after
Dirac. But during the next year, when the string theory was
rapidly developing, attempts to develop the competitive idea of
membranes ended without success. The situation has changed in 1986
when a supersymmetric membrane was discovered by Hughes, Lu and
Polchinski \cite{HLP}. But the real breakthrough was paradoxically
made on a ground of the string theory, when Polchinski \cite{Pol2}
has shown that superstrings necessarily have branes in the
spectrum as sources of Ramond-Ramond charges in the theory.

%%%%%%%%%%%%%%%%%%%%%%%%%%%%%%%%%%%%%%%%%%%%%%%%%%%%%%%%%%%%%%%%%
%%%%%%%%%%%%%%%%%%%%%%%%%%%%%%%%%%%%%%%%%%%%%%%%%%%%%%%%%%%%%%%%%
\subsection{Fundamental $p$-branes.}

\label{p_branes}

\akapit Constructing the brane theory it is natural to follow the
derivation of string theory. One should then start with the
Nambu--Goto-like action:
 \be
  S_{NG} = -T_d \int d^d \xi
  \sqrt{ \left| \det \left( \de_a X^M \de_b X_N \eta_{MN} \right)\right|}
  \label{pbrane_NG_action} \\
 \ee
or the Polyakov-like action \cite{Polyakov, HT} of a
$p$-dimensional object sweeping out in space-time a $d=p+1$
dimensional worldvolume:
 \be
  S_{P} = -\frac{T_d}{2} \int d^d \xi \sqrt{|\gamma|}
  \left(\gamma^{ab} \de_a X^M \de_b X^N \eta_{MN} - (d-2) \right).
   \label{pbrane_P_action}
 \ee
The extended objects that after supersymmetrization and
quantization would be described by such a theory are known as
$p$-branes. The theory of superstrings should appear in this
picture as a theory of $1$-branes and be only a specific case of
the theory. However it turns out that the superstring $p=1$ case
is specific and very difficult to generalize to arbitrary $p$. A
first evidence of the difficulties is the worldvolume cosmological
term $\sqrt{|\gamma|}(d-2)$ in (\ref{pbrane_P_action}) which
vanishes for strings but survives and breaks Weyl symmetry 
(\ref{Weyl_inv_1} - \ref{Weyl_inv_2}) in other cases.

As for string theory there are two ways of introducing
supersymmetry -- by imposing worldvolume supersymmetry and further
requirement of spacetime supersymmetry (a spinning brane) or by
imposing spacetime supersymmetry from the very beginning
(Green-Schwarz construction).

In the case of a spinning brane \cite{HT, BST1} the presence of the worldvolume
cosmological term precludes starting from (\ref{pbrane_P_action})
(so called no-go theorem for spinning membranes). The modified
action can be introduced \cite{DDI} which exhibits Weyl invariance and is
classically still equivalent to (\ref{pbrane_NG_action}) and
(\ref{pbrane_P_action}):
 \be
  S_{W} = -T_d \int d^d \xi \sqrt{|\gamma|}
   \left( \frac{1}{d}\gamma^{ab} \de_a X^M \de_b X^N \eta_{MN} \right)^{d/2}.
   \label{pbrane_W_action}
 \ee
But even this Weyl invariant action does not lead to a fully
successful theory of a spinning membrane.

The second possibility of supersymmetrisation is the Green-Schwarz
spacetime supersymmetry \cite{HLP, BST2}. This construction is an extension of 
(\ref{GS+WZ_str_action}), so similarily as in that case there should be
introduced the supercoordinates $Z^{\bf M} = (X^M, \theta^\alpha)$,
the supervielbein $E_{\bf M}^{\overline{\bf M}}$ with
$\overline{\bf M} = (\overline{M}, \alpha)$
labeling tangent space coordinates and $E_a^{\overline{\bf
M}}=\de_a Z^{\bf M} E_{\bf M}^{\overline{\bf M}}$.
Then the action is:
 \be
  S_{GS} = T_d \int d^d \xi \left[ \frac{\sqrt{|\gamma|}}{2}  \left(
   - \gamma^{ab} E_a^{\overline{M}} E_b^{\overline{N}}
   \eta_{\overline{M}\overline{N}}\! +\! (d\!-\!2) \right)
   + \frac{1}{d!} \epsilon^{a_1 \ldots a_d} E_{a_1}^{\overline{\bf M}_1}
   \!\!\ldots E_{a_d}^{\overline{\bf M}_d}
     C_{\overline{\bf M}_d \ldots \overline{\bf M}_1 }
   \right]
  \label{pbrane_GS+WZ_action}
 \ee
and it is invariant under $\kappa$ symmetry:
 \be
  \delta_\kappa Z^{\bf M} E_{\bf M}^{\overline{M}} = 0,
  \qquad
  \delta_\kappa Z^{\bf M} E_{\bf M}^{\alpha} =
  (1+\Gamma)^\alpha_\beta \kappa^\beta(\xi),
 \ee
 where
 \be
  \Gamma^\alpha_\beta
   = \frac{(-1)^{d(d-3)/4}}{d!\sqrt{|\gamma}|} \epsilon^{a_1 \ldots a_d}
   E_{a_1}^{\overline{\bf M}_1} \ldots E_{a_d}^{\overline{\bf M}_d}
  (\Gamma_{\overline{M}_1 \ldots \overline{M}_d})^\alpha_\beta.
 \ee
The action preserves worldvolume supersymmetry only for certain
triplets of $D,\ d$ and $N$.

When the super-$p$-brane is moving in a spacetime it sweeps out a
$d$-dimensional worldvolume. It is convenient to set spacetime
coordinates as:
 \be
  X^M(\xi)=\left( X^a(\xi),Y^m(\xi) \right),
  \qquad \mbox{where} \qquad X^a(\xi)=\xi^a.
 \ee
The super-$p$-brane has therefore exactly $D-d$ bosonic degrees of
freedom. To count fermionic degrees of freedom we introduce $n$ as
the number of worldvolume supersymmetries and $m$ as the number of
real components of an irreducible spinor in a given worldvolume
dimension. Then the number of fermionic degrees freedom on--shell
is $mn/2$. But calculating the same for spacetime fermions one
should take $MN/4$ (where $N$ is a number of spacetime
supersymmetries and $M$ a number of real components of an
irreducible spinor in a given spacetime dimension) because the
$\kappa$ symmetry halves a number of physical degrees of freedom.
The condition of equality of the number of bosonic and fermionic
degrees of freedom for the Green-Schwarz super-$p$-branes is
therefore:
 \be
  D-d=\frac12 mn = \frac14 MN.
  \label{brane_scan_cond}
 \ee
Using the known numbers of dimensions of irreducible spinor
representations with Lorentzian signature we find that the
condition is satisfied by four fundamental solutions \cite{AETW}:
\begin{itemize}
 \item "octonionic" branes with $D=11$, $d=3$, $m=2$, $n=8$,
 $M=32$, $N=1$
 \item "quaternionic" branes with $D=10$, $d=6$, $m=8$, $n=1$,
 $M=16$, $N=1$
 \item "complex" branes with $D=6$, $d=4$, $m=4$, $n=1$,
 $M=8$, $N=1$
 \item "real" branes with $D=4$, $d=3$, $m=2$, $n=1$,
 $M=4$, $N=1$
\end{itemize}
The fundamental solutions are maximal ones in four series and
can be denoted as $(D_{max}, d_{\max})$. Other members of the
series can be obtained by a double reduction of $k$ dimensions
$(D,d)=(D_{max}\!-\!k, d_{\max}\!-\!k)$ for $k=1,\ldots,d_{\max}\!-\!1$. Note
that for $d>2$ all found super-$p$-branes have $N=1$.

The case $d=2$ is special and requires more detailed analysis. The
reduction of the fundamental solutions gives superstrings in four
possible spacetime dimensions: $D=3,\ 4,\ 6,\ 10$. All of them
have $N=2$, so they are type II superstrings. But for $d=2$ the
(\ref{brane_scan_cond}) is not the only possibility. In this case
it is allowed to treat left and right moving modes independently
and apply supersymmetry only to one of them. Then instead of
(\ref{brane_scan_cond}) another condition should be satisfied:
 \be
  D-2= n = \frac12 MN
  \label{brane_scan_cond2}
 \ee
leading to $D=3,4,6,10$ solutions with $N=1$ corresponding to
heterotic superstrings.

It is worth to note, that the maximal spacetime dimension obtained
in this procedure is $D=11$ being in excellent agreement with the
analogous result derived before for supergravity theories.
Moreover in the super-$p$-branes the condition $D\le 11$ is
derived without the restriction that spin is not bigger than $2$.

The conditions (\ref{brane_scan_cond}) and
(\ref{brane_scan_cond2}) are valid only in the case when it is
assumed that the worldvolume fields form scalar multiplets. One
can relax this assumption and introduce on the worldvolume other
supersymmetry representations (i.e. vector or tensor
supermultiplets) with additional fields. Then one can define more
super-$p$-branes of various kinds and the table \ref{brane_scan}
shows the result of such a search \cite{Duff3}. However it should be noted,
that matching fermionic and bosonic degrees of freedom is only a
necessary but not a sufficient condition that a supersymmetric
theory exists in a general case. To prove that the presumable
super-$p$-brane really exists, an analog of
(\ref{pbrane_GS+WZ_action}) should be written and examined in each
case.

\begin{table}[htbp]
 \begin{tabular}{c|ccccccccccc}
  D  &   &   &   &   &   &   &   &   &   &   &   \\
  \hline
  11 & . & . & S & . & . & T & . & . & . & . & . \\
  10 & V &S,V& V & V & V &S,V& V & V & V & V &   \\
  9  & S & . & . & . & S & . & . & . & . &   &   \\
  8  & . & . & . & S & . & . & . & . &   &   &   \\
  7  & . & . & S & . & . & T & . &   &   &   &   \\
  6  & V &S,V& V &S,V& V & V &   &   &   &   &   \\
  5  & S & . & S & . & . &   &   &   &   &   &   \\
  4  & V &S,V&S,V& V &   &   &   &   &   &   &   \\
  3  &S,V&S,V& V &   &   &   &   &   &   &   &   \\
  2  & S & . &   &   &   &   &   &   &   &   &   \\
  1  & . &   &   &   &   &   &   &   &   &   &   \\
  \hline
  d  & 1 & 2 & 3 & 4 & 5 & 6 & 7 & 8 & 9 & 10& 11
 \end{tabular}
 \caption{The brane scan. S - scalar; V - vector; T - tensor.}
 \label{brane_scan}
\end{table}

Several cases in the table seem to be especially interesting. Let
us take $N\!\!=\!\!1$ super-$5$-brane in $D=10$. The action
(\ref{pbrane_GS+WZ_action}) of the super-$5$-brane has to contain
an antisymmetric tensor potential of rank $6$ with $7$-form field
strength. The case is very interesting because there is a dual
formulation of $(1,0)$ supergravity where the $3$-form field
strength is replaced by a $7$-form \cite{Chamse}. Both these supergravity
theories are equivalent and anomaly free when coupled to $SO(32)$
or $E(8) \otimes E(8)$ super-Yang-Mills. For the $3$-form
formulation the theory is a low energy limit of the heterotic
superstrings with $(1,0)$ supersymmetry. The other formulation
suggests that there should be a "heterotic" super-$5$-brane theory
dual to the theory of heterotic superstrings \cite{Duff2}.

Even more exciting possibility is pointed out by super-$2$-brane
in $D=11$. It can be checked, that $\kappa$ symmetry requires that
fields $g_{MN}$ and $C_{MNR}$ appearing in
(\ref{pbrane_GS+WZ_action}) satisfy constraints equivalent with
the equations of motion of the eleven dimensional supergravity \cite{HullT, BST2}.
Moreover double dimensional reduction procedure applied to the
super-$2$-brane gives the type IIA superstring which is S-dual to
M-theory. This fact suggests to put forward a hypothesis that
M-theory could be a quantum theory of super-$2$-branes in a
similar way as superstring theory is a quantum theory of
super-$1$-branes.

A quantization \cite{Hoppe, Duff3}, of super-$p$-branes for $p > 1$ is significantly
more difficult than in the case of superstrings ($p=1$). The main
problem is that for $p > 1$ there are not enough symmetries to
gauge away all internal degrees of freedom so in contradistinction
to string worldsheet there is no classical meaning of a distance
on the worldvolume. Therefore there is no evidence that a brane
theory is finite or even renormalizable (but on the other hand
neither there is evidence to the contrary). Some attempts to
quantize the most promising $p=2$ brane in $D=11$ have shown that
the resulting theory should be in some aspects similar to
super-Yang-Mills theory defined in $D-1$ dimensions with an exotic
gauge group $SU(\infty)$. The result was extended to other
super-$2$-branes and it was also suggested that cases with $p>2$
correspond to analogs of gauge theory where gauge vectors are
replaced by higher rank antisymmetric tensors \cite{BST3}. Next it was checked
that the super-$2$-brane is anomaly free only in $D=11$ \cite{BarsPS, DIPSS}. However
all these studies are rather tests of various possibilities, so
M-theory understood as a membrane theory is still more a
conjecture than a fact.

%%%%%%%%%%%%%%%%%%%%%%%%%%%%%%%%%%%%%%%%%%%%%%%%%%%%%%%%%%%%%%%%%
%%%%%%%%%%%%%%%%%%%%%%%%%%%%%%%%%%%%%%%%%%%%%%%%%%%%%%%%%%%%%%%%%
\subsection{Branes as sources of antisymmetric tensor fields.}
\label{b_antytensor}

\akapit In the previous section it was observed that a $p$-brane is
accompanied by a field of $(p+1)$-forms. This observation is a
part of a more fundamental rule: a theory with a $(p+1)$-form
potential is connected with the existence of a $(p+1)$-dimensional
charged objects: branes. Let us examine it more carefully.

Consider a model with an antisymmetric tensor field $A_{[n-1]}$ of rank $n-1$.
A general action of the model is given by:
 \bea
  S &=& S_{kin} + S_{CS} + S_{int}, \\
  S_{kin} &=& \int_{\calM_D} \sqrt{{\rm g}} \; d^DX e^{a\phi}
   \left( -\frac{1}{2n!} F^{M_1 \ldots M_n} F_{M_1 \ldots M_n} \right) \nn \\
   &=& \int_{\calM_D} e^{a\phi}
   \left( -\frac{1}{2} F_{[n]} \wedge (\ast F)_{[D-n]} \right)
   = \int_{\calM_D} e^{a\phi}
   \left( -\frac{1}{2} \left|F_{[n]}\right|^2 \right),
   \label{dform_action}
 \eea
where $\phi$ is a dilatonic scalar field, $a$ -- a constant.
$S_{int}$ depends at most linearly on $A_{[n-1]}$, but has an
arbitrary dependence on any other fields, in particular it has to
contain kinetic terms for the dilaton $\phi$ and graviton
$g_{MN}$. The remaining $S_{CS}$ is Chern--Simons term trilinear
in $A_{[n-1]}$. For simplicity of the discussion let us put in
this section $S_{CS}=0$. The $F$ is given by:
 \be
  F_{[n]} = d A_{[n-1]}. \label{F_def1}
 \ee
With this relation the field $F$ has to satisfy the Bianchi
identity:
 \be
  d F_{[n]} = dd A_{[n-1]} = 0.  \label{Bianchi_1}
 \ee
The theory exhibits gauge invariance under:
 \be
  A_{[n-1]} \rightarrow A_{[n-1]}+ d \chi_{[n-2]}
 \ee
 and the equations of motion derived form (\ref{dform_action}) are:
 \be
  d \left( e^{a\phi}(\ast F)_{[D-n]} \right) = (\ast J_e)_{[D-n+1]},
  \label{dform_eom}
 \ee
where the $J_{e[n-1]}$ is the conserved Noether current of the
theory and its shape is specified by the $S_{int}$.

This picture is quite similar to that one of electrodynamics but
now the current $J_e$ is in general $(n-1)$-form (and not a
vector) and (if it possesses a nonzero timelike component) it
defines not a worldline of electrically charged point particle,
but $(n-1)$-dimensional worldvolume of $p=n-2$ dimensional object
propagating in time. The object carries an elementary charge of
the field $F$, in other words the Noether charge associated with the
conserved Noether current $J_e$, which by analogy to
electrodynamics is very often called an electric charge. The
object is known as an electric (elementary) $p$-brane. If
$S_{int}$ has a form of (\ref{pbrane_GS+WZ_action}) then the
$p$-brane can be identified with the super-$p$-brane introduced in
the section \ref{p_branes}.

We can calculate the electric charge of the brane in the usual way as:
 \be
  Q_{e[n-1]}=\int_{\calM_{D-n+1}} (\ast J_e)_{[D-n+1]} =
  \int_{S^{D-n}} e^{a\phi} (\ast F)_{[D-n]},
  \label{el_charge}
 \ee
where the $\calM_{D-n+1}$ is a subspace transversal to the
electric brane and the $S^{D-n}$ is a sphere surrounding a
point-like image of the brane under a projection of the $\calM_D$
on the $\calM_{D-n+1}$.

It is possible to construct also a magnetic (solitonic) brane.
Then (\ref{F_def1}) is replaced by
 \be
  F_{[n]} = d A_{[n-1]} + \eta_{[n]}, \label{F_def2}
 \ee
where $\eta_{[n]}$ is an arbitrary not exact $n$-form, so the
Bianchi identity is then:
 \be
  d F_{[n]} = (\ast J_m)_{[n+1]}
  \label{Bianchi_2}
 \ee
instead of (\ref{Bianchi_1}), where $d \eta_{[n]} = (\ast
J_m)_{[n+1]}$. In this way via the modified Bianchi identity a
$p=D-n-2$ dimensional object can be introduced to the theory -- a
magnetic (solitonic) $p$-brane. The brane carries a magnetic
(called also solitonic or topological) charge defined as:
 \be
  Q_{m[D-n-1]} = \int_{\calM_{n+1}} (\ast J_m)_{[n+1]} =
  \int_{S^n} F_{[n]},
  \label{mag_charge}
 \ee
where $\calM_{n+1}$ is a subspace transversal to the magnetic
brane and the $S^n$ is a sphere surrounding an image of the brane
under a projection of $\calM_D$ on $\calM_{n+1}$.

When $n=D/2$ both the electric and the magnetic branes have the
same dimension and can even coincide. If additionally the field
$F_{[n]}$ is selfdual or anti-selfdual, i.e. it obeys:
 \be
  F_{[n]} = \pm \ast F_{[n]},
 \ee
the branes necessarily coincide so we then have a single brane
carrying simultaneously electric and magnetic charges. This
category of branes is called dyonic branes.

Some features of the electric and magnetic branes are strongly
correlated.
For example:
 \be
  d_e + d_m = D-2,
 \ee
where $d_e$ and $d_m$ are dimensions of electric and magnetic
brane's worldvolumes:
 \be
  d_e = n-1 \qquad d_m = D-n-1.
 \ee
Because of that it is convenient to define a mapping described by
the symbol of tilde $\quad \tilde{} \quad$ and acting on integer
numbers as follows:
 \be
  \tilde{d} = D-d-2.
  \label{tilde_map}
 \ee
In the case of the brane worldvolume dimensions it gives:
 \be
  \td_e = d_m \qquad \td_m = d_e.
 \ee

Another interesting fact is that if the discussed theory is
quantum then the charges (\ref{el_charge}) and (\ref{mag_charge})
have to satisfy the Dirac's quantization condition:
 \be
  Q_{e[d_e]} Q_{m[d_m]} = 2 \pi N,
  \label{Dirac_quant}
 \ee
where $N$ is an integer. The above quantization rule was
originally derived by Dirac in the case of electromagnetic theory
in $D=4$. To prove (\ref{Dirac_quant}) in this case consider a
hypothetic magnetic monopole (i.e. a magnetically charged
particle) with a charge $Q_m$ located at the origin of the
coordinate system. Let the coordinates be spherical
$(r,\phi,\theta)$ where $r\in (0,\infty)$, $\phi \in [0,2\pi)$ and
$\theta \in [-\pi/2,\pi/2]$. The magnetic monopole is a source of
an electromagnetic field whose strength $F_{[2]}$ has to satisfy
$\int_{S^2} F_{[2]}= Q_m$, so:
 \be
  F_{[2]} = \frac{Q_m}{4\pi} \cos \theta d\theta \wedge d\phi.
 \ee
Solving $F_{[2]}=dA_{[1]}$ one finds the corresponding vector
potential. But the potential cannot be expressed by a single
formula globally. It is necessary to introduce at least two maps
covering together the whole space with different $A_{[1]}$'s and
different choice of gauge on each. For example:
 \bea
  A^+_{[1]} = \frac{Q_m}{4\pi} \left( \sin\theta +1 \right)
  d\phi \quad \mbox{where} \quad \theta \neq +\pi/2, \\
  A^-_{[1]} = \frac{Q_m}{4\pi} \left( \sin\theta -1 \right)
  d\phi \quad \mbox{where} \quad \theta \neq -\pi/2.
 \eea
It gives $(A^+ - A^-)_{[1]}= (Q_m/2\pi) d\phi$. If an electrically
charged particle with charge $Q_e$ moves in the field of the
magnetic monopole its wave function should also be given by two
sections: $\psi^+$ and $\psi^-$. Both the wave functions differ by a
phase:
 \be
  \exp\left( i \frac{Q_e Q_m}{2\pi} \phi \right).
 \ee
But because shifting the $\phi$ coordinate by $2\pi$ gives the
same point, the phase has to be unchanged by such an operation and
this requirement leads to $Q_e Q_m = 2 \pi N$.

Considering general branes, the Dirac monopole has to be replaced
by a magnetic $(D-d-3)$-brane and the electric point particle by
an electric $(d-1)$-brane, where the branes are orthogonal one to
the other and not intersecting. Making a projection in the
$D$-dimensional space-time parallel to all space-like directions
that are parallel to any of the branes one gets a $(1+3)$
dimensional picture identical to the one described before. So, the
condition (\ref{Dirac_quant}) has to be true not only for point
particles but also for branes.

It is crucial to observe that if we reformulate the theory
defining as a fundamental field $\tilde{F} = \ast F$, then the
equation of motion (\ref{dform_eom}) is replaced by the Bianchi
(\ref{Bianchi_2}) identity and vice versa, but the kinetic term in
(\ref{dform_action}) does not change:
 \be
  F_{[n]} \wedge (\ast F)_{[D-n]} =
  (\ast F)_{[D-n]} \wedge (\ast \ast F)_{[n]}
   = \tilde{F}_{[D-n]} \wedge (\ast \tilde{F})_{[n]},
 \ee
where the identity $\ast^2 = (-1)^{1+n(D-n)}$ valid when applied
to $n$-forms was used.
This is called electric/magnetic duality.
But of course in a general case nothing guarantee
that the duality is an exact symmetry of the theory
not only an interesting coincidence.

However in 1970s it was noted that in some supersymmetric gauge
theories electric and magnetic charges and masses of all particles
described by the theory have to obey a universal relation:
\be
 M^2 = \alpha^2 \left(Q_e^2+Q_m^2\right),
\ee where $\alpha$ is a constant. Therefore if roles of the
electric and magnetic charges are exchanged the mass is still
preserved. This inspired Montonen and Olive to conjecture
\cite{MO} that the electric/magnetic duality could be a real
symmetry of the whole quantum theory. Let us discuss this
conjecture in more detail. Consider a quantum state which carries
electric and magnetic charges and the electric charge is quantized
such that $Q_e=n_e q$, where $n_e$ is an integral quantum number
and $q$ is a fundamental quanta of the electric charge. Then the
magnetic charge has to be quantized as well, but by the Dirac
quantization rule (\ref{Dirac_quant}) it should be given by
$Q_m=n_m (2\pi/q)$. So the electric/magnetic duality cannot by
described just by a replacing of the quantum numbers $n_e$ with
$n_m$ and vice versa. Simultaneously also the coupling $q$ has to
be inversed, what means that the duality connects weak and strong
coupled sectors of the theory. This idea has an interesting
continuation on the ground of string theory where S-duality was
discovered \cite{FILQ, Rey}.

Up to now we have not specified for what range of $p$ it is
reasonable to define a brane. The simple model described in this
section works properly if $p=0,1,\ldots,D-3$ what corresponds to
$d=1,2,\ldots,D-2$. For each case there is a well defined
antisymmetric tensor potential $A_{[n-1]}$, its strength
$F_{[n]}$ and a dual strength $(\ast F)_{[D-n]}$. In
particular, $p=0$ reduces to a point particle and $p=1$ -- to a
string. But it is possible to extend the definition of branes
beyond that range.

$p=-1$ makes no formal problems. It is a dual to $(D-3)$-brane and
is given by a scalar potential and vector strength field. However,
some features of the $(-1)$-brane can seem quite odd. Because its
worldvolume is zero-dimensional it cannot propagate and exists
only in one moment in time. This class of objects was previously
discovered in electrodynamics and are called instantons.

Of course it is possible to extend definition of the instantons,
to objects which are not necessary points in spacetime. If we
relax the condition, that the currents $J_e$ or $J_m$ must have
nonvanishing timelike component, we allow a situation where they
span purely spatial hypersurfaces of nonzero dimension. Such
hypersurfaces are called S-branes \cite{Hull, GutS, CGG}. 
We will not discuss them in this
work and concentrate on the branes evolving with time. It is
however worth noting that S-branes can be especially interesting
in the context of cosmological models where they can play a role
of an initial singularity.

A $(D-2)$-brane is usually called a domain wall and it is described
by a rank $D-1$ potential $A_{[D-1]}$ and a rank $D$ strength
$F_{[D]}$. Such a field appears for example in massive $(1,1)$
supergravity (\ref{sugra_IIAmass_ac}). Without external sources it
can be deduced from equation (\ref{dform_eom}) that $(\ast
F)_{[0]}$ and therefore $F_{[D]}$ have to be constants and have no
propagating states. Thus, the kinetic term $|F_{[D]}|^2$
contributes to a lagrangian effectively as a cosmological term.

A brane with $p=D-1$ is very special. Because its antisymmetric
potential has to be proportional to the volume element it should
be interpreted just as the whole spacetime. A strength field
related to the $(D-1)$-brane obviously vanishes since it is a form
of rank $D+1$.

Branes with $p \leq -2$ or $p \geq D$ cannot exist because there
is no possibility to introduce an antisymmetric potential in these
cases.

%%%%%%%%%%%%%%%%%%%%%%%%%%%%%%%%%%%%%%%%%%%%%%%%%%%%%%%%%%%%%%%%%
%%%%%%%%%%%%%%%%%%%%%%%%%%%%%%%%%%%%%%%%%%%%%%%%%%%%%%%%%%%%%%%%%
\subsection{Branes in superstring and M-theory.}

\akapit Knowing that branes are closely related to antisymmetric
forms one may note that the massless sector of every superstring
theory contains several fields of this kind. So it is natural to
expect that the theories could admit existence of brane-like
objects. However there is a question what are these objects in
this context and whether they are necessary or only theoretically
possible elements of string theories. The answer is more
surprising than one could expect. The branes are not only an
intrinsic part of any string theory but they also give an
excellent tool for forecasting and examining features of the
M-theory.

%%%%%%%%%%%%%%%%%%%%%%%%%%%%%%%%%%%%%%%%%%%%%%%%%%%%%%%%%%%%%%%%%
\subsubsection{D${}_p$-branes in bosonic strings.}

\akapit Consider bosonic oriented closed string theory with one,
say the twenty sixth, dimension compactified on a circle of radius
$R$. A quantum states' spectrum of such theory is different than
the spectrum of free theory (\ref{s_spectrum}). However, besides
Kaluza-Klein modes which are naturally expected by analogy with
the compactification procedure of point particle theories, states
of other kind can also appear. They are labelled by so called
winding numbers $w$ counting how many times a closed string is
wound around the compact dimension. Because closed strings with
different $w$ are topologically inequivalent they have to form
different states. The full spectrum is:
 \be
  m^2 = \left( \frac{n}{R}+\frac{wR}{\alpha'} \right)^2 +
  \frac{4}{\alpha'} (N-1),
 \ee
where $n$ numbers Kaluza-Klein excitation levels. A crucial
observation is, that the spectrum and the whole theory is
invariant under:
 \be
  R \rightarrow R' = \frac{\alpha'}{R},
  \qquad w \rightarrow n, \qquad n \rightarrow w.
 \ee
In other words the theory compactified on a small circle $R$ is
equivalent to a theory compactified on a big circle
$R'=\frac{\alpha'}{R}$ if roles of the winding and the
Kaluza-Klein states are simultaneously interchanged. In the limit
$R\rightarrow 0$ a theory dimensionally reduced by one is dual to
a theory on a noncompact space $R\rightarrow\infty$. It can be
checked that the duality also effectively reverses sign of the
right-moving modes, so in terms of coordinates it is given by:
 \bea
  X^{26}(\xi^1,\xi^2) &=& X^{26}_L(\xi^2 - \xi^1) +
  X^{26}_R(\xi^2 + \xi^1) \rightarrow  \nn \\
  &\rightarrow&
  {X'}^{26}(\xi^1,\xi^2) \;= \;X^{26}_L(\xi^2 - \xi^1) -
  X^{26}_R(\xi^2 + \xi^1).
  \label{X_T_dual}
 \eea
This is an example of T-duality and can be extended to more
general cases of toroidal compactifications.

One can wonder what is happening in a similar situation with open
strings which cannot have preserved winding numbers. The spectrum
of such compactified theory is just:
 \be
  m^2 = \frac{n^2}{R^2} + \frac{4}{\alpha'} (N-1),
 \ee
what evidently is not invariant under $R\rightarrow
\frac{\alpha'}{R}$. This seems to lead to a contradiction if one
remembers that theories with interacting open strings have to
contain also closed strings. The contradiction is however only
apparent. A difference between open and closed strings lays in the
endpoints of open strings not in their interior. So, a naive
consideration gives a prediction, that a dual to open string
theory compactified on a circle with radius $R\rightarrow 0$
should be a theory with open strings having endpoints confined on
a $25$-dimensional hypersurface. Actually, applying
(\ref{X_T_dual}) to the Neumann boundary conditions of open string
(\ref{b_c_open}) one really obtains Dirichlet boundary conditions:
 \be
  \frac{\de}{\de \xi^1} {X'}^{26} (\xi^1,0) = 0 =
  \frac{\de}{\de \xi^1} {X'}^{26} (\xi^1,l),
  \label{b_c_open_D}
 \ee
defining a hypersurface called Dirichlet brane, D-brane or
D${}_p$-brane, where $p$ is a number of spatial dimensions of the
hypersurface \cite{Pol2, Pol3}.

Of course T-duality can be applied not only to the "original"
theory where all open strings satisfy only Neumann boundary
condition but also to a theory with D${}_p$-brane of arbitrary
$p$. If one T-dualize $k_1+k_2$ dimensions, $k_1$ tangent and
$k_2$ orthogonal to the D${}_p$-brane, then one obtains a theory
with D${}_{(p-k_1+k_2)}$-brane. In this picture the "original"
string theory is a theory with D${}_{25}$-branes filling the whole
space.

Additional properties are revealed when one takes into account
Chan-Patton states. It can be shown that $U(N)$ oriented open
string theory after T-dualization gives a theory with exactly $N$
parallel D${}_p$-branes. A state dual to $|i,j>$ where $i,j$ are
indices of the gauge symmetry group is then realized by a string
having one end glued to the $i$-th brane and the second to the
$j$-th brane. If all branes are separated, a gauge symmetry group
of the dual theory is $U(1)^N$. If some of the branes coincide,
for example if there are $n$ distinct locations for the branes
with $k_i$ branes at each ($\sum_{i=1}^n k_i = N$) then the gauge
group is $\otimes_{i=1}^n U(k_i)$. Maximally the original $U(N)$
group can be restored.

A little different situation occurs for unoriented strings. The
starting symmetry is then $SO(N)$ or $Sp(N)$ and the T-dual space
is not compactified on a circle or a torus but rather on an
orientifold $S^k/{\bf Z}_2$. If $N$ is even then all the branes
are grouped into $N/2$ pairs with partners living at points
related by ${\bf Z}_2$ symmetry, so effectively one sees maximally
$N/2$ branes on the orientifold. If $N$ is odd, then there is an
additional brane with no partner which has to be localized at
${\bf Z}_2$ fixed plane. A gauge group related to a pair of branes
is $U(1)$ and $U(k)$ in a case of $k$ coincident pairs. But if the
branes coincide at one of the fixed planes then the symmetry is
$SO(2k)$ or $Sp(2k)$ instead of $U(k)$. So in the extreme case
when all the branes are at the fixed plane the original symmetry
$SO(N)$ or $Sp(N)$ is restored again. Note, that the above results
are consistent with the previous observations that a theory with
open strings in a flat empty space can be equivalently interpreted
as a theory in a space filled out by $N$ D${}_{25}$-branes.

Consider a low energy effective physics on a worldvolume of a
single D${}_p$-brane. It has to be given by a vector field
$A_\mu(\xi)$ involved with the gauge group of the Chan-Patton
states and fields induced on the brane by the background of string
massless states:
 \bea
  g_{\mu\nu}(\xi) &=& \frac{\de X^M}{\de \xi^\mu}
  \frac{\de X^N}{\de \xi^\nu}g_{MN}(X(\xi)), \\
  C_{\mu\nu}(\xi) &=& \frac{\de X^M}{\de \xi^\mu}
  \frac{\de X^N}{\de \xi^\nu}C_{MN}(X(\xi)), \\
  \phi(\xi) &=& \phi(X(\xi)).
 \eea
As an appropriate action describing a dynamics on the brane one
usually postulates Born-Infeld action \cite{Leigh, CGMNW}:
 \be
  S_p = - T_p \int d \xi^{p+1} e^{-\phi(\xi)}
   \left\{ - \det \left[ g_{\mu\nu}(\xi) + C_{\mu\nu}(\xi) +
   2 \pi \alpha' F_{\mu\nu}(\xi) \right] \right\}^{1/2},
   \label{D_brane_ac}
 \ee
where $T_p$ describes a tension of the brane when an expectation
value of the dilaton $\phi$ vanishes. More precisely a physical
tension of the brane in an arbitrary background is:
 \be
  \tau_p = T_p e^{-<\phi>}.
  \label{brane_mass_1}
 \ee
It is interesting that for D${}_p$ branes of various $p$ the
following identity holds:
 \be
  \tau_p = \frac{\tau_{p-1}}{2\pi \sqrt{\alpha'}}.
  \label{brane_mass_2}
 \ee

The action (\ref{D_brane_ac}) describes a kind of
$(p+1)$-dimensional gauge theory, where the gauge group and other
features of the theory depend directly on the brane configuration
and on the type of the string theory. So, we can study gauge
theories not only as completely separate models but also in
connection with string theory or (in a limit) with supergravity
theories.

%%%%%%%%%%%%%%%%%%%%%%%%%%%%%%%%%%%%%%%%%%%%%%%%%%%%%%%%%%%%%%%%%
\subsubsection{D${}_p$-branes in superstrings.}

\akapit
Up to now we were considering D-branes in bosonic string theories,
but it should be remembered that five consistent, interacting
string theories are necessarily supersymmetric. Fortunately, the
bosonic sector discussed so far can be simply adjusted -- for
example one should replace 26 dimensions by 10, set for the
tension:
 \be
  \tau_p = \frac{1}{(2\pi)^p e^\phi (\alpha')^{\frac{p+1}{2}}},
  \label{brane_mass_3}
 \ee
add fermionic terms to (\ref{D_brane_ac}), and then everything
what was written for bosonic D-branes is still valid for D-branes
in superstring theories. Moreover, new important features can be
detected.

Let us focus on the type I $SO(32)$ superstring theory. The theory
describes unoriented open and closed strings. After T-dualisation
on $9-p$ toroidally compactified dimensions one obtains a theory
with open superstrings having endpoints at $16$ D${}_p$-branes and
closed superstrings propagating in a bulk. There is some
evidence that T-duality is an exact symmetry of superstrings, so
the dual theory with branes is well defined. Since in a limit when
the distance grows to infinity it gives either type IIA or type
IIB theory and since changing the distance between two adjacent
D-branes is a continuous operation,  there are strong conjectures
that all the theories at finite distances should also be
consistent. In this model the type I, IIA, IIB theories and the
theories with the 16 branes are only special limits of more
general superstring theory (and further M-theory) and one should
rather treat them as different states of the same theory than as
separate theories.

It is important to note that similar arguments lead to a
prediction that a general superstring state does not need to be
given by a set of exactly $16$ parallel D-branes of the same
dimension. For example one can move to infinity and neglect only
some branes but not all of them and get a state with $N<16$
branes. It is also possible to rotate a brane and then encounter
branes intersecting at some angles (even orthogonally in the
extreme case). Consider then two orthogonal D${}_p$-branes.
Applying T-duality to the configuration in a direction tangent to
only one of them one obtains a state with D${}_{(p+1)}$-brane
perpendicular to D${}_{(p-1)}$-brane. In a similar way many other
states allowing various number of D-branes of various dimensions
intersecting at various angles can be constructed.

There is no evidence that all possible states of the string theory
have to be built upon a flat space or a space with branes only.
Probably other backgrounds also can exists. Unfortunately methods
of construction of quantum interacting string theories in a
nontrivial background are not yet developed. So, the only possible
opportunity to attain a knowledge of the "not flat" states is to
study the branes.

In the section \ref{b_antytensor} it was shown that a theory with
antisymmetric tensor fields admits an existence of branes --
extended charged objects. It was proved that D-branes carry in
string theory Ramond-Ramond charges \cite{Pol2}, so they are spun
by fields from the Ramond-Ramond sector. In the R-R sector of type
IIA theory there are antisymmetric tensors $F_{[2]}$, $F_{[4]}$
and their duals $(\ast F)_{[6]}$ and $(\ast F)_{[8]}$. Then the
potentials of the fields have ranks $1,\ 3,\ 5,\ 7$ and spun
D${}_p$-branes of $p=0,\ 2,\ 4,\ 6$ respectively. In the same way
$p=-1,\ 1,\ 3,\ 5,\ 7$ branes can be related to antisymmetric fields
$F_{[1]}$, $F_{[3]}$, $F_{[5]}=(\ast F)_{[5]}$, $(\ast F)_{[7]}$
and $(\ast F)_{[9]}$ of type IIB theory. But there are also
D${}_8$ and D${}_9$-branes which cannot be directly connected with
any massless field of the type IIA or the type IIB theory.

The D${}_8$-brane has to be related to the antisymmetric field
strength of rank $10$. Such a field has no propagating degrees of
freedom in ten dimensions, so including such a field is not
obvious from the field theory point of view. But besides the usual
$N=(1,1)$ supergravity identified as a low energy limit of the
type IIA theory there is also its extension - the massive
supergravity (\ref{sugra_IIAmass_ac}), which contains the field
$F_{[10]}$. On the ground of type IIA superstrings, the existence
of D${}_9$ brane can be interpreted as a cosmological constant of
arbitrary value \cite{Pol2}.

By analogy to other branes one could expect the D${}_9$-brane in
type IIB superstrings. The appropriate coupling of the brane to
the background fields is:
 \be
  n Q \int A_{[10]},
  \label{b9}
 \ee
where $n$ is a number of branes. But the variation of (\ref{b9})
in the action with respect to $A_{[10]}$ leads to $n=0$.
Fortunately in type I superstring with $SO(32)$ gauge symmetry
group there are additional terms which modify (\ref{b9}) replacing
$n$ by $n\!-\!32$ \cite{PC}. So we can interpret the type I theory
with its open strings having free ends as a theory defined on 16
coincident pairs of D${}_9$-branes. All the other D-branes can be
then derived from the model by T-duality.

%%%%%%%%%%%%%%%%%%%%%%%%%%%%%%%%%%%%%%%%%%%%%%%%%%%%%%%%%%%%%%%%%%
\subsubsection{BPS and non-BPS, stable and unstable D-brane
configurations.}

\akapit Remembering that five consistent superstring states
defined in flat space are supersymmetric and hence stable, it is
very interesting to study analogous properties of states with
branes.

The type IIA D${}_{2p}$-branes and the type IIB
D${}_{2p+1}$-branes always break precisely half of
supersymmetries. To see this take a state with $16$ parallel
branes which is T-dual to type I superstrings. The duality
requires that the state must have $N=1$ supersymmetry. But if a
distance between the adjacent branes is growing the state in the
bulk tends to type II state which obviously is $N=2$
supersymmetric. So, half of possible supersymmetries are broken by
the branes. Writing more formally, if $Q_L$ and $Q_R$ are
supercharges corresponding to left and right moving modes of the
theory, then on the D${}_p$-brane:
 \be
  Q_L+ \prod_m \beta^m Q_R
  \label{susy_charge_pres}
 \ee
is conserved, where $m$ denotes $p$ spatial directions
perpendicular to the brane and:
 \be
  \beta^m = \Gamma^V \Gamma^m,
 \ee
where $\Gamma^V$ is the chiral operator in ten
dimensions.

The states which preserve part of supersymmetry are usually called
BPS-states because they saturate so called
Bogomolnyi-Prasad-Sommerfield (BPS) inequality:
  \be
   \calE \geq Q,
   \label{BPS_1}
  \ee
where $\calE$ is the energy (mass) density and $Q$ -- the charge density.
For a single p-brane it can be equivalently written as:
 \be
  \tau_p \geq Q.
 \ee

The formula (\ref{susy_charge_pres}) shows that the number of
broken supersymmetries in a given state depends on geometrical
properties of configuration of the branes. In particular, an
arbitrary number of parallel D${}_p$-branes with the same $p$
breaks the same number of supersymmetries as a single
D${}_p$-brane. But if the branes are not parallel, they usually break 
together more supersymmetries and for some configurations of 
D-branes no supersymmetry can be preserved at all. However even 
if two branes are situated at nonzero relative angle it is
possible to find such specific value of the angle for which
the same amount of supersymmetries is preserved as by the
parallel configuration \cite{BDL}. 

The non-BPS states also have to be present in string theory. A
proof for it is simple. The world we know is nonsupersymmetric, so
it has to be a low energy limit of some non-BPS state. An
important task consists in finding such a nonsupersymmetric state
and test its properties (first of all to check if it is stable).

Probably the simplest example of a non-BPS-state is a system of
two parallel D-branes with opposite R-R charges or in other words
a system of a D-brane and an anti-D-brane. A key ingredient of the
construction is that by (\ref{susy_charge_pres}) each of the
objects breaks the complementary half of supersymmetry. In terms
of (\ref{BPS_1}) one can see that for the brane-anti-brane system
the total mass is twice the mass of a single brane, but the
effective charge vanishes. However such a brane-anti-brane
configuration is unstable due to tachyons living on the brane
worldvolumes \cite{BanksS}.

Another non-BPS-state can be produced from a system of coincident
D${}_{2p}$-brane and anti-D${}_{2p}$-brane in type IIA theory with
the projection operator $(-1)^{F_L}$ acting on it \cite{Sen}. The
$(-1)^{F_L}$ changes sign of the left moving fermions, so it
brings the type IIA superstrings in the bulk to the type IIB. It
also removes half of states on the brane-anti-brane system,
specifically this half which is related to degrees of freedom
describing possibility of disjoining the branes. So, the object is
effectively a single D${}_{2p}$-brane in the type IIB theory. An
analogous construction leads to type IIA D${}_{2p+1}$-branes. Both
the classes of branes are unstable and their masses are by $\sqrt{2}$
bigger than masses of corresponding type IIA or IIB BPS-branes:
 \be
  \tau_{2p, IIB} = \sqrt{2} \tau_{2p, IIA}, \qquad
  \tau_{2p+1, IIA} = \sqrt{2} \tau_{2p+1, IIB}.
 \ee
Applying the $(-1)^{F_l}$ to the non-BPS type IIB D${}_{2p}$-brane
or type IIA D${}_{2p+1}$-brane once again changes the bulk theory
and projects out next part of states living on the brane. The
result of the operation is the already known BPS IIA
D${}_{2p}$-brane or IIB D${}_{2p+1}$-brane respectively.

But we are still looking for the stable non-BPS states. They are
interesting for several reasons:
\begin{itemize}
\item They belong to the string theory spectrum,
so they are necessary to fully describe the theory.
\item By (\ref{BPS_1}) they are objects whose masses are not
bounded by demand of supersymmetry, but still possible to
calculate for various values of the string perturbative  coupling
constant. Therefore they provide an opportunity to study string
theory at finite coupling.
\item The worldvolume theory on the stable non-BPS brane should
belong to nonsupersymmetric gauge theories which are much less
understood than the supersymmetric ones.
\end{itemize}

The usual procedure leading to construction of a stable non-BPS
brane is to apply to one of IIA or IIB unstable non-BPS states an
orbifolding or orientifolding projection which removes the
tachyonic states \cite{Sen}. In this way, acting with the worldsheet parity
operator $\Omega$ on the IIB D${}_0$-brane the type I stable
D${}_0$-brane can be found. This example is very interesting
because it can serve as an illustration to the S-duality between
type I and heterotic $SO(32)$ theories. At the first massive level
the heterotic theory possesses non-BPS states which are in the
spinor representation of the $SO(32)$ gauge group. But because
these states are the lightest in such representation they could
not decay without violating the quantum numbers conservation law.
Hence, they are stable. Now one can identify the states as the
dual partners of the stable type I D${}_0$-branes.

The stable non-BPS brane states can be also studied in
type IIA theory compactified on orbifold $T^4/{\calI}^4$, where
${\calI}^4$ is the spacetime parity which changes signs of the
four compactified coordinates. The model contains for example
D${}_1$-branes. By T-dualisation it gives type IIB theory on
$T^4/(-1)^{F_L}{\calI}^4$ which is furthermore dual to the IIB on
$T^4/\Omega{\calI}^4$.
For more examples see also \cite{Sen, LerdaR, Mukhi}.

%%%%%%%%%%%%%%%%%%%%%%%%%%%%%%%%%%%%%%%%%%%%%%%%%%%%%%%%%%%%%%%%%
\subsubsection{NS-branes.}

\akapit There is one more antisymmetric field, common for massless
limits of all five basic superstring theories and not correlated
to any of the D${}_p$-brane. This is the rank three tensor
$H_{[3]}$ belonging to the NS-NS sector as it was shown in
(\ref{sugra_IIA_ac}), (\ref{sugra_IIB_ac}) and (\ref{sugra_I_ac}).
The antisymmetric potential of the field couples to the
fundamental superstring, so we can interpret the superstring as an
electric NS${}_1$-brane.

Furthermore an existence of a magnetic NS${}_5$-brane, can be
predicted. Such an object is a solitonic solution of classical
equations of motion. It will be shown later that there is also an
additional argument based on S-duality that the magnetic NS-branes
should be an intrinsic part of the general superstring theory.

First consider a fundamental string and a D${}_1$-brane in type
IIB theory. The objects are similar but not identical. Both are
$1$-branes and both have the same massless quantum excitations,
but their tensions are different and obey the relation:
 \be
  \frac{\tau_{F1}}{\tau_{D1}} = e^\phi.
 \ee
The above relation shows that in the weakly coupled limit where
$e^\phi << 1$ the fundamental string is much lighter than the
D${}_1$-brane but this is no longer true when the coupling is
getting stronger. For a strongly coupled case with
${e^\phi}'=e^{-\phi}$ the situation is opposite. The symmetry of
inversing the coupling constant with simultaneous exchange of
F-string and D${}_1$-brane is called weak-strong duality and is a
simplest example of S-duality. So, both states should be thought
of in general as different manifestations of the same object.
Similarly it can be checked that type IIB D${}_3$-brane is
S-selfdual.

Now let us take D${}_5$-brane in type IIB theory. It is
electric/magnetic dual to the D${}_1$ brane, which is S-dual to
the fundamental string, which is in turn electric/magnetic dual to
the NS${}_5$-brane. Therefore to preserve consistency,
D${}_5$-brane and NS${}_5$-brane should be related by S-duality:
 \be
  \begin{array}{ccccc}
   F_1         & \leftarrow & S & \rightarrow & D_1 \\
   \uparrow    &            &   &             & \uparrow \\
   el/mag      &            &   &             & el/mag \\
   \downarrow  &            &   &             & \downarrow \\
   NS_5        & \leftarrow & S & \rightarrow & D_5
  \end{array}
 \ee

Recall that a fundamental open string has its ends attached to
D${}_p$-branes. But if one object is joined with another, their
duals have to be tied up by the same relation. Therefore if we
have F-string with its endpoints on a D${}_5$-brane, we should
expect also a D${}_1$-brane glued to NS${}_5$-brane. Analogously
from a composition of a F-string and a D${}_3$-brane, a system
where a D${}_1$-brane ends on a D${}_3$-brane can be deduced.
Applying to the system various T-dualities one obtains a
D${}_p$-brane joined to D${}_q$-brane where $p$ and $q$ are
arbitrary.

%%%%%%%%%%%%%%%%%%%%%%%%%%%%%%%%%%%%%%%%%%%%%%%%%%%%%%%%%%%%%%%%%
\subsubsection{M-branes.}

\akapit Even more spectacular results follow when S-duality acts
on objects in the type IIA theory, especially on D${}_0$-branes.
Because the D${}_p$-brane sweeps out a $p+1$ dimensional
worldvolume and its tension is given by (\ref{brane_mass_3}), the
mass scale related to the D${}_p$-brane is of order:
 \be
  m_p \approx \left( e^{-\phi}
  \right)^{\frac{1}{p+1}} {\alpha'}^{-1/2}.
  \label{brane_mass_4}
 \ee
Therefore at strong coupling the lightest states are built of the
D${}_0$-branes and total mass of a system consisting $n$ objects
of this type is equal to:
 \be
  \frac{n}{ e^\phi {\alpha'}^{1/2}}.
 \ee
If the coupling $e^\phi$ tends to infinity, a split between states
with $n$ and $(n+1)$ branes tends to zero and in the limit the
spectrum becomes continuous. This picture looks identically as a
model with the eleventh dimension compactified on a circle of
radius:
 \be
  R_{11} = e^\phi {\alpha'}^{1/2},
 \ee
when the radius grows up. This is the essence of the argument that
the M-theory S-dual to type IIA should be eleven dimensional.

The M-theory is largely unknown, but because we assume that its
low energy limit is described by $D=11$ supergravity, we can
predict some properties of it. In particular, because in the
action (\ref{D11sugra}) a third rank antisymmetric potential is
present, we expect an existence of so called M${}_2$ and
M${}_5$-branes carrying respectively electric and magnetic charges
of $A_{[3]}$. Furthermore we can describe the branes of type IIA
superstrings in terms of M-theory with compactified eleventh
dimension:
 \begin{itemize}
  \item F-string is the M${}_2$-brane wrapped on the eleventh dimension,
  \item NS${}_5$-brane is the M${}_5$-brane transverse
  to the compactified dimension,
  \item D${}_0$-brane is a state carrying Kaluza-Klein electric charge,
  \item D${}_2$-brane is the M${}_2$-brane transverse to the
  compactified dimension,
  \item D${}_4$-brane is the M${}_5$-brane wrapped on the
  eleventh dimension,
  \item D${}_6$-brane is a Kaluza-Klein magnetic monopole,
%  \item D${}_8$-brane ???.
 \end{itemize}
Summarizing these considerations one can conjecture that in the
unified theory there should be essentially only one type of
electric brane and one type of magnetic brane.

%%%%%%%%%%%%%%%%%%%%%%%%%%%%%%%%%%%%%%%%%%%%%%%%%%%%%%%%%%%%%%%%%
%%%%%%%%%%%%%%%%%%%%%%%%%%%%%%%%%%%%%%%%%%%%%%%%%%%%%%%%%%%%%%%%%
%%%%%%%%%%%%%%%%%%%%%%%%%%%%%%%%%%%%%%%%%%%%%%%%%%%%%%%%%%%%%%%%%
\newpage
\section{Branes in supergravity.}
\label{sugra_branes}

\akapit Most of supergravity theories have in its field content
one or more antisymmetric tensors. Therefore as observed in the
previous section \ref{b_antytensor}, one should expect that among
possible classical solutions there exist also brane solutions.
Such solutions can be found by at least two equivalent methods.
The first one consists in solving the system of equations with an
inhomogeneity given by a delta function (source term in
(\ref{pbrane_GS+WZ_action})) coinciding with the location of a
brane. The second method consists in solving the homogenous
equations of motion and imposing boundary conditions appropriate
to given brane configuration. Integration constants (or their
combinations) describing the solution should be identified with
physical quantities like tension or charge.

The problem of finding and classifying all (or at least as many as
possible) classical solutions for a given model is of fundamental
importance not only because these solutions characterize the basic
features of the model but also because they serve as the building
blocks for constructing Hilbert space in quantum theory. The
importance of the issue in the case of higher dimensional
supergravities is strongly amplified since they are identified as
the low energy limit of the respective superstring states (which
are largely unknown except in the simplest cases). One can
therefore expects that supergravity description of the branes
should be instructive also from the superstring theory point of
view. First because it gives information about the underlying
classical geometry and second because supergravity solutions
should give us insight into the superstring spectrum. Furthermore
to solve the supergravity equations of motion exact methods can be
used while in superstring theory only perturbative methods are
known. Therefore the results obtained in these two ways can be in
some aspects complementary.

There are however several problems which have to be stressed.
Supergravity is only an effective low energy limit of the
superstring theory, so there is no guarantee that the exact
solution derived with constraints describing a given brane
configuration is still a good classical approximation of any exact
solution in the quantum theory. It has to be checked case by case
whether the constraints imposed agree with conditions leading to
the low energy limit \cite{HS, CHS}. More precisely, the
supergravity action is the limit of the string theory where
$\alpha' \rightarrow 0$. Because a tension of a p-brane
(\ref{brane_mass_3}) is of order ${\alpha'}^{-(p+1)/2}$, the
relation between a given string state and its low energy
(supergravity) description should be well defined when the total
mass of the respective brane configuration goes to infinity. In
the case of BPS branes, due to  vanishing force theorem, one can
prove that a solution describing $n$ branes is a direct
superposition of $n$ single brane solutions with the total mass:
 \be
  M_{n,p} = n \sqrt[p+1]{\tau_p},
 \ee
which tends to infinity for $n\rightarrow \infty$. But this simple
picture usually does not work for non-BPS states and despite some
partial results \cite{BDFLMR, BL}, the question of classical
description of non-BPS states is still open.

Another problem has a technical character. While there are
mathematical theorems stating that functions satisfying the
equations of motion with given boundary conditions always exist,
they do not provide us these solutions in explicit form. In other
words, we are able to find only a limited number of solutions,
usually only if some additional assumptions leading to
simplification of the starting equations are made. Therefore each
essentially new solution is a significant achievement.

%%%%%%%%%%%%%%%%%%%%%%%%%%%%%%%%%%%%%%%%%%%%%%%%%%%%%%%%%%%%%%%%
\subsection{Single charge solution in the harmonic gauge.}
\label{single_charge_harm}

\akapit Let us consider at the beginning one of the simplest
models (in detail discussed in \cite{Stelle}): $D$-dimensional
theory describing graviton $g_{MN}$, dilaton $\phi$ and a single
antisymmetric potential $A_{[n-1]}$ with its field strength
$F_{[n]}$. We examine only a consistent bosonic truncation of the
theory, i.e. with no terms with fermionic fields taken into
account. Equivalent formulation of the condition is an assumption
that vacuum values of the fermionic fields are identically equal
to zero. Additionally the contribution from the Chern-Simons term
is neglected.

The action is then:
 \be
  S= \int_\calM dX^M \sqrt{|g|}
 \left( R- \de_M\phi \de^M\phi - \frac{1}{2} e^{a\phi}|F_{[n]}|^2 \right)
 \label{1_brane_ac}
 \ee
and the equations of motion:
 \bea
  R_{MN} &=& \frac{1}{2} \de_M \phi \de_N \phi + S_{MN},
  \label{1_eqm_11} \\
  0 &=& \nabla_M \left( e^{a\phi} F^{N_1 \ldots N_n} \right),
  \label{1_eqm_12} \\
  \nabla^2 \phi &=& \frac{a}{2} e^{a\phi} |F_{[n]}|^2,
  \label{1_eqm_13}
 \eea
where:
 \be
  S_{MN} = \frac{1}{2}e^{a\phi}
  \left(\frac{1}{(n-1)!} F_{[n]MR_1\ldots R_{n-1}}
  F_{[n]N}{}^{R_1\ldots R_{n-1}}
   - \frac{n-1}{D-2} |F_{[n]}|^2 g_{MN} \right).
   \label{1_S_0}
 \ee

We search for a single charge solution, what means that the
$A_{[n-1]}$ has only one independent non-zero component which
carries electric or magnetic charge but not both (the case of
dyonic brane is excluded). The brane enforces a split of the whole
space-time $\calM$ into a multiplication of two mutually
orthogonal subspaces \footnote{For simplicity we identify a
submanifold with its tangent bundle.} $V \otimes V_\emptyset$,
where $V$ is a worldvolume of the brane and $V_\emptyset =
\calM/V$. We assume that the solution is maximally space-time
symmetric what is equivalent to breaking by the brane of the
Poincar\'{e} symmetry $ISO(D-1,1)$ to subgroup $ISO(d-1,1)\times
SO(D-d)$ where $d$ is the dimension of the brane worldvolume.

Under these assumptions we can write the following ans\"atze for
the metric tensor and the dilaton:
 \bea
  ds^2(X) &=& e^{2A(r)} dx^\mu dx^\nu \eta_{\mu\nu}+
  e^{2B(r)} dy^m dy^n \delta_{mn},
  \label{g_ansatz_1} \\
  \phi(X) &=& \phi(r),
 \eea
where $x^\mu$ with $\mu=1,\ldots d$ are coordinates in directions
parallel to the brane, $y^m$ with $m=d+1,\ldots D$ -- in
transversal and:
 \be
  r=\sqrt{y^m y^n \delta_{mn}}.
 \ee
For the antisymmetric tensor we need to distinguish two separate
cases. The electric brane corresponds to $F$ with the only
nonvanishing component:
 \be
  F_{m \mu_1 \ldots \mu_d} (X) = \sigma
  \epsilon_{\mu_1 \ldots \mu_d} \de_m \exp (C(r)),
  \label{F_el_ansatz_1}
 \ee
where $\sigma=\pm 1$, but the solitonic brane:
 \be
  F_{m_1 \ldots m_{\td+1}} (X) =
  \epsilon_{ m_1 \ldots m_{\td+1} n} \frac{\lambda y^n}{r^{\td+2}},
  \label{F_mag_ansatz_1}
 \ee
where $\lambda$ is a real constant.

Substituting (\ref{g_ansatz_1} -- \ref{F_mag_ansatz_1}) into
(\ref{1_eqm_11} -- \ref{1_eqm_13}) we can rewrite the equations of
motion as:
 \bea
  A''+d(A')^2+\td A'B' +\frac{\td+1}{r} A' &=&
  \frac{\td}{2(D-2)} (S')^2,
  \label{1_eqm_21} \\
  B''+\td (B')^2+ d A'B' +\frac{2\td+1}{r} B' +
  \frac{d}{r} A' &=& -\frac{d}{2(D-2)} (S')^2, 
 \eea
 \bea
  \td B''+\! dA''-\!2dA'B'+\!d(A')^2-\!\td 
(B')^2-\!\frac{\td}{r}B'-\!\frac{d}{r}A' +
  \!\frac{1}{2}(\phi')^2 &=& \frac{1}{2} (S')^2, \\
  \phi''+d A' \phi' +\td B'\phi' +\frac{\td+1}{r} \phi'
  &=& - \frac{\varsigma a}{2} (S')^2, \\
  C'' + C' \left( C' + \frac{\td+1}{r} -dA' +\td B' +a \phi' \right) &=& 0,
  \label{1_eqm_25}
 \eea
where the prime denotes a derivation with respect to $r$ and:
\bea
  \varsigma &=& \left\{ \begin{array}{ll}
   +1 & \mbox{(electric),} \\
   -1 & \mbox{(magnetic),} \end{array} \right. \\
  S' &=& \left\{ \begin{array}{ll}
   \sigma \exp(\frac{1}{2} a \phi - d A )(e^C)' & \mbox{(electric),} \\
   & \\
   \exp(\frac{1}{2} a \phi - \td B)
   \frac{\lambda }{r^{\td+1}} & \mbox{(magnetic).}
  \end{array} \right.
  \label{1_S_1}
 \eea

A crucial observation is that the equations are drastically
simplified when additional assumptions about the solution are
made:
 \bea
  dA' + \td B' &=& 0,
  \label{1_lincon_11} \\
  \td \phi' + \varsigma a (D-2) A' &=& 0,
  \label{1_lincon_12}
 \eea
where:
 \be
  \td > 0, \qquad a \neq 0.
 \ee
The (\ref{1_lincon_11}) is sometimes called the harmonic gauge
because it always leads to solutions where the metric tensor is
expressed via harmonic functions \cite{Gibbons, IM2}. In this
case, the solution reads:
 \bea
  ds^2 &=& H(r)^{\frac{-4\td}{\Delta(D-2)}} dx^\mu dx^\nu \eta_{\mu\nu}
        +  H(r)^{\frac{4d   }{\Delta(D-2)}} dy^m dy^n \delta_{mn},
        \label{1_sol_1} \\
  e^\phi &=& H(r)^{\frac{2\varsigma a}{\Delta}},
  \label{1_sol_2} \\
  F^{electric}{}_{m \mu_1 \ldots \mu_d} &=& \sigma
  \epsilon_{\mu_1 \ldots \mu_d} \de_m \left( H(r)^{-1} \right),
    \label{1_sol_3} \\
  F^{magnetic}{}_{m_1 \ldots m_{\td+1}} &=&
  \epsilon_{m_1 \ldots m_{\td+1}n} \de^n H(r),
  \label{1_sol_4} \\
  H(r) &=& 1+ \frac{k}{r^\td},
  \label{1_sol_5} \\
  \Delta &=& \frac{2d\td}{D-2} + a^2,
  \label{1_sol_6}
 \eea
where $k$ is a positive integration constant -- in the magnetic
case:
 \be
  k = \frac{\sqrt{\Delta} \lambda}{2 \td}
  \label{k_lambda}.
 \ee
To arrive at a more symmetric relation between electric and
magnetic cases, it is convenient to use this identity as a
definition of $\lambda$ for the electric brane.
Note, that the constant $\Delta$ can be equivalently written as:
 \be
  \Delta = \frac{2}{\frac{1}{d}+\frac{1}{\td}} + a^2, \label{1_sol_6b}
 \ee
so it is just harmonic average of $d$ and $\td$ enlarged by $a^2$.
Setting $\phi=0$ and $a=0$ in the above solution we obtain a
solution for a model without the dilaton. If we include a dilaton
but still keep $a=0$ it is necessary to replace (\ref{1_sol_2})
by:
 \be
  e^\phi = H(r) \label{1_sol_7}.
 \ee

In some special cases, the solution (\ref{1_sol_1}-\ref{1_sol_7})
describes:
\begin{itemize}
 \item M${}_2$-brane \cite{DS} if $D=11$, $d=3$ and $\phi = 0$,
 \item M${}_5$-brane \cite{Guven} if $D=11$, $d=6$ and $\phi = 0$,
 \item NS${}_1$-brane \cite{DGHR} if $D=10$, $d=2$ and $a = -1$,
 \item NS${}_5$-brane \cite{DL} if $D=10$, $d=6$ and $a = -1$.
\end{itemize}

%%%%%%%%%%%%%%%%%%%%%%%%%%%%%%%%%%%%%%%%%%%%%%%%%%%%%%%%%%%%%%%%%%%%%%%%
\subsubsection{Geometry of the solution.}
\label{geo_sol_susy}

\akapit The positivity of $k$ of the previous subsection is a
consequence of an extra requirement that we want to avoid a
singularity for $r>0$. Other restrictions imposed on integration
constants to derive the above solutions are: $A,B,\phi \rightarrow
0 $ when $r\rightarrow \infty$, what means that at the infinity
the solution asymptotically goes to Minkowski spacetime. Near
$r=0$ the geometry is curved and corresponds to $AdS_{d+1} \times
S^{\td+1}$.

Consider the solution without the dilaton. More careful analysis
shows that the point $r=0$ is rather a horizon than a singularity
and the solution can be extended beyond it \cite{GT,DGT,GHT}. It
is convenient to introduce the so called interpolating coordinates
where $r$ is replaced by $r_{int}$ satisfying:
 \be
  r^\td = \frac{k r_{int}^d}{1-r_{int}^d}.
  \label{r_r_int_change}
 \ee
So, the "flat infinity" $r \rightarrow \infty$ corresponds to
$r_{int} \rightarrow 1$ and the horizon $r=0$ to $r_{int}=0$. The
horizon is sometimes called a degenerate one, because its
properties are not the same as for the horizon in the classical
Schwarzschild solution. Particularly for the brane solution the
$g_{tt}$ component of the metric tensor does not change its sign
at the horizon as it happens at the Schwarzschild horizon. In
other words light-cones cannot flip over inside the horizon.

The interpolating coordinates are well defined also for
$r_{int}<0$. If $d$ is odd it can be checked that at
$r_{int}\rightarrow -\infty$ the solution describes a geometry
near a naked singularity and the singularity can be therefore
identified with a localization of a fundamental brane. If $d$ is
even, the formula (\ref{r_r_int_change}) is invariant under
$r_{int} \leftrightarrow - r_{int}$, so the area described by
negative $r_{int}$ is a mirror of the area where $r_{int}$ is
positive and there is no singularity at all. Of course the
construction is only one of many possible analytical
continuations. However its virtue is that in $D=11$ supergravity
it allows to identify the electric $2$-brane with the singular
solution as it should be for the elementary object of the theory,
while the magnetic non-singular $5$-brane solution can be
interpreted as a soliton.

If we consider the solution with the dilaton the situation is a
little different. The field $\phi$ is scalar, so it is invariant
under any coordinate change and the points where it tends to
infinity are necessary singularities of the whole solution. By
(\ref{1_sol_2}) it means that the horizon $r=0$ coincides with the
singularity.

The solution (\ref{1_sol_1}-\ref{1_sol_7}) can be extended to
cases $\td=0$ or $\td=-1$ when the factor $k/r^\td$ is replaced by
$k \ln r$ or $kr$ respectively. But for such solutions it is not
true that at $r \rightarrow \infty$ they describe Minkowski
spacetime or (if $\td=0$) that there is no singularity of the
metric for positive $r$.

%%%%%%%%%%%%%%%%%%%%%%%%%%%%%%%%%%%%%%%%%%%%%%%%%%%%%%%%%%%%%%%%%%%%%%
\subsubsection{Charges and tension.}

\akapit Let us calculate a number of degrees of freedom of the solution
(\ref{1_sol_1} -- \ref{1_sol_6}). The ans\"atze were expressed in
terms of four scalar function $A,B,C,\phi$ which appear in the
equations of motion (\ref{1_eqm_21} -- \ref{1_eqm_25}) with their
second derivatives. Integrating the equations, one should expect 8
integration constants. But, since there are four function and five
equations, one of them can by used as a constraint reducing the
number of free parameters by one. The linearity conditions
(\ref{1_lincon_11} -- \ref{1_lincon_12}) cancel next two and the
requirement of $A,B,\phi \rightarrow 0 $ at $r\rightarrow \infty$
additional three. One of the remaining two illustrates the fact
that $A_{[n-1]}$ is determined up to an additive constant. The
last one is the parameter $k$ and being the only essential
parameter it has to determine both the energy and charge of a
solution.

We can calculate the electric and magnetic charges associated with
the solution from the identities (\ref{el_charge}) and
(\ref{mag_charge}). If we use the relation (\ref{k_lambda}) then
we get:
 \be
  Q_{e/m} = \lambda \Omega_{\td+1} =
  \frac{2 k \td}{\sqrt{\Delta}} \Omega_{\td+1},
 \ee
where $\Omega_{\td+1}$ is a surface of unit $(\td+1)$-sphere.

Next, thanks to the ADM mass formula which reads \cite{MP, Maldacena}:
 \be
  g_{tt} = -1 + \frac{\tau_{d-1}}{\Omega_{\td+1}(D-2)r^\td}+
  o\left(\frac{1}{r^\td}\right),
   \qquad \mbox{when} \quad r \rightarrow \infty,
 \ee
or equivalently \cite{Stelle}:
 \be
  \tau_{d-1} = \int_{\de V_\emptyset} d \Omega_{\td+1} r^\td y^m
  \left( \de^n h_{mn} - \de_m h^i_i \right),
 \ee
where:
 \be
  g_{MN} = \eta_{MN} + h_{MN}
 \ee
and the indices $i$ run over all $D-1$ spatial directions, we find
the tension (or the energy density) of the solution:
 \be
  \tau_{d-1} = \frac{2}{\sqrt{\Delta}} \lambda \Omega_{\td+1}.
  \label{1_energ_1}
 \ee

Another way leading to the same result is to find the stress-energy
density pseudotensor:
 \be
  T_{MN} = T(A)_{MN} + T(\phi)_{MN},
 \ee
where $T(A)_{MN}$ and $T(\phi)_{MN}$ are contribution to the total
stress-energy density coming from the antisymmetric potential and
the dilaton respectively. {}From (\ref{1_eqm_11}) we see, that:
 \bea
  T(A)_{MN} &=& S_{MN} - \frac12 S_R^R g_{MN}
   = \frac{e^{a\phi}}{2(n-1)!}
   F_{[n]MR_1\ldots R_{n-1}} F_{[n]N}{}^{R_1\ldots R_{n-1}},
   \label{1_s_en_2} \\
  T(\phi)_{MN}  &=& \frac12 \de_M \phi \de_N \phi -
  \frac14 \de_R \phi \de^R \phi g_{MN}.
  \label{1_s_en_3}
 \eea
And the energy density corresponding to the $T_{tt}$ component of
$T_{MN}$, is:
 \be
  T_{tt} = T(A)_{tt} + T(\phi)_{tt} = (S')^2 + (\phi')^2.
 \ee
The energy of the intersecting branes system is then:
 \be
  \calE = \int_\calM d^DX \sqrt{|\det g|} T(A)^t{}_t
    = \Omega_{\td+1} |V| \int_0^\infty dr r^{\td+1} (S')^2
    = \frac{2 \lambda}{\sqrt{\Delta}} \Omega_{\td+1} |V|
    \label{1_energ_3}
 \ee
and it is precisely equal to (\ref{1_energ_1}) multiplied by the
brane worldvolume.

In $D=11$ supergravity the electric brane solution is
characterized by $d=3$, $\td=6$, $a=0$ and the magnetic one by
$d=6$, $\td=3$, $a=0$. So, for both $\Delta=4$. At $D=10$
NS-branes have $d=2$, $\td=6$ or $d=6$, $\td=2$ and $a=-1$ what
gives $\Delta=4$ again. Also, the analogous solutions related to
D-branes have the same value of $\Delta$. Therefore all of them
saturate the BPS bound (\ref{BPS_1}) and are supersymmetric. Also
in lower dimensions most antisymmetric fields and then the branes
corresponding to them are described by $\Delta=4$ so the solutions
in harmonic gauge preserve half of the original supersymmetries.
This general rule is a consequence of a fact that a value of the
$\Delta$ is preserved under Kaluza-Klein dimensional reduction
\cite{LPSS} and all supergravity theories in lower dimensions can
be derived with this procedure from $D=11$ supergravity or type
$N=2$, $D=10$ supergravity. However other values of $\Delta$ are
also possible. It can happen when the antisymmetric tensor in the
considered supergravity theory is a linear combination of several
fields of the same rank obtained by the dimensional reduction
procedure \cite{LP2}. Such solution observed from the higher
dimensional point of view describes rather several coinciding
branes.

%%%%%%%%%%%%%%%%%%%%%%%%%%%%%%%%%%%%%%%%%%%%%%%%%%%%%%%%%%%%%%%%%%%%
\subsubsection{Supersymmetry.}
\label{susy_break}

\akapit The supersymmetry preserving condition is in general given
by the requirement that for all fermionic fields $\Psi$ in the
theory the vacuum value of its supersymmetric transformation
$<\delta_\eta \Psi >$ is vanishing. We take as an example an
electric 2-brane in $D=11$ supergravity. Because the theory
possesses only one fermionic field $\Psi_M$, the supersymmetry
preserving condition is just:
 \be
  \delta_\eta \Psi_M = 0,
  \label{1_D11_susy_con_1}
 \ee
where $\delta_\eta \psi_M$ is given by (\ref{D11sugra_psi_eta}).
The elementary brane splits the $11$-dimensional spacetime into
$3+8$, hence $11$-dimensional Dirac matrices $\Gamma_M$ have to
decompose as:
 \bea
  \Gamma_M &\rightarrow& \left( e^A \Gamma_\mu, e^B \Gamma_m \right),\\
  \Gamma_\mu   &=& \gamma_\mu \times \Sigma_V, \\
  %\qquad \mbox{for} \quad \mu = 1,2,11, \\
  \Gamma_m     &=& Id \times \Sigma_m,
  % \qquad \mbox{for} \quad m = 3,\ldots, 10,
 \eea
where $\gamma_\mu$ are Dirac matrices in $3$ dimensions,
$\Sigma_m$ --- Dirac matrices in $8$ dimensions and $\Sigma_V$ is
the $8$-dimensional chiral operator. It is convenient to
introduce:
 \be
  \Gamma_V=Id \times \Sigma_V, \label{gamma_v}
 \ee
so the following identities hold:
 \bea
  \Gamma^{\mu\nu\rho}\epsilon_{\mu\nu\rho} &=& 6\Gamma_V, \\
  \Gamma^{\mu\nu}\epsilon_{\mu\nu\rho} &=& 2\Gamma_\rho \Gamma_V.
 \eea

Substituting (\ref{g_ansatz_1}) and (\ref{F_el_ansatz_1}) into
(\ref{1_D11_susy_con_1}) we get:
 \bea
  \delta_\eta \Psi_\mu &=& \frac{1}{6} e^{-2A-B} \Gamma_\mu\Gamma^m\de_m
  \left(e^{3A}+\sigma e^C\Gamma_V\right) \eta,
  \label{susy_el_psi1} \\
  \delta_\eta \Psi_m &=& \left(\de_m N +
  \frac{\sigma}{6} e^{C-3A} \de_m C \Gamma_V \right) \eta
   + \frac{1}{2}\Gamma_m{}^n \left(\de_n B - \frac{\sigma}{6} e^{C-3A}
  \de_n C \Gamma_V \right) \eta.
  \label{susy_el_psi2}
 \eea
where:
 \be
  \eta(r)=e^{N(r)} \eta_0
 \ee
and $\eta_0$ is a constant parameter. So, we see, that the
condition (\ref{1_D11_susy_con_1}) is satisfied only
if simultaneously:
 \bea
  2B' + A' &=& 0,
  \label{susy_unb1_11e} \\
  \frac{d}{dr} e^{3A} &=& \sigma \frac{d}{dr} e^C,
  \label{susy_unb2_11e} \\
  2N' &=& A',
  \label{susy_unb3_11e} \\
  \Gamma_V \eta_0 &=& -\sigma \eta_0.
  \label{susy_unb4_11e}
 \eea
The first of these conditions is equivalent to (\ref{1_lincon_11})
and the second is a consequence of (\ref{1_lincon_11}) and
(\ref{1_lincon_12}). This shows explicitly that harmonic gauge is
a necessary condition for preserving supersymmetry. The third
condition states that the preserved supersymmetry is rigid and the
fourth that the spinorial parameters corresponding to the
preserved part of the supersymmetry are chiral in the $8$
dimensions transversal to the brane. We see also that
(\ref{susy_unb4_11e}) is invariant under a transformation:
 \be
  \sigma \rightarrow - \sigma
 \ee
describing a reversal of the electric charge and a reversal of the
parameter $\eta$ chirality, at once. In other words, the brane and
anti-brane preserve opposite chirality parts of supersymmetry.

For a magnetic $5$-brane in the $D=11$ supergravity an equivalent
derivation to the previously described can be conducted giving
analogous conclusions.
However, in this case instead of
(\ref{susy_el_psi1} -- \ref{susy_el_psi2}) one has:
 \bea
  \delta_\eta \Psi_\mu &=&
  \frac{1}{2} e^{A-B} \Gamma_\mu \Gamma^m \frac{y_m}{r}
  \left(A'+\frac{1}{6} e^{-3B} \frac{\lambda}{r^4} \Gamma_V \right) \eta,
  \label{susy_mag_psi1} \\
  \delta_\eta \Psi_m &=& \frac{y_m}{r}
  \left(N' + \frac{1}{12} e^{-3B} \frac{\lambda}{r^4} \Gamma_V \right) \eta
   + \Gamma_m{}^n \frac{y_m}{r}
  \left(\frac{1}{2} B' - \frac{1}{6} e^{-3B} \frac{\lambda}{r^4}
  \Gamma_V \right) \eta, \label{susy_mag_psi2}
 \eea
with:
 \be
  \Gamma^{mnrst}\epsilon_{mnrst} = 120 \, \Gamma_V.
 \ee
The decomposition of the Dirac matrices gives now:
 \bea
  \Gamma_\mu &=& \gamma_\mu \otimes \Sigma_{11}, \\
  \Gamma_m   &=& Id \otimes \Sigma_m,
  \qquad \mbox{for} \quad m=7,\ldots,10, \\
  \Gamma_{11} &=& \gamma_V \otimes \Sigma_{11}, 
 \eea
where $\gamma_\mu$ ($\mu=1,\ldots, 6$) are the Dirac matrices
defined on the worldvolume of the brane with $\gamma_V$ the six
dimensional chiral operator constructed of them and $\Sigma_m$
($m=7,\ldots,11)$ are the Dirac matrices on the remaining five
dimensions. So, for the operator $\Gamma_V$ we have:
 \be
  \Gamma_V = \gamma_V \otimes Id.
 \ee

%%%%%%%%%%%%%%%%%%%%%%%%%%%%%%%%%%%%%%%%%%%%%%%%%%%%%%%%%%%%%%
\subsubsection{Multi center solution.}

\akapit The crucial trick applied to find the solution
(\ref{1_sol_1} -- \ref{1_sol_6}) was to make additional
assumptions simplifying the equations of motion (\ref{1_eqm_11} --
\ref{1_eqm_13}) which allowed to reduce it to:
 \be
  \nabla^2 H = \left( \de^2_r + \frac{\td+1}{r} \de_r \right) H = 0.
  \label{harm_eq}
 \ee
But because the operator $\nabla^2$ is linear it means that if any
two $H_1$ and $H_2$ obey (\ref{harm_eq}) their sum $H_1+H_2$ is
also a good solution of the equation. Therefore in general one can
replace (\ref{1_sol_5}) by:
 \be
  H(r) = 1 + \sum_{i=1}^N \frac{k_i}{|\vec{y}+ \vec{y}_{0i}|^\td}.
 \ee
A natural interpretation of this so-called multi-center solution
is to assume that it describes a set of $N$ parallel identically
oriented (i.e. characterized by the same sign of the R-R charge)
branes, each localized at $y^m = y^m_{0i}$. Physically, a
possibility of such a solution follows from a fact that in such
configuration of branes attractive forces carried by the graviton
$g_{MN}$ and the dilaton $\phi$ are precisely cancelled by
repulsive forces of identically oriented antisymmetric fields. The
whole brane configuration breaks the same amount of supersymmetry
as each of its components and the total charge denisty and energy 
density is given by:
 \be
  Q = \Omega_{\td+1} \sum_i \lambda_i, \qquad \calE =
  \Omega_{\td+1} \frac{2}{\sqrt{\Delta}} \sum_i \lambda_i.
 \ee

%%%%%%%%%%%%%%%%%%%%%%%%%%%%%%%%%%%%%%%%%%%%%%%%%%%%%%%%%%%%%
%%%%%%%%%%%%%%%%%%%%%%%%%%%%%%%%%%%%%%%%%%%%%%%%%%%%%%%%%%%%%
\subsection{Nonsupersymmetric single-charge solutions.}

\akapit The solution given in the previous section can be
generalized if some of the constraints imposed on the model are
relaxed. One possibility is to drop the harmonic gauge condition
(\ref{1_lincon_11}). A class of solutions obtained in this way was
initially discussed in \cite{LPX, Tollsten} and a complete
solution was presented in \cite{ZZ1}. We will discuss those cases
in more detail in chapter \ref{nonharm_branes}. Let us only
mention here that since harmonic gauge is not imposed, the
equations of motion cannot be reduced to (\ref{harm_eq}) and the
solutions are not still governed by harmonic functions and are
nonsupersymmetric in general. In a connection with string theory
they can be used as the classical description of the
brane-anti-brane system \cite{BMO}.

%%%%%%%%%%%%%%%%%%%%%%%%%%%%%%%%%%%%%%%%%%%%%%%%%%%%%%%%%%%%%%%%%%%%%%
\subsubsection{Black branes.}

\akapit Another possibility is to relax the demand of the $ISO(d-1,1)$
symmetry on the brane worldvolume. Branes of this type are called
black branes \cite{Guven, HS, DL, DLP} because of their similarity
to the Schwarzschild black hole\footnote{ More precisely a black
hole can be interpreted as a black $0$-brane.}. The simplest case
occurs when instead of the $ISO(d-1,1)$ we have $SO(d-1)$
symmetry. The appropriate ansatz for the metric tensor is then:
 \be
  ds^2(X) = - e^{2A_t(r)}dt^2 + e^{2A_x(r)}
  dx^{\tilde{\mu}} dx_{\tilde{\mu }} + e^{2B(r)} dy^m dy_m,
 \ee
where the indices $\tilde{\mu}$ run through the $d-1$ spatial
directions parallel to the brane. In such a case the harmonic
gauge condition (\ref{1_lincon_11}) has to be rewritten in the
form:
 \be
  A_t'+(d-1)A_x'+\td B' = 0.
 \ee
It is convenient to write the black brane solution in
Schwarzschild coordinates, where the Schwarzschild radial
coordinate $r_s$ is related to the original isotropic $r$ by:
 \bea
  r &=& r_s \left( \frac{\sqrt{H_+(r_s)}+
  \sqrt{H_-(r_s)}}{2}\right)^{\frac{2}{\td}}, \\
  H_\pm (r_s) &=& 1- \left( \frac{r_\pm}{r_s} \right)^\td.
 \eea
The spacetime interval takes then a form:
 \bea
  ds^2 &=& H_-(r_s)^{\frac{4\td}{\Delta(D-2)}}
        \left( - \frac{H_+(r_s)}{H_-(r_s)} dt^2 +
        dx^{\tilde{\mu}} dx_{\tilde{\mu }} \right) \nn \\
       &+& H_-(r)^{\frac{2a^2}{\Delta \td}-1}
        \left( \frac{1}{H_+(r_s) H_-(r_s)} dr_s^2 +
        r_s^2 d\Omega^2_{\td+1} \right)
 \label{2_sol_1}
 \eea
and the scalar field:
 \be
  e^\phi = H_-(r_s)^{\frac{2\varsigma a}{\Delta}}.
  \label{2_sol_2}
 \ee
The solution depends on two nonnegative parameters $r_+$ and
$r_-$. The first describes a localization of an event horizon and
the second of an inner horizon.
Both the horizons are nongenerate, it means that similarly as for
the horizon in the Schwarzschild black hole solution signs of the
$g_{tt}$ and $g_{rr}$ components of the metric tensors are reversed
when one goes with the solution through the horizon.
If $r_+=r_-$ both the sing reversions cancel one with the other
and one obtains the degenerate horizon known for the supersymmetric
brane solution discussed in the paragraph \ref{geo_sol_susy}.
The horizon at $r_s=r_+$ is nonsingular, but the horizon at
$r_s=r_-$ usually coincides with a singularity what is a
consequence of a fact that the dilaton depends on $H_-$.

The two parameters $r_+$ and $r_-$ describe also the energy density and
the charge density of the black brane:
 \bea
  \calE &=& \Omega_{\td+1} \td \left( r_+ r_- \right)^{\td/2}, \\
  Q &=& \Omega_{\td+1} \left( (\td+1)r_+^\td - r_-^\td \right).
 \eea
So, we see that in general the solution does not saturate the BPS
bound and is not supersymmetric in spite of the fact that it is
expressed in terms of harmonic functions $H_+$ and $H_-$. This
explicitly illustrates that the harmonic gauge leads to harmonic
functions in the solution but it is not sufficient to preserve
supersymmetry. However in a special situation when $r_+=r_-$ the
solution becomes supersymmetric and reproduces the isomorphic
solution (\ref{1_sol_1} -- \ref{1_sol_6}).

Finally it is also possible to find a brane solution without
imposing the harmonic gauge and Poincar\'{e} worldvolume symmetry.
This kind of generalized black branes was given in \cite{ZZ2} and
discussed in \cite{BMO}.

%%%%%%%%%%%%%%%%%%%%%%%%%%%%%%%%%%%%%%%%%%%%%%%%%%%%%%%%%%%%%%%%%%%%
%%%%%%%%%%%%%%%%%%%%%%%%%%%%%%%%%%%%%%%%%%%%%%%%%%%%%%%%%%%%%%%%%%%%
\subsection{Solutions with many branes.}

\akapit The procedure of relaxing some constraints imposed on the
equations of motion gives a very rich collection of various
solutions if we allow for configurations describing many branes
instead of the single brane. Such configurations can be introduced
to the model in two fundamental ways (with possible combinations
of the two).

The first is to consider a theory with several, say $N_A$,
antisymmetric tensors $F^i_{[n_i]}$ and assume that each of the
fields still supports only one brane. Because the ranks $n_i$ can
take various values, the branes have dimensions potentially
completely uncorrelated with one another. An action relevant for
such model can be built starting from the single brane action
(\ref{1_brane_ac}) by replacing the single kinetic terms of
antisymmetric fields by a sum:
 \be
  \frac{1}{2} \sum_{i=1}^{N_A} e^{a_i \phi} \left| F^i_{[n_i]} \right|^2.
 \ee
In the equations of motion instead of $S_{MN}$ we obtain a sum
$\sum_i S^i_{MN}$ where each component describes the contribution
from different $F^i_{[n_i]}$ and thus different branes. The system
of equations is also enlarged because (\ref{1_eqm_12}) now appears
in $N_A$ copies, one for each $F^i_{[n_i]}$.

In the second method we still have only one antisymmetric tensor
but allow it to be the source of several branes. In this case the
branes are of only two possible kinds: one electric and one
magnetic. The action of the model is the same as
(\ref{1_brane_ac}), all differences are encoded in the form of the
assumed solution. Of course a natural generalization is to combine
these two ways and study brane configurations related to systems
with many antisymmetric fields each describing many independent
branes. The situation where many branes are given by one
antisymmetric field is known in literature as composite branes.

The composite branes of a single antisymmetric tensor can be in
some cases equivalently described in terms of several single
charged antisymmetric fields. This possibility arises if the first
component of the $S_{MN}$ (\ref{1_S_0}) tensor has diagonal form
or equivalently when the stress--energy tensor $T(A)_{MN}$
(\ref{1_s_en_2}) is diagonal. Then:
 \be
  F_{[n]MR_1\ldots R_{n-1}} F_{[n]N}{}^{R_1\ldots R_{n-1}}=0,
  \qquad \mbox{if} \qquad M\neq N. \label{s_en_diag}
 \ee
This means that there are no direct interactions among independent
elements of the $F^i_{[n]}$ \cite{IM1}.

In supergravity theories there can also exist different scalar
fields so the multi-scalar brane models should be also considered.
But in supergravities there are two kinds of scalar fields each
having different features. One type consists of dilaton fields
which in lagrangians appear as exponential factors multiplying the
kinetic terms of the antisymmetric fields. The other type consists
of rank zero antisymmetric potentials which can be responsible for
the existence of instantonic branes.

For a single brane it is always possible to choose Cartesian
coordinate system with $d$ directions parallel to the brane and
$\td+2=D-d$ perpendicular to it and describe localization of the
brane by constraints $X^m=X_0^m$ where $m$ runs from $d+1$ to $D$.
This possibility is very convenient because it allows to identify
the brane worldvolume with a single independent element of
$A_{[n-1]}$ antisymmetric potential. But for two or more branes
the feature has in general no simple extension. If we orient the
Cartesian coordinate system to be in an agreement with one brane
nothing can a priory guarantee that the other is either parallel
or perpendicular to the directions of the coordinate system. In
other words the branes can be oriented at any angle one to the
other.

But it this work we restrict ourselves only to configurations of
orthogonally or paralelly oriented branes. Consider then two
branes of this kind: a $p_1$-brane and a $p_2$-brane. Such a
configuration induces a split of the spacetime directions into
four segments: $\{1,2\}$, $\{1\}$, $\{2\}$ and $\emptyset$. The
$\{1,2\}$ contains directions parallel to both of the branes. It
is necessarily not trivial because the time always belongs to it.
The $\{1\}$ segment is characterized as tangent to the first and
perpendicular to the second brane. Analogously the $\{2\}$ segment
-- parallel to the second but transverse to the first. Finally the
$\emptyset$ segment contains directions normal to both the branes.
The directions in the $\{1,2\}$ segment are usually called common
tangent, in the $\{1\}$ and $\{2\}$ -- relative transverse and in
the $\{\emptyset\}$ -- overall transverse. Further we can
decompose the set of the coordinates $\{ X^M \}$ into four
subsets:
 \be
  \{ X^M \} \rightarrow \{ x^{\mu_{\{1,2\}}}, x^{\mu_{\{1\}}},
  x^{\mu_{\{2\}}}, y^m \}.
 \ee
A spacetime localization of the branes is completely determined
when the following constraints are imposed:
 \be
  \begin{array}{ccc}
   x^{\mu_{\{2\}}} = x^{\mu_{\{2\}}}_1, & y^m = y^m_1,
   & \mbox{for the first brane,} \\
   x^{\mu_{\{1\}}} = x^{\mu_{\{1\}}}_2, & y^m = y^m_2,
   & \mbox{for the second brane.}
  \end{array}
 \ee
where $x^{\mu_{\{1\}}}_2, x^{\mu_{\{2\}}}_1, y^m_1, y^m_2$ are
constants. The splitting procedure can be extended to arbitrary
$N_A$ and always gives one common tangent
segment, no more then one overall transverse and maximally
$2^{N_A}-2$ relative transverse segments.

%%%%%%%%%%%%%%%%%%%%%%%%%%%%%%%%%%%%%%%%%%%%%%%%%%%%%%%%%%%%%%%%
\subsubsection{Intersecting branes.}

\akapit The two branes as described before intersect
orthogonally\footnote{ Of course in general branes can intersect
not only orthogonally but also at arbitrary angles. See for example
\cite{BDL, S-J, GGPT}} when $y^m_1 = y^m_2$ and the intersection 
is a subspace given by:
 \be
   x^{\mu_{\{1\}}} = x^{\mu_{\{1\}}}_2, \quad
   x^{\mu_{\{2\}}} = x^{\mu_{\{2\}}}_1, \quad
   y^m = y^m_1 = y^m_2.
 \ee

A description of branes intersections becomes more complicated
when the number of branes is bigger than two because intersections
of any two different pairs of branes may have no common point.
However, in a special situation there exists nonempty subspace
common for all the $N_B$ branes. It is naturally to name the case
as commonly intersecting branes however it is usually called just
"intersecting branes" \cite{Gauntlett} what is shorter but less
precise. The (commonly) intersecting brane configurations are very
special and have many features distinguishing them from the
others, what is a consequence of a fact that they have relatively
more spacetime symmetries. Let us discuss it in some detail.

Each $p$-brane breaks the original $ISO(D-1,1)$ symmetry of flat
empty spacetime to $ISO(p,1)\times SO(D-p-1)$. But when we have two
or more branes the symmetry is usually only a local approximation,
because in globally each brane can break a different part of
$ISO(D-1,1)$. Therefore a group of the commonly preserved symmetry
is much smaller and in the extreme case can be reduced to a
translation in time exclusively. But the situation is different if
the branes are commonly orthogonally intersecting because then the
preserved symmetry has always a form:
 \be
  ISO(d_{\{1,2,\ldots,N_A\}}-1,1) \times
  \left( \bigotimes_{\tilde{I}} SO(d_{\tilde{I}}) \right)
  \label{ort_com_int_sym}
 \ee
where $d_{\{1,2,\ldots,N_A\}}$ is a dimension of the intersection
common for the all $N_A$ branes and $d_{\tilde{I}}$ are dimensions
of the other segments of the spacetime distinguished by the
branes.

For a solution corresponding to a brane configuration which is
characterized by (\ref{ort_com_int_sym}) it is natural to assume
that it depends on variables $r_{\tilde{I}}$, where
$r_{\tilde{I}}$ is the radial coordinate in the ${\tilde{I}}$-th
segment and where the origin of the coordinate system lies on the
common intersection. But when such ans\"{a}tze are imposed the
equations of motion form a quite complicated system of partial
differential equations with second order derivatives with respect
to the $r_{\tilde{I}}$'s. It is not easy to solve that system in
full generality. Additional restrictions simplifying the equations
are needed.

Usually it is assumed that the solution has to depend only on $r$
-- radial coordinate in the overall transverse space. This kind of
solutions are called delocalized, averaged or smeared branes,
because they do not determine at what points the branes are
situated in the relative transverse directions. One can wonder if
the solution derived under such assumption can have any physical
meaning. It turns out that there is one important application --
if the relative transverse dimensions are small in comparison to
the common tangent and overall transverse ones since then we can
use the intersecting branes configurations as a background for
compactification models (with the relative transverse dimensions
compactified). We have to note that the notion of intersection of
delocalized branes has a rather imprecise meaning -- if they are
not localized we cannot be sure that they really intersect. But
because such terminology is common, we will use it here too.

Since a set of orthogonally intersecting branes enforces a split
of the spacetime into several orthogonal segments it is natural to
assume that an ansatz for the metric tensor should admit
independent factors to encode possibly different length scales for
each of the segments. Call $e^{A_{\{1,\ldots N_A\}}},
e^{A_{\tilde{I}}}, e^B$ the factors for the common tangent, the
relative transverse and the overall transverse directions
respectively. This dramatically increases the number of degrees of
freedom of the solution and complicates its derivation. Therefore
one usually imposes some set of linearity conditions as in the
case of the single brane. One of them is the generalized harmonic
gauge:
 \be
  d_{\{1,2,\ldots,N_A\}} A_{\{1,\ldots N_A\}} +
  \sum_{\tilde{I}} d_{\tilde{I}} A_{\tilde{I}} + \td B = 0,
  \label{gen_harm_gauge}
 \ee
where
 \be
  \td = D-d_{\{1,2,\ldots,N_A\}} - \sum_{\tilde{I}} d_{\tilde{I}} - 2.
 \ee
Applying it as an additional requirement on the solution has
analogous effects as for the single brane -- it enforces the
solution to be constructed of harmonic functions only and is
necessary but not sufficient for preserving the supersymmetry. It
is also possible to impose more restrictive conditions:
 \be
  A_{\tilde{I}} = A_{\tilde{J}} \quad \mbox{for all}
  \quad \tilde{I}, \tilde{J},
  \label{2_lin_con_12}
 \ee
what means that all relative transverse dimensions are governed by
the same factor, or even:
 \be
  A_{\tilde{I}} = A_{\{1,\ldots N_A\}} \quad
  \mbox{for all} \quad \tilde{I}.
  \label{2_lin_con_13}
 \ee

The multibrane systems are classified as BPS or non-BPS states
where the criterion is respectively saturation or non-saturation
of the multibrane version of the BPS inequality (\ref{BPS_1}). But
while the total energy is just a sum of component energies:
 \be
  \calE = \sum_i \calE_i,
 \ee
the charge is in general rather ''vector-like'':
 \be
  Q^2 = \sum_i Q_i^2.
 \ee
The BPS states fall into two categories: marginal and non-marginal
\cite{Tseytlin}. For the marginal (called also threshold) states
the BPS bound is degenerated and can be written as:
 \be
  \calE = \sum_i Q_i.
 \ee

Examples of intersecting branes solutions which satisfy
(\ref{gen_harm_gauge}) are solutions constructed as orthogonal
superposition of some number of the single charge supersymmetric
brane solutions (\ref{1_sol_1} -- \ref{1_sol_6}) with the use of
so called harmonic function rule \cite{Gauntlett, Tseytlin2, GKT}.
The rule for a nondilatonic solution tells, that if:
 \be
  H_i(r)=1+\frac{k_i}{r^\td},
 \ee
for $i=1,\ldots N_A$, and $D_i$ is a dimension of the $i$-th
brane worldvolume, then:
 \bea
  e^{2A_{\{1,\ldots N_A\}}} &=& \prod_{i=1}^{N_A} H_i^{\frac{D_i}{D-2}-1},
  \label{harm_fun_rule_1} \\
  e^{2A_{\tilde{I}}} &=& \prod_{i\in tan(\tilde{I})} H_i^{\frac{D_i}{D-2}-1}
                         \prod_{j\in trans(\tilde{I})} H_j^{\frac{D_j}{D-2}},
  \label{harm_fun_rule_2} \\
  e^{2B} &=& \prod_{i=1}^{N_A} H_i^{\frac{D_i}{D-2}},
  \label{harm_fun_rule_3}
 \eea
where $tan(\tilde{I})$ (respectively $trans(\tilde{I})$) is a set
of such indices $i$ which correspond to the branes for which the
$\tilde{I}$-th segment of the metric contains directions tangent
(transversal) to the brane worldvolume.

Wider class of solutions can be obtained when the multibrane
equations of motion are solved directly. It is interesting that it
is possible to reduce the equations of the intersecting branes to
a Toda-like system which is an extension of the Liouville
equation. A possibility of the reduction was observed in
\cite{LPX} under assumption of (\ref{2_lin_con_12}), in
generalized harmonic gauge \cite{IK, CGM} and further it was
proved in a general case without any linearity condition assumed
\cite{FM}. We will analyze the last case with details in the next
chapter. But of course all the examples given already and later do
not run short a set of possible but still solvable generalizations
of the brane problem. For some others see \cite{Gauntlett, Stelle, LP1, IM2, Smith}.

%%%%%%%%%%%%%%%%%%%%%%%%%%%%%%%%%%%%%%%%%%%%%%%%%%%%%%%%%%%%%%%%%
%%%%%%%%%%%%%%%%%%%%%%%%%%%%%%%%%%%%%%%%%%%%%%%%%%%%%%%%%%%%%%%%%
%%%%%%%%%%%%%%%%%%%%%%%%%%%%%%%%%%%%%%%%%%%%%%%%%%%%%%%%%%%%%%%%%
\newpage
\section{Branes without harmonic gauge.}
\label{nonharm_branes}

\akapit In the chapter \ref{sugra_branes} a short introduction to the
subject of branes in supergravity was given. It was noted that in
the case of single brane as well as in the case of many
intersecting branes a rich class of nonsupersymmetric solution
could be derived when the condition of the harmonic gauge was
dropped. Now we turn our attention to this kind of solutions and
to a method leading to them.

%%%%%%%%%%%%%%%%%%%%%%%%%%%%%%%%%%%%%%%%%%%%%%%%%%%%%%%%%%%%%%%%%%%%
%%%%%%%%%%%%%%%%%%%%%%%%%%%%%%%%%%%%%%%%%%%%%%%%%%%%%%%%%%%%%%%%%%%%
\subsection{Commonly orthogonally intersecting non-composite
delocalised branes.}
\label{intersecting_b_base}

\akapit Consider a $D$ dimensional theory having after consistent
bosonic truncation a following action:
 \bea
  S &=& \int_{\cal M} d^DX \sqrt{|\det g|}
   \left(R - \frac{1}{2} \sum_{\alpha=1}^{N_\phi}
   \de_M \phi_\alpha \de^M \phi_\alpha
   - \sum_{i=1}^{N_A} \frac{e^{\sum_{\alpha=1}^{N_\phi}
   a_{i\alpha} \phi_\alpha}}{2}
   \left|F^i \right|^2 \right),
 \label{lagrangian_1}
 \eea
where $F^i$ are antisymmetric $n_i$-forms, $\phi^\alpha$ -- scalar
fields, $a_{i\alpha}$ -- constants, $\calM$ -- a manifold of
dimension $D$ and $(X^M)$ coordinates on it. A bosonic sector of
most of supergravity theories like (\ref{D11sugra}),
(\ref{sugra_IIA_ac}), (\ref{sugra_I_ac}) is well described by the
above action if additional assumptions leading to cancellation of
Chern-Simons term are made.

The equations of motions derived from (\ref{lagrangian_1}) are:
 \bea
  R_{MN} &=& \frac{1}{2} \sum_\alpha \de_M \phi_\alpha \de_N \phi_\alpha
   + \sum_i \frac{e^{\sum_\alpha a_{i\alpha} \phi_\alpha}}{2(n_i-1)!} S^i_{MN},
   \label{eqm_11} \\
  0 &=& \nabla_M \left( e^{\sum_\alpha a_{i\alpha} \phi_\alpha}
  F^{iM R_1\ldots R_{n_i-1}} \right),
   \label{eqm_12} \\
  \nabla^2 \phi_\alpha &=& \sum_i \frac{a_{i\alpha}}{2{n_i}!}
   e^{\sum_\beta a_{i\beta} \phi_\beta} F^i_{R_1 \ldots R_{n_i}}
   F^{iR_1\ldots R_{n_i}},
   \label{eqm_13}
 \eea
 with:
 \be
  S^i_{MN}= F^i_{M R_1\ldots R_{n_i-1}} F^i_{N}{}^{R_1\ldots R_{n_i-1}}
   - \frac{n_i-1}{n_i(D-2)}F^i_{R_1\ldots R_{n_i}} F^{iR_1\ldots R_{n_i}}g_{MN},
 \ee
where $\sum_i$ and $\sum_\alpha$ are sums over all possible values
of $i=1,\ldots, N_A$ and $\alpha=1,\ldots, N_\phi$.

%%%%%%%%%%%%%%%%%%%%%%%%%%%%%%%%%%%%%%%%%%%%%%%%%%%%%%%%%%%%%%%%%%%%
\subsubsection{The model.}
\label{model}

\akapit We search for a solution which allows $N_A$ commonly
orthogonally intersecting electric or magnetic non-composite
branes. Let $V_i$ be a worldvolume of the $i$-th brane. Then each
$V_i$ is supported by a potential of a different $F^i$ or $\ast
F^i$. We define indices $I, J, \ldots$ running through the set of
all non-empty subsets of $\{1,\ldots,N_A\}$ and 
$V_I$ with $I=\{i_1,\ldots,i_k\}$ as a subspace
spun by vectors simultaneously parallel to all $V_{i_1}, \ldots,
V_{i_k}$ and transversal to all $V_{i_{k+1}}, \ldots, V_{i_{N_A}}$.
Next, let $V$ be a cartesian product of common tangent and all
relative transverse directions, $V_\emptyset$ -- the overall
transverse space and $\hat{V}_i$ -- a subspace of $V$ transverse
to $V_i$. Volumes of the subspaces $V_i$, $V_I$, $\hat{V}_i$ will
be denoted respectively by $|V_i|$, $|V_I|$ and $|\hat{V}_i|$.
If all $V_i$ intersect at the point $\{0\}$ 
(the center of the coordinate system) we can write:
 \bea
  V_I &=& \{0\} \cup \left(\left( \bigcap_{i\in I} V_i \right) \setminus
  \left( \sum_{j \unin I} V_j \right)\right), \\
  V &=& \sum_i V_i=\oplus_I V_I, \\
  V_\emptyset &=& \calM/V, \\
  \hat{V}_i &=& \sum_{I:i\unin I} V_I = V/V_i.
 \eea

Introduce numbers describing dimensions of the subspaces:
 \bea
  d_I &=& \dim V_I, \label{d_I} \\
  D_i &=& \dim V_i=\sum_{I:i\in I}d_I, \label{d_i} \\
  \hat{D}_i &=& \dim \hat{V}_i=\sum_{I:i\unin I}d_I,
  \label{d_hat_i} \\
  d &=& \dim V=\sum_I d_I.
  \label{d_no}
 \eea
For particular brane configurations some of $V_I $ can be
zero-dimensional. For example, if $V_i \subset V_j$ for some
$i,j$, then $V_{\{i\}}=\{0\}$. We will use also the mapping
\hskip7pt $\tilde{}$ \hskip7pt defined by (\ref{tilde_map}).

All the branes are assumed to propagate in time, so the
instantonic solution is a priori disallowed. Note that because the
branes are delocalized in the relative transverse directions, the
solution would be trivial giving all fields constant when
$V_\emptyset=\{0\}$. So we not consider the case. Therefore, the
dimensions (\ref{d_I} -- \ref{d_no}) have to obey:
 \bea
  1 &\leq& d_{\{1,\ldots,N_A\}} \leq D-1, \\
  -1 &\leq& \td \leq D-3, \\
  0 &\leq& d_I \leq D-2 \mbox{   for   } I \neq \{1,\ldots,N_A\}.
 \eea

%%%%%%%%%%%%%%%%%%%%%%%%%%%%%%%%%%%%%%%%%%%%%%%%%%%%%%%%%%%%%%%%%%%%%%%%
\subsubsection{Ans\"{a}tze.}

\akapit Let us now set ans\"{a}tze consistent with all assumed
conditions. We call by $(x^{\mu^i})$ the coordinates on $V_i$, by
$(x^{\hat{\mu}^i})$ -- the coordinates on $\hat{V}_i$, by
$(x^{\mu^I})$ -- on $V_I$, by $(y^m)$ -- on $V_\emptyset$ and we
use sum convention for all indices enumerating coordinates but not
for indices like $i$ or $I$. Because we search for a delocalized
solution all the fields should depend nontrivially only on
$r=\sqrt{y^m y^n \delta_{mn}}$ -- a radial coordinate in the
overall transverse space. Derivatives with respect to the $r$ are
denoted by primes $f'=df/dr$.

The simplest is an ansatz for the scalar fields:
\be
\phi_\alpha(X) = \phi_\alpha (r).
 \label{phi_ansatz}
\ee

More complicated is a case of the metric tensor. It has to be
divided into $N_g$ segments related to $V_\emptyset$ and those
$V_I$ which are at least one-dimensional ($2 \leq N_g \leq
2^{N_A}$):
 \be
  ds^2(X) = \sum_I e^{2A_I(r)} dx^{\mu^I} dx^{\nu^I} \eta_{\mu^I \nu^I}
  + e^{2B(r)} \left( (dr)^2 + r^2 d\Omega^2_{(\td+1)} \right),
  \label{interval_1}
 \ee
where $d\Omega_{\td+1}^2$ is the space interval of the $\td+1$
dimensional unit sphere and $\eta_{\mu^I \nu^I}=\delta_{\mu^I
\nu^I}$ if $I\neq\{1,\ldots,N_A\}$. It has to be explained that in
the formula (\ref{interval_1}) and any formulae below by
$\sum_{I:r(I)}$ we denote a sum over those $I$ for which $V_I$ are
at least one-dimensional and which satisfy the restriction $r(I)$.

For the antisymmetric tensor fields $F^i$, two cases should be
distinguished. If the $i$-th brane is electric, the only nonzero
components of $F^i$ have the form:
 \be
  F^i_{m \mu^i_1 \ldots \mu^i_{D_i}} (X)
  = \sigma_i \epsilon_{\mu^i_1 \ldots \mu^i_{D_i}} \de_m \exp (C_i(r)),
  \label{el_Fansatz}
 \ee
when for the $i$-th brane being magnetic only:
 \be
  F^i_{\hat{\mu}^i_1 \ldots \hat{\mu}^i_{\hat{D}_i}
  m_1 \ldots m_{\td+1}} (X)
  = \epsilon_{\hat{\mu}^i_1 \ldots \hat{\mu}^i_{\hat{D}_i}
  m_1 \ldots m_{\td+1} n}\frac{\lambda_i y_n}{r^{\td+2}}
  \label{mag_Fansatz}
 \ee
do not vanish. In the above $\lambda_i$ is an arbitrary nonzero
real constant and $\sigma_i$ takes discrete values $+1$ or $-1$.
In the both cases the relative opposite signs illustrate
possibility of an exchanging in the model the given brane with its
anti-brane.

Now it is possible to derive explicit forms of the Ricci tensor
and other objects appearing in the equations of motion. But first
we need to find a vielbein, which is defined as:
 \bea e_{M\bar{N}}e_{R\bar{S}} \eta^{\bar{N}\bar{S}} &=& g_{MR}, \\
      e_{M\bar{N}}e_{R\bar{S}} g^{MR}       &=& \eta_{\bar{N}\bar{S}},
 \eea
where the bared indices enumerate the local Lorentz coordinates
which are lifted by $\eta^{\bar{M}\bar{N}}$. So in our case from
(\ref{interval_1}) we have:
 \be
  e_M^{\bar{M}} = \left( \begin{array}{ll}
   {\rm diag}_I \left( \exp(A_I(r))
   {\rm e}_{\mu^I}^{\bar{\mu}^I} \right) & 0 \\
   0 & \exp(B(r)) {\rm e}_m^{\bar{m}}
  \end{array} \right),
 \ee
where
 \be
  {\rm e}_{M\bar{N}}{\rm e}_{R\bar{S}} \eta^{\bar{N}\bar{S}}
  = \eta_{MR}, \qquad
  {\rm e}_{M\bar{N}}{\rm e}_{R\bar{S}} \eta^{MR}
  = \eta_{\bar{N}\bar{S}}.
 \ee

With the vielbein we can calculate a spin connection
$\omega_{M\bar{N}\bar{R}}$ from:
 \bea
  \omega_{M\bar{N}\bar{R}} &=& \frac12 (e_{\bar{N}}^S
  \Omega_{MS\bar{R}}-e_{\bar{R}}^S \Omega_{MS\bar{N}}
  - e_{\bar{N}}^S e_{\bar{R}}^T e_M^{\bar{U}} \Omega_{ST\bar{U}}), \\
  \Omega_{MN\bar{R}} &=& \de_M e_{N\bar{R}} - \de_N e_{M\bar{R}}.
 \eea
We find that the only nonzero components of the field $\Omega$ are:
 \bea
  \Omega_{\mu^I n\bar{\rho}^I} &=&
  - (\de_n A_I) e^{A_I} {\rm e}_{\mu^I\bar{\rho}^I}, \\
  \Omega_{mn\bar{r}} &=& e^B( (\de_mB) {\rm e}_{n\bar{r}}
  - (\de_nB) {\rm e}_{m\bar{r}}),
 \eea
so the spin connection is described as:
 \bea
  \omega_{\mu^I\bar{\nu}^I\bar{r}} &=&
  e^{A_I-B} (\de_{\bar{r}}A_I) {\rm e}_{\mu \bar{\nu}}, \\
  \omega_{m \bar{n}\bar{r}} &=&
  (\de_{\bar{r}}B) {\rm e}_{m \bar{n}}-(\de_{\bar{n}}B) {\rm e}_{m \bar{r}}
 \eea
and has all other elements vanishing.

The Riemann tensor can be derived from the spin connection because
it obeys:
 \be
  R_{MNRS} = 2 e_{R\bar{R}} e_{S\bar{S}}
  \left( \de_{[M} \omega_{N]}{}^{\bar{R}\bar{S}} +
  \omega_{[M}{}^{\bar{R}\bar{T}} \omega_{N]\bar{T}}{}^{\bar{S}} \right),
 \ee
so in our case its nonzero components are given as:
 \bea
  R_{\mu^I\nu^I\rho^I\sigma^I} &=& e^{4A_I-2B} (A'_I)^2
  \left( \eta_{\mu^I\sigma^I} \eta_{\nu^I\rho^I}
    - \eta_{\mu^I\rho^I} \eta_{\nu^I\sigma^I} \right), \\
  R_{\mu^I n \rho^I s} &=&
    - e^{2A_I} \left[ \left(A''_I+(A'_I)^2 - 2A'_I B' -
    \frac{1}{r}A'_I \right) \frac{y_n y_s}{r^2} + \right. \nonumber \\
    &+& \left. \left( A'_I B' + \frac{1}{r}A'_I \right) \delta_{ns} \right]
    \eta_{\mu^I\nu^I}, \\
  R_{\mu^I\nu^J\rho^I\sigma^J} &=& - e^{2A_I+2A_J-2B} A'_I A'_J
  \eta_{\mu^I\rho^I} \eta_{\nu^J\sigma^J}, \qquad
    \mbox{where} \quad I\neq J, \\
  R_{mnrs} &=& e^{2B} \left[ \left( (B')^2+\frac{2}{r} B' \right)
    (\delta_{nr}\delta_{ms} - \delta_{mr}\delta_{ns}) + \right. \\
    &+& \!\left. \left( B''\!-(B')^2\!-\frac{1}{r}B' \right)
    \left( \frac{y_m y_s}{r^2}\delta_{nr} +
    \frac{y_n y_r}{r^2}\delta_{ms}
    - \frac{y_m y_r}{r^2}\delta_{ns} -
    \frac{y_n y_s}{r^2}\delta_{mr} \right) \right]. \nonumber
 \eea

The Ricci tensor can be calculated from:
 \be
  R_{NS}=R_{MNRS}g^{MR}
 \ee
and all its elements are zero except:
\bea
  R_{\mu^I\nu^I} &=& - e^{2(A_I-B)} \left( A''_I +
  \frac{\td+1}{r} A'_I + \sum_J d_J A'_J A'_I + \td A'_I B' \right)
   \eta_{\mu^I\nu^I},\\
  R_{mn} &=&
   - \left[ B''+\frac{2\td+1}{r}B'+\td(B')^2+
   \sum_I d_I A'_I \left( B'+\frac{1}{r} \right) \right] \delta_{mn} + \\
   &-& \left[ \td B'' - \frac{\td}{r}B' - \td(B')^2 +
   \sum_I d_I \left( A''_I - 2A'_I B' + (A'_I)^2 -
   \frac{1}{r}A'_I \right) \right] \frac{y_my_n}{r^2}. \nonumber
 \eea
And finally the Ricci scalar:
 \bea
  R &=& R_{MN} g^{MN} \nonumber \\
    &=& -e^{-2B} \left[ 2 \sum_I d_I
     \left( A''_I + (A'_I)^2 + A_I \sum_J d_J A'_J +
     \frac{2(\td+1)}{r}A'_I + 2\td A'_I B' \right) \right. + \nonumber \\
    &+& \left. \frac{2(\td+1)^2}{r}B' + \td(\td+1)(B')^2 +
    2(\td+1)B'' \right].
 \eea

Let us rewrite the equations of motion
(\ref{eqm_11}--\ref{eqm_13}) in terms of the functions $A_I$, $B$,
$\phi_\alpha$ and (only for the elementary branes) $C_i$
introduced by (\ref{phi_ansatz}--\ref{mag_Fansatz}):
 \bea
  A''_I + A'_I \left( \sum_J d_J A'_J  + \td B' + \frac{\td+1}{r} \right)
   &=& \frac{\sum_{i\in I} \tD_i(S'_i)^2 -
   \sum_{i\unin I} D_i(S'_i)^2}{2(D-2)},
   \label{eqm_21} \\
  B'' +\! \td(B')^2 +\! \frac{2\td\!+\!1}{r}B' +
  (B'\!+\!\frac{1}{r}) \sum_I d_I A'_I
   &=& - \frac{\sum_i D_i(S'_i)^2}{2(D-2)},
   \label{eqm_22} \\
  \phi''_\alpha + \phi'_\alpha \left( \sum_I d_I A'_I + \td B' +
  \frac{\td+1}{r} \right)
   &=& -\frac{1}{2}\sum_i \varsigma_i a_{i\alpha} (S'_i)^2,
   \label{eqm_23}
 \eea
 \bea
  \td B'' - \td(B')^2 - \frac{\td}{r}B' + \sum_I d_I
  \left(A''_I -\frac{1}{r}A'_I-2A'_I B'+ (A'_I)^2 \right)
   &=& \nonumber \\
   = \, \frac{1}{2}\sum_i (S'_i)^2 \!\! &-& \!\! \frac{1}{2}\sum_\alpha (\phi'_\alpha)^2,
   \label{eqm_24} \\
   C'_i \left( C'_i - \sum_{I:i\in I} d_I A'_I +
   \sum_{I:i \unin I} d_I A'_I + \td B'
   + \sum_\alpha a_{i\alpha} \phi'_\alpha + \frac{\td+1}{r} \right)
   &=& - C''_i,
   \label{eqm_25}
 \eea where:
 \bea
  \varsigma_i &=& \left\{ \begin{array}{ll}
   +1 & \mbox{(electric),} \\
   -1 & \mbox{(magnetic),} \end{array} \right. \\
  S'_i &=& \left\{ \begin{array}{ll}
   \sigma_i (e^{C_i})'
   \exp(\frac{1}{2}\sum_\alpha a_{i\alpha}\phi_\alpha-
   \sum_{I:i\in I} d_I A_I)
   & \mbox{(electric),} \\
   & \\
   \frac{\lambda_i}{r^{\td+1}}
   \exp(\frac{1}{2}\sum_\alpha a_{i\alpha}\phi_\alpha -
   \sum_{I:i\unin I} d_I A_I -\td B)
   & \mbox{(magnetic).} \end{array} \right.
  \label{S_1}
 \eea

%%%%%%%%%%%%%%%%%%%%%%%%%%%%%%%%%%%%%%%%%%%%%%%%%%%%%%%%%%%%%%%%%
\subsubsection{Harmonic gauge.}
\label{s_harm_gauge}

\akapit Assuming that $\td \neq 0$ define a function:
 \be
  \chi = \sum_I d_I A_I + \td B\,.
  \label{lincon_1}
 \ee
Condition $\chi=0$ is equivalent to the generalized harmonic gauge
(\ref{gen_harm_gauge}). Imposing it enormously simplifies the 
equations
(\ref{eqm_21} -- \ref{eqm_25}) and leads to a solution expressed in
terms of harmonic functions on $V_\emptyset$ \cite{IM2, Gibbons}.
In case of supergravity theories the condition  is necessary but not
sufficient for preserving supersymmetry. Here we do not make any a
priori assumption on $\chi$, so the results presented below remain
valid in more general classes of nonsupersymmetric solutions and
solutions not governed by harmonic functions.

Summing (\ref{eqm_21}--\ref{eqm_22}) one can see that $\chi$ has
to satisfy the following equation:
 \be
  \chi''+(\chi')^2+\frac{2\td +1}{r}\chi' = 0,
 \ee
 which can be solved and gives:
 \be
  \chi(r)=\ln \left| \frac{c_\chi-1/r^{2\td}}{c_\chi-c_0} \right|
  +\epsilon_\chi(c_0),
  \label{chi1}
 \ee
where $c_\chi$ and $c_0$ are constants taking real as well as
infinite values and $\epsilon_\chi$ is a function of $c_0$. Of
course it is only one of many possible variants how the $\chi$ can
be parameterized. But with this one we have:
 \be
  \lim_{c_\chi\rightarrow +\infty} \chi =
  \lim_{c_\chi\rightarrow -\infty} \chi = \epsilon_\chi(c_0).
 \ee
So points $c_\chi=+\infty$ and $c_\chi=-\infty$ can be identified
in the parameter space and then the space is compact in $c_\chi$
direction. Harmonic gauge is restored when $c_\chi=\pm \infty$ and
$\epsilon_\chi=0$.

See, that because of arbitrariness of $\epsilon_\chi$ the
parameterization used in (\ref{chi1}) is not unique but rather
constitutes a class of parameterizations labelled by $c_0$. Each
choice of $c_0$ gives different parameterization, but each is
singular at $c_\chi=c_0$. In the discussion below we choose
$c_0=1$, but we should remember that it can be generalized to
arbitrary $c_0$.

It is very convenient to introduce instead of $c_\chi$ new
parameters: $R \in [0,+\infty]$ and $s_\chi \in \{-1,+1\}$ such
that:
 \be
  s_\chi R^{2\td}=1/c_\chi,
  \label{R_def}
 \ee
so if $R=0$ or $R=\infty$ both possible signs of $s_\chi$ describe
the same point in the parameter space. Rewriting $\chi$ with these
parameters one obtains:
 \be
  \chi(r)=\ln \left|
  \frac{1-s_\chi \left(R/r \right)^{2\td}}{1-s_\chi R^{2\td}}
  \right|+\epsilon_\chi.
  \label{chi2}
 \ee
For $R=0$ (harmonic gauge) it simplifies to $\chi = \epsilon_\chi$
and for $R=\infty$ to $\chi = -2\td \ln r + \epsilon_\chi$.

%%%%%%%%%%%%%%%%%%%%%%%%%%%%%%%%%%%%%%%%%%%%%%%%%%%%%%%%%%%%%%%%
\subsubsection{$\vartheta$ coordinate.}
\label{s_vtheta}

\akapit Looking at the equations of motion (\ref{eqm_21} --
\ref{eqm_25}) one can see that the equations with second
 derivative of $A_I$ or $\phi_\alpha$ have a form:
 \be
  f''+\left( \chi' + \frac{\td+1}{r} \right) f' = \const (S')^2.
  \label{harm_chi}
 \ee
The left hand side is just $\nabla^2 f$, so (\ref{harm_chi}) is
curved space harmonic equation where a contribution from the
curvature is given by $\chi'$. Under a redefinition of variable as
$r \rightarrow \vartheta$ (\ref{harm_chi}) transforms to:
 \be
  \ddot{f}(\vartheta')^2 +\left( \vartheta'' + \chi'\vartheta' +
  \frac{\td+1}{r}\vartheta' \right) \dot{f}
   = \const (\dot{S}\vartheta')^2,
 \ee
where the "dots" describe derivatives with respect to $\vartheta$.
And this convinces that it would be very convenient to work with
such variable $\vartheta$ for which:
 \be
  \vartheta'' + \chi'\vartheta' + \frac{\td+1}{r}\vartheta' = 0.
  \label{th_eqm_1}
 \ee

A function which satisfies the above equation is:
 \be
  \vartheta(r) = \left\{ \begin{array}{ll}
  \frac{1}{2\td}\left(\frac{1}{R^\td}+R^\td \right)
  \left(\arctan((\frac{R}{r})^\td) - \arctan(R^\td) \right),
  & s_\chi=-1, \\
  & \\
  \frac{1}{2\td}\left|\frac{1}{R^\td}-R^\td \right|
  \left(\Arth((\frac{R}{r})^\td) - \Arth(R^\td) \right),
  & s_\chi=+1,
  \end{array} \right.
  \label{th_1}
 \ee
 where:
 \be
 \Arth(x) = \left\{ \begin{array}{ll}
   \artanh (x), & \mbox{   for   } |x|<1, \\
   -\arcoth (x), & \mbox{   for   } |x|>1.
   \end{array} \right. \label{Arth}
 \ee
Of course this is not the most general solution of
(\ref{th_eqm_1}) which can be generated from the function
$\vartheta$ given in(\ref{th_1}) as $a\vartheta+b$ where the
integration constants $a$ and $b$ can be independently set in two
areas separated by points where $\chi$ is singular. One of the
areas is $r<R$, $s_\chi=+1$ and the second is the remaining part
of the spacetime.

The function $\vartheta$ obeys following conditions:
 \bea
  \lim_{R\rightarrow \infty} \vartheta (r;R,s_\chi=\pm 1)
  &=& -\frac{1}{2\td}\left(r^\td - 1 \right), \\
  \lim_{R\rightarrow 0} \vartheta (r;R,s_\chi=\pm 1) &=&
  \frac{1}{2\td}\left(\frac{1}{r^\td} - 1 \right).
 \eea
So in harmonic gauge ($R=0$) and only then $\vartheta$ is a flat
space harmonic function of $r$. And the parameter $R$ (or $c_\chi$
equivalently) can be then treated as a measure how distant is a
given case from the harmonic one.

The function $\vartheta$ can be understood as a space coordinate
instead of $r$ and the coordinate change is singular only at
$r=R$, $s_\chi=+1$. Further, in the section \ref{diag_delta_sol}
we will see, that the replacing of $r$ with $\vartheta$ is not
only a simple coordinate change, because the solution formulated
with use of $\vartheta$ is well defined also for such values of
$\vartheta$ which cannot be related to any $r$ by (\ref{th_1}). In
other words, $\vartheta$ covers wider area of spacetime than $r$.

The space-time interval expressed in terms of $\vartheta$ is:
 \be
  ds^2(\vartheta) = \sum_I e^{2A_I(\vartheta)} dx^{\mu^I} dx_{\mu^I}
   + e^{2\Btheta(\vartheta)} \left( d\vartheta^2 +
   \rho(\vartheta)^2 d\Omega^2 \right),
  \label{interval_2}
 \ee
where
 \bea
  e^B &=& e^\Btheta |\vartheta'|, \label{Btheta_0} \\
  \rho &=& |\vartheta'| r
  \label{rho_0}
 \eea
and the coordinate change factor $\vartheta'$ because of
(\ref{th_1}) has to be equal to:
 \be
  \frac{d\vartheta}{d r} = - \frac{1}{2r^{\td+1}}
  \exp \left(-\chi(r)+\epsilon_\chi \right).
 \ee
The identity (\ref{rho_0}) allows us to write $\rho$ explicitly in
terms of the variable $r$ or $\vartheta$:
 \bea
  \rho(r) &=& \left\{ \begin{array}{ll}
   \frac{1}{2} \left| \frac{(1/R)^{\td}+R^{\td}}{(r/R)^{\td}+(R/r)^{\td}} \right|,  & s_\chi=-1,\\
   & \\
   \frac{1}{2} \left| \frac{(1/R)^{\td}-R^{\td}}{(r/R)^{\td}-(R/r)^{\td}} \right|,  & s_\chi=+1, \\
  \end{array} \right.
  \label{rho_2} \\
  \rho(\vartheta) &=& \left\{ \begin{array}{ll}
    \frac{1}{4} \left|(1/R)^{\td}+R^{\td} \right|
    \left| \sin \left(\frac{4\td\vartheta}{(1/R)^{\td}+R^{\td}}+
    2\arctan R^\td\right)\right|,
   & s_\chi=-1,\\
   & \\
    \frac{1}{4} \left|(1/R)^{\td}-R^{\td} \right|
    \left| \sinh\left(\frac{4\td\vartheta}{\left|(1/R)^{\td}-R^{\td}\right|}
    +2\Arth R^\td
    \right)\right|,
   & s_\chi=+1.
  \end{array} \right.
  \label{rho_1}
 \eea

{}From the definition of function $\chi$ (\ref{lincon_1}) we have:
 \be
  e^{B(r)}=\exp \left( \frac{1}{\td}\chi(r) -
  \sum_I \frac{d_I}{\td} A_I(r) \right), \label{b_a}
 \ee
 and:
 \be
  e^{\Btheta(\vartheta)} =
  \exp\left(-\sum_I \frac{d_I}{\td} A_I(\vartheta) +
  \frac{\epsilon_\chi}{\td}\right)
  \frac{\rho(\vartheta)^{-(1+\frac{1}{\td})}}{2},
  \label{Btheta_1}
 \ee
Interesting, that if one substitutes (\ref{Btheta_0}) into
(\ref{eqm_22}) one then obtains:
 \be
  \Btheta''+\Btheta'\left( \chi'+\frac{\td+1}{r} \right) -
  \frac{\td(\td+1)}{r^2} = - \frac{\sum_i D_i(S'_i)^2}{2(D-2)}.
 \ee
The left hand side of the equation is similar to (\ref{harm_chi})
but in addition the $\td (\td+1) /r^2$ component appears. The
component is the origin for the $\rho^{-(1+1/\td)}$ factor in
(\ref{Btheta_1}).

Thanks to (\ref{b_a}, \ref{lincon_1}) and the already known
function $\chi$ we are able to reduce a number of unknown
functions in the equations of motion i.e. to replace $B$ (or
$B_\theta$) by certain combination of $A_I$'s and to drop the
equation (\ref{eqm_22}) which does not contain any independent
piece of information. The remaining equations
(\ref{eqm_21}--\ref{eqm_25}) can be translated to a system
depending on the $\vartheta$ variable:
 \bea
  \ddot{A}_I &=& \frac{\sum_{i\in I} \tD_i(\dot{S}_i)^2 -
  \sum_{i\unin I} D_i(\dot{S}_i)^2}{2(D-2)}, \label{eqm_31} \\
  \ddot{\phi}_\alpha &=&
  -\frac{1}{2}\sum_i \varsigma_i a_{i\alpha} (\dot{S}_i)^2,
  \label{eqm_32}\\
  \ddot{S}_i &=& - \left(\frac12 \sum_\alpha a_{i\alpha} \dot{\phi}_\alpha -
  \sum_{I:i\in I} d_I \dot{A}_I \right)\dot{S}_i,
  \mbox{  (electric)}
  \label{eqm_34}
 \eea
 \be
  \frac{1}{\td} \left( \sum_I d_I \dot{A}_I \right)^2 +
  \sum_I \left( d_I (\dot{A}_I)^2 \right)
  + \frac{1}{2}\sum_\alpha(\dot{\phi}_\alpha)^2 + \Lambda_\chi =
  \frac{1}{2}\sum_i (\dot{S}_i)^2 ,
  \label{eqm_33}
 \ee
 where:
 \bea
  \dot{S}_i &=& \left\{ \begin{array}{ll}
   \sigma_i (e^{C_i})\dot{}
   \exp(\frac{1}{2}\sum_\alpha a_{i\alpha}\phi_\alpha -
   \sum_{I:i\in I} d_I A_I), & \mbox{(electric),} \\
   & \\
   -2\lambda_i e^{-\epsilon_\chi}
   \exp(\frac{1}{2}\sum_\alpha a_{i\alpha}\phi_\alpha+
   \sum_{I:i\in I} d_I A_I),
   & \mbox{(magnetic),}
  \end{array} \right. \label{S_2} \\
  \Lambda_\chi &=& -\frac{16\td(\td+1)c_\chi}{(c_\chi-1)^2}.
  \label{kosmo_1}
 \eea

The system (\ref{eqm_31}--\ref{kosmo_1}) together with
(\ref{rho_1}) and (\ref{Btheta_1}) carries complete information
originally contained in (\ref{eqm_21}--\ref{eqm_25}, \ref{S_1}).
Since the $r$-variable system(\ref{eqm_21}--\ref{eqm_25},
\ref{S_1}) drastically simplifies when harmonic gauge is imposed,
it is interesting what happens to (\ref{eqm_31} -- \ref{kosmo_1},
\ref{Btheta_1}, \ref{rho_1}) in an analogous situation. If one
treats the $\vartheta$ as a fundamental coordinate then all
dependence of the solution on the parameter $R$ (so also all
differences between the harmonic and a general cases) enters only
in two places: in the $\rho$ function (\ref{rho_1}) which
influences a form of the $\Btheta$ function (\ref{Btheta_1}) and
in the $\Lambda_\chi$ constant (\ref{kosmo_1}) appearing in
(\ref{eqm_33}). However (\ref{eqm_33}) is not a dynamic equation
but rather a constraint decreasing by one a number of integration
constants.

%%%%%%%%%%%%%%%%%%%%%%%%%%%%%%%%%%%%%%%%%%%%%%%%%%%%%%%%%%%%%%%%%
\subsubsection{The $\Delta$ matrix and reduction to Toda-like system.}

\akapit Define:
 \be
  \omega_i = \exp \left( \frac12 \sum_\alpha
  \varsigma_i a_{i\alpha} \phi_\alpha-\sum_{I:i\in I} d_I A_I \right).
  \label{omega_def}
 \ee
With these functions one can find that (\ref{eqm_34}) leads to:
 \be
  \dot{S}_i = p_i/\omega_i, \label{S_3}
 \ee
 where $p_i$ are nonzero real integration constants.
Simultaneously (\ref{S_2}) for magnetic branes gives:
 \be
  \dot{S}_i=\frac{-2\lambda_i e^{-\epsilon_\chi}}{\omega_i},
 \ee
 so, after the identification:
 \be
  p_i=-2\lambda_i e^{-\epsilon_\chi}, \qquad \sigma_i =
  \sgn \lambda_i, \label{p_lambda}
 \ee
the relation (\ref{S_3}) is valid for electric as well as for
magnetic branes.

It can be checked from (\ref{eqm_31}--\ref{eqm_32}) that
$\omega_i$ have to satisfy the following system of equations:
 \be
  \frac{d^2}{d\vartheta^2} (\ln |\omega_i|) =
  - \sum_j \Delta_{ij} \frac{p_j^2}{4\omega_j^2}, \label{Toda_1}
 \ee
what is equivalent to a Toda-like system. A proof of the
equivalence, together with a short introduction to a theory of
Toda systems and methods of solving such kind of differential
equations will be given in the section \ref{Ts_and_sol}.

Elements of $\Delta$ matrix are:
 \be
  \Delta_{ij} = \frac{2}{D-2}
  \left( \sum_{\bar{I}:i,j\in \bar{I}} d_{\bar{I}}
   \sum_{\bar{J}:i,j\unin \bar{J}} d_{\bar{J}} -
   \sum_{\bar{I}:i \in \bar{I},j \unin \bar{I}} d_{\bar{I}}
   \sum_{\bar{J}:i \unin \bar{J},j \in \bar{J} }d_{\bar{J}} \right)
   + \sum_\alpha \varsigma_i a_{i\alpha} \varsigma_j a_{j\alpha},
   \label{delta_m_1}
 \ee
where indices $\bar{I}, \bar{J}$ run through all values allowed
for $I, J$ and additionally $\emptyset$, and by $d_\emptyset$ is
understood $\td$ (but not $\dim V_\emptyset$). By the definition
(\ref{delta_m_1}) the matrix $\Delta$ is symmetric. Its diagonal
elements obey:
 \be
  \Delta_{ii}=\frac{2 D_i \tD_i}{D-2} + \sum_\alpha a^2_{i\alpha}
 \ee
and the non-diagonal ones are bounded by:
 \be
  \Delta_{ij} \leq \frac12 (\Delta_{ii}+\Delta_{jj}).
 \ee

If $\det(\Delta) \neq 0$ then it is possible to express all
functions $A_I$, $\Btheta$ (\ref{interval_2}), $\phi_\alpha$
(\ref{phi_ansatz}), $C_i$ (\ref{el_Fansatz}) in terms of
$\omega_i$:
 \bea
  \exp(A_I(\vartheta)) &=& E_I \left( \prod_i
   \omega_i(\vartheta)^{\gamma^i_I} \right)
   \exp(c_I\vartheta),
   \label{sol_11} \\
  \exp(\Btheta(\vartheta)) &=& E_B \left(
   \prod_i \omega_i(\vartheta)^{\gamma^i_B} \right)
   \exp(c_B\vartheta) \rho(\vartheta)^{-(1+1/\td)}
   \label{sol_12}, \\
  \exp(\phi_\alpha(\vartheta)) &=&
   E_\alpha \left( \prod_i \omega_i(\vartheta)^{\gamma^i_\alpha}
   \right)\exp(c_\alpha\vartheta),
   \label{sol_13} \\
  \frac{d}{d\vartheta} \exp(C_i(\vartheta))&=&
  \frac{p_i\sigma_i}{\omega_i(\vartheta)^2},
   \label{sol_14}
 \eea
where $\gamma^i_I$, $\gamma^i_B$ and $\gamma^i_\alpha$ have to
satisfy:
 \bea
  \frac{D-2}{2} \sum_i \Delta_{ij} \gamma^i_I &=&
  \left\{ \begin{array}{ll}
     -\tD_j & \mbox{if    }  j \in I, \\
        D_j & \mbox{if    }  j \in \!\!\!\!\! / \;I,
     \end{array} \right. \\
  \sum_i \Delta_{ij} \gamma^i_\alpha &=& 2 a_{j \alpha}, \\
  \gamma^i_B &=& - \frac{1}{\td} \sum_I d_I \gamma^i_I.
 \eea
It can be checked that if all $a_{i \alpha}$ vanish, the numbers
$\gamma^i_I$ and $\gamma^i_B$ are in an excellent agreement with
the harmonic function rule presented in (\ref{harm_fun_rule_1} --
\ref{harm_fun_rule_3}). Values of real constants
$c_I$, $c_B$, $c_\alpha$ and positive
constants $E_I$, $E_B$, $E_\alpha$ are restricted by:
 \bea
  0 &=& \frac12 \sum_\alpha \varsigma_i a_{i\alpha} c_\alpha -
  \sum_{I:i\in I} d_I c_I, \label{sol_const_1} \\
  \prod_{I:i\in I}E_I^{d_I} &=&
  \prod_\alpha E_\alpha^{\frac12\varsigma_i a_{i\alpha}},
  \label{sol_const_2} \\
  0 &=& \sum_I d_I c_I +\td c_B, \label{sol_const_3} \\
  \frac{e^{\epsilon_\chi}}{2} &=& \left( \prod_I E_I^{d_I} \right) E_B^\td.
  \label{sol_const_4}
 \eea
So the problem of finding brane solution in gravity coupled to an
arbitrary number of antisymmetric tensors and scalar fields
without assumption of harmonic gauge can be reduced to solving the
Toda-like system (\ref{Toda_1}) with a constraint derived from
(\ref{eqm_33}):
 \be
  \sum_{ij} (\Delta^{-1})_{ij} \frac{\dot{\omega}_i}{\omega_i}
  \frac{\dot{\omega}_j}{\omega_j}
  + \frac12 (\Lambda_\chi + \Lambda_c) =
  \sum_i \frac{p_i^2}{4 \omega_i^2},
  \label{eq_en}
 \ee
where:
 \be
  \Lambda_c = \sum_I d_I c_I^2 + \td c_B^2 +
  \frac12 \sum_\alpha c_\alpha^2.
 \ee

Note, that if the model describes a single brane without dilaton
the solution is constructed only of $e^\Btheta$ and one $e^{A_I}$,
the constraints (\ref{sol_const_1}) and (\ref{sol_const_3})
enforce the constants $c_I$ and $c_B$ to vanish. Consequently the
identity $\Lambda_c=0$ holds.

%%%%%%%%%%%%%%%%%%%%%%%%%%%%%%%%%%%%%%%%%%%%%%%%%%%%%%%%%%%%%%%%%
\subsubsection{Solution for $\td=0$.}

\akapit Defining the function $\chi$ (\ref{lincon_1}) we imposed
to the discussed model additional constraint that $\td \neq 0$
what means that the overall transverse space cannot be two
dimensional. Now let us turn our attention to the opposite case,
i.e. we assume now that $\td=0$. In such situation the equations
of motion expressed in terms of the scalar functions $A_I, B, C_i,
\phi_\alpha$ (\ref{eqm_21} -- \ref{eqm_25}) simplify to:
 \be
  A''_I + A'_I \left( \sum_J d_J A'_J + \frac{1}{r} \right)
   = \frac{\sum_{i\in I} \tD_i(S'_i)^2 -
   \sum_{i\unin I} D_i(S'_i)^2}{2(D-2)}, \label{td0_eqm_21} 
 \ee
 \bea
  B'' + B' \left( \sum_I d_I A'_I + \frac{1}{r} \right) +
  \frac{1}{r} \sum_I d_I A'_I
   &=& - \frac{\sum_i D_i(S'_i)^2}{2(D-2)}, \label{td0_eqm_22} \\
  \phi''_\alpha + \phi'_\alpha \left( \sum_I d_I A'_I +
  \frac{1}{r} \right)
   &=& -\frac{1}{2}\sum_i \varsigma_i a_{i\alpha} (S'_i)^2,
   \label{td0_eqm_23} \\
  \sum_I d_I \left(A''_I -\frac{1}{r}A'_I-2A'_I B'+ (A'_I)^2 \right)
   &=& \frac{1}{2}\sum_i (S'_i)^2 -
   \frac{1}{2}\sum_\alpha (\phi'_\alpha)^2, \label{td0_eqm_24} \\
   C'_i \left( C'_i -\! \sum_{I:i\in I} d_I A'_I +
   \!\sum_{I:i \unin I} d_I A'_I
   + \!\sum_\alpha a_{i\alpha} \phi'_\alpha +
   \!\frac{1}{r} \right) &=& - C''_i, \label{td0_eqm_25}
 \eea
 where:
 \bea
  \varsigma_i &=& \left\{ \begin{array}{ll}
   +1 & \mbox{(electric),} \\
   -1 & \mbox{(magnetic),} \end{array} \right. \\
  S'_i &=& \left\{ \begin{array}{ll}
   \sigma_i (e^{C_i})'
   \exp(\frac{1}{2}\sum_\alpha a_{i\alpha}\phi_\alpha-
   \sum_{I:i\in I} d_I A_I)
   & \mbox{(electric),} \\
   & \\
   \frac{\lambda_i}{r^{\td+1}}
   \exp(\frac{1}{2}\sum_\alpha a_{i\alpha}\phi_\alpha -
   \sum_{I:i\unin I} d_I A_I)
   & \mbox{(magnetic).} \end{array} \right.
  \label{td0_S_1}
 \eea

Furthermore the definition of the $\chi$ function (\ref{lincon_1})
gives:
 \be
  \chi = \sum_I d_I A_I
  \label{td0_lincon_1}
 \ee
and the function has to satisfy:
 \be
  \chi''+(\chi')^2 + \frac{1}{r}\chi' = 0.
 \ee
Solving the above equation one immediately obtains:
 \be
  \chi(r) = \ln \left| \frac{c_\chi-\ln(1/r)}{c_\chi -c_0}\right| +
  \epsilon_\chi(c_0). \label{td0_chi1}
 \ee
We see that the factor $1/r^{2\td}$ appearing in (\ref{chi1}) is
replaced here by $\ln(1/r)$. The parameter $c_\chi$ takes not only
real but also infinite values and the limits of $\chi$ for
$c_\chi=\pm \infty$ are equal:
 \be
  \lim_{c_\chi \rightarrow \pm \infty} \chi(r;c_\chi) = \epsilon_\chi,
 \ee
so we can identify these points in the space of parameters
analogously as it was done for the $\td \neq 0$ case. We can also
introduce the parameter $R$, which in this situation instead of
(\ref{R_def}) has to obey:
 \be
  c_\chi = \ln\left(\frac{1}{R}\right), \label{td0_R_def}
 \ee
what after setting $c_0=0$ allows us to rewrite (\ref{td0_chi1})
as:
 \be
  \chi(r) = \ln \left|\frac{\ln(R/r)}{\ln R}\right|+\epsilon_\chi.
 \ee
Note that in distinction to (\ref{R_def}) it is not necessary to
introduce discrete parameter $s_\chi$ because for positive $R$ the
logarithm $\ln R$ takes positive as well as negative values.

The equation for $\vartheta$ is now:
 \be
  \vartheta''+\chi'\vartheta'+\frac{1}{r}\vartheta' = 0.
 \ee
so it is satisfied in general by $\vartheta = a \chi + b$ for
arbitrary real constants $a$ and $b$ which again can be
independently chosen for areas where $r>R$ and $r<R$ respectively.
Take then:
 \be
  \vartheta(r) = \left\{ \begin{array}{ccc}
   -|\ln R| \ln \left|\frac{\ln(R/r)}{\ln R}\right| & \mbox{for}, &  r<L, \\
   & & \\
    |\ln R| \ln \left|\frac{\ln(R/r)}{\ln R}\right| & \mbox{for}, &  r>L,
  \end{array} \right.
 \ee
what preserves a possibility of the identification
$c_\chi=+\infty$ ($R=0$) with $c_\chi=-\infty$ ($R=\infty$)
because gives:
 \be
  \lim_{R\rightarrow 0} \vartheta(r;R) =
  \lim_{R\rightarrow \infty} \vartheta(r;R) = \ln \frac{1}{r}.
  \label{theta_lim}
 \ee

If $R\neq 1$ it is possible to make a coordinate change which
replaces $r$ with $\vartheta$. But in distinction to the situation
of $\td \neq 0$, in this case $r$ usually covers wider area of
spacetime then $\vartheta$. More precisely for $R\in(0,\infty)$,
when $r$ ranges from $0$ to $R$, the coordinate $\vartheta$ runs
over all real values and makes it one again when $r \in (R,\infty)$.
So the change of the coordinates is well defined only if we restrict
the discussion to one of the areas. But if $R=0$ or $R=\infty$ then
from (\ref{theta_lim}) we have $\vartheta = -\ln r$ what works
properly for all positive $r$.

The spacetime interval in the coordinates $\vartheta$ is the same
as (\ref{interval_2}) and $\rho$ and $\Btheta$ are still defined
by (\ref{rho_0}) and (\ref{Btheta_0}) respectively, so:
 \be
  \rho = \exp \left( - \frac{\vartheta}{|\ln R|} \right) =
  \left| \frac{\ln R}{\ln (R/r) }\right|.
 \ee
But for $\Btheta$ we are not able to write an analog of
(\ref{Btheta_1}) because this function does not contribute to
$\chi$ if $\td=0$. The equations of motion in terms of $\vartheta$
are then:
 \bea
  \ddot{A}_I &=& \frac{\sum_{i\in I} \tD_i(\dot{S}_i)^2 -
  \sum_{i\unin I} D_i(\dot{S}_i)^2}{2(D-2)}, \label{td0_eqm_31} \\
  \ddot{\phi}_\alpha &=&
  -\frac{1}{2}\sum_i \varsigma_i a_{i\alpha} (\dot{S}_i)^2,
  \label{td0_eqm_32}\\
  \ddot{B}_\vartheta &=& - \frac{\sum_i D_i(\dot{S}_i)^2}{2(D-2)},
  \label{td0_eqm_33} \\
  \ddot{S}_i &=& - \left(\frac12 \sum_\alpha a_{i\alpha}
  \dot{\phi}_\alpha -
  \sum_{I:i\in I} d_I \dot{A}_I \right)\dot{S}_i, \qquad
  \mbox{(electric)},
  \label{td0_eqm_34}
 \eea
 \be
  \frac{1}{c_\chi^2} - \frac{2 \sgn(r-R)}{c_\chi} \dot{B}_\vartheta +
  \sum_I \left( d_I (\dot{A}_I)^2 \right)
  + \frac{1}{2}\sum_\alpha(\dot{\phi}_\alpha)^2 =
  \frac{1}{2}\sum_i (\dot{S}_i)^2 .
  \label{td0_eqm_35}
 \ee

The system can be solved analogously as in the case of $\td \neq
0$ by introducing the functions $\omega_i$ (\ref{omega_def}). But
because for $\td=0$ the Toda-like system (\ref{Toda_1}) does not
carry information necessary to determine a form of the $\Btheta$
we have to solve additionally (\ref{td0_eqm_33}). So as a result
of solving the Toda-like system, if $\det \Delta \neq 0$ we obtain
for the functions $A_I$ and $\phi_\alpha$:
 \bea
  \exp(A_I(\vartheta)) &=& E_I \left( \prod_i
   \omega_i(\vartheta)^{\gamma^i_I} \right)
   \exp(c_I\vartheta),
   \label{td0_sol_11} \\
   \exp(\phi_\alpha(\vartheta)) &=&
   E_\alpha \left( \prod_i \omega_i(\vartheta)^{\gamma^i_\alpha}
   \right)\exp(c_\alpha\vartheta),
   \label{td0_sol_12}
 \eea
 where $\gamma^i_I$ and $\gamma^i_\alpha$ have to satisfy:
 \bea
  \frac{D-2}{2} \sum_i \Delta_{ij} \gamma^i_I &=& \left\{ \begin{array}{ll}
     -\tD_j & \mbox{if    }  j \in I, \\
        D_j & \mbox{if    }  j \in \!\!\!\!\! / \;I,
     \end{array} \right. \\
  \sum_i \Delta_{ij} \gamma^i_\alpha &=& 2 a_{j \alpha}
 \eea
and values of real constants $c_I$, $c_\alpha$ and positive
constants $E_I$, $E_\alpha$ are restricted by:
 \bea
  0 &=& \frac12 \sum_\alpha \varsigma_i a_{i\alpha} c_\alpha -
  \sum_{I:i\in I} d_I c_I, \label{constr_1} \\
  \prod_{I:i\in I}E_I^{d_I} &=&
  \prod_\alpha E_\alpha^{\frac12\varsigma_i a_{i\alpha}},
  \label{constr_2} \\
  0 &=& \sum_I d_I c_I, \label{constr_3} \\
  \frac{e^{\epsilon_\chi}}{2} &=& \left( \prod_I E_I^{d_I} \right).
  \label{constr_4}
 \eea
But for the remaining functions $C_i$ and $\Btheta$ we have to
solve the following equations:
 \bea
  \frac{d^2}{d\vartheta^2} \ln \Btheta(\vartheta) &=&
  - \frac{1}{2(D-2)} \sum_i \frac{D_i p_i^2}{\omega_i(\vartheta)^2},
   \label{td0_sol_13} \\
  \frac{d}{d\vartheta} \exp(C_i(\vartheta))&=&
  \frac{p_i\sigma_i}{\omega_i(\vartheta)^2}
   \label{td0_sol_14}
 \eea
and to remove one integration constant by use of the constraint
(\ref{td0_eqm_35}).

Note also that combining the system (\ref{Toda_1}) with the
equation (\ref{td0_eqm_33}) we again obtain a Toda-like system:
 \be
  \frac{d^2}{d\vartheta^2} \ln \omega_{\bar{i}} =
   - \sum_{\bar{j}} {\bar{\Delta}}_{\bar{i}\bar{j}}
   \frac{p_{\bar{j}}^2}{4\omega_{\bar{j}}^2}, \label{td0_Toda_1}
 \ee
where the indices $\bar{i}, \bar{j}$ run through $1, \ldots, N_A$
as $i,j$ but additionally can take value $0$ and:
 \bea
  p_0 &=& 0, \\
  \omega_0 &=& \ln \Btheta, \\
  \bar{\Delta}_{\bar{i}\bar{j}} &=& \left( \begin{array}{cc}
   \Delta_{00}, & \Delta_{0j} \\
   \Delta_{i0}, & \Delta_{ij}
   \end{array} \right), \\
  \Delta_{00} &=& 0, \\
  \Delta_{0j} &=& \frac{2 D_j}{D-2}, \\
  \Delta_{i0} &=& 0.
 \eea

%%%%%%%%%%%%%%%%%%%%%%%%%%%%%%%%%%%%%%%%%%%%%%%%%%%%%%%%%%%%%%%%%
\subsubsection{Charges and masses}

\akapit Each of the branes in the system under consideration carry
some Ramond-Ramond charge. For the magnetic branes the charge is
defined by (\ref{mag_charge}) what immediately gives:
 \be
  Q_m^i = \int_{\hat{V}_i \times \de V_\emptyset} F^i =
  \Omega_{(\td+1)} |\hat{V}_i| \lambda_i.
 \ee
For the elementary branes we have to count the electric charge
from (\ref{el_charge}) and obtain:
 \be
  Q_e^i = \int_{\hat{V}_i \times \de V_\emptyset}
  \exp \left(\sum_\alpha a_{i \alpha}\phi_\alpha \right) \ast F^i
   = \Omega_{(\td+1)} |\hat{V}_i| q_e^i,
 \ee
where the quantity $q_e$ is defined by a formula which seems to be
very complicated:
 \be
  q_e^i = \sigma_i \exp \left(\sum_\alpha a_{i \alpha}\phi_\alpha -
  \sum_{I: i \in I} d_I A_I
        + \sum_{I: i \unin I} d_I A_I + \td B \right) (e^{C_i})' r^{\td+1}.
 \ee
but after more detailed analysis proves to be a constant number:
 \be
  q_e^i = \lambda_i.
 \ee

We can also calculate energy of the intersecting branes system.
Let us first see that for the gravitational tensors like Ricci
tensor, Ricci scalar and Einstein tensor:
 \be
  G_{MN} = R_{MN}-\frac{1}{2}g_{MN}R,
 \ee
thanks to the equations of motion (\ref{eqm_21} -- \ref{eqm_25})
is it possible to write them in terms of the functions $S'_i$
(\ref{S_1}):
 \bea
  R_{\mu^I\nu^I} &=& - e^{2(A_I-B)} \frac{\sum_{i\in I} \tD_i(S'_i)^2 -
  \sum_{i\unin I} D_i(S'_i)^2}{2(D-2)}
   \eta_{\mu^I\nu^I}, \label{riccit1} \\
  R_{mn} &=& \frac{\sum_i D_i(S'_i)^2}{2(D-2)} \delta_{mn} -
  \frac{1}{2} \sum_i (S'_i)^2 \frac{y_my_n}{r^2}
   + \frac{1}{2} \sum_\alpha (\phi'_\alpha)^2 \frac{y_my_n}{r^2}, \\
  R &=& e^{-2B} \left( \frac{\sum_i (D_i-\tD_i)(S'_i)^2}{2(D-2)}+
  \frac{1}{2}\sum_\alpha (\phi'_\alpha)^2 \right), \\
  G_{\mu^I\nu^I} &=& - \frac{1}{4}e^{2(A_I-B)}
  \left(\sum_{i\in I} (S'_i)^2 - \sum_{i \unin I} (S'_i)^2
   + \sum_\alpha (\phi'_\alpha)^2 \right) \eta_{\mu^I\nu^I},
   \label{einst1} \\
  G_{mn} &=& \left( \sum_I (S'_i)^2 -
  \sum_\alpha (\phi'_\alpha)^2 \right)
   \left( \frac{1}{4}\delta_{mn}-\frac{y_my_n}{2r^2} \right),
  \label{einst2}
 \eea
So, the component with double timelike indices of the stress-energy tensor is:
 \be
  T^t{}_t=G^t{}_t= e^{-2B}\left(\sum_i (S'_i)^2 +
  \sum_\alpha (\phi'_\alpha)^2 \right) \label{energ_dens}
 \ee
and an integral of it:
 \bea
  \calE &=& \int_\calM d^DX \sqrt{|\det g|} T^t{}_t \\
  % &=& \int_\calM d^DX e^{\chi} \left(\sum_i (S'_i)^2 + \sum_\alpha (\phi'_\alpha)^2 \right) \\
   &=& \Omega_{(\td+1)} |V| \int dr r^{\td+1} e^{\chi}
       \left(\sum_i (S'_i)^2 + \sum_\alpha (\phi'_\alpha)^2 \right)
       \label{energ_tot_1} \\
   &=& -\frac{e^{\epsilon_\chi}}{2} \Omega_{(\td+1)} |V| \int d\vartheta
       \left(\sum_i (\dot{S}_i)^2 +
       \sum_\alpha (\dot{\phi}_\alpha)^2 \right) \\
   &=& -\frac{e^{\epsilon_\chi}}{2} \Omega_{(\td+1)} |V| \int d\vartheta
       \left(\sum_i \left(\frac{p_i}{\omega_i} \right)^2 +
       \sum_\alpha \left( \sum_i \gamma_\alpha^i
       \frac{\dot{\omega}_i}{\omega_i} + c_\alpha \right)^2 \right).
       \label{energ_tot_3}
 \eea
The above integration should be taken over the whole space to
cover the whole energy density which can contribute to the total
energy. However it will be shown later (see the paragraph
\ref{sol_nd_vr}) that "the whole space" seen by a given observer
is not necessary the same as the area described by $r \in
(0,\infty)$ or $\vartheta \in \left(\vartheta(r=0),
\vartheta(r=\infty) \right)$ and to determine the area properly
some additional analysis is needed.

%%%%%%%%%%%%%%%%%%%%%%%%%%%%%%%%%%%%%%%%%%%%%%%%%%%%%%%%%%%%%%%%%
%%%%%%%%%%%%%%%%%%%%%%%%%%%%%%%%%%%%%%%%%%%%%%%%%%%%%%%%%%%%%%%%%
\subsection{Toda and Toda-like systems.}
\label{Ts_and_sol}

\akapit The Toda system is a system of second order differential
equations which can be written in a form \cite{AC}:
 \be
  \ddot{x}_i = -\sum_j K_{ij} \exp(-x_j).
  \label{Toda_2}
 \ee
where $K_{ij}$ is a Cartan matrix i.e. a matrix satisfying:
 \bea
  K_{ii} &=& 2, \label{Cartan_m_11} \\
  K_{ij} &\leq& 0, \qquad \mbox{for} \quad i\neq j,
  \label{Cartan_m_12} \\
  K_{ij}=0 &\Leftrightarrow& K_{ji}=0.
  \label{Cartan_m_13}
 \eea
After the substitution:
 \be
  q_i = \left( K^{-1} \right)_{ij} x_j,
 \ee
the Toda system is reformulated as:
 \be
  \ddot{q}_i = - \exp\left(-\sum_j K_{ij} q_j \right).
  \label{Toda_3}
 \ee
The above equations can be derived from a following lagrangian:
 \be
  L = -\frac{1}{2} \sum_{ij} K_{ij} \dot{q}_i \dot{q}_j
      - \sum_i \exp \left( -\sum_j K_{ij} q_j \right).
 \ee
This type of systems was studied in the context of gauge theories
on lattices \cite{Toda1,Toda2}, where the matrix $K_{ij}$ was
understood as a Cartan matrix of a simple Lie algebra
corresponding to a given compact Lie group describing the
considered gauge symmetry. There are four infinite series of Lie
algebras $A(n)$, $B(n)$, $C(n)$ and $D(n)$ and five exceptional
examples $E(6)$, $E(7)$, $E(8)$, $F(3)$ and $G(2)$. Let us review
them and their Cartan matrices.
\begin{itemize}
\item $A(n)$ or $sl(n+1,\bf{C})$:
 \be
  K_{A(n)}=\left( \begin{array}{rrrcrrr}
    2 &-1 & 0 & \cdots & 0 & 0 & 0 \\
   -1 & 2 &-1 & \cdots & 0 & 0 & 0 \\
    0 &-1 & 2 & \cdots & 0 & 0 & 0 \\
    \vdots & \vdots & \vdots & \ddots & \vdots & \vdots & \vdots \\
    0 & 0 & 0 & \cdots & 2 &-1 & 0 \\
    0 & 0 & 0 & \cdots &-1 & 2 &-1 \\
    0 & 0 & 0 & \cdots & 0 &-1 & 2
  \end{array} \right),
 \ee
\item $B(n)$ or $so(2n+1)$:
 \be
  K_{B(n)}=\left( \begin{array}{rrrcrrr}
    2 &-1 & 0 & \cdots & 0 & 0 & 0 \\
   -1 & 2 &-1 & \cdots & 0 & 0 & 0 \\
    0 &-1 & 2 & \cdots & 0 & 0 & 0 \\
    \vdots & \vdots & \vdots & \ddots & \vdots & \vdots & \vdots \\
    0 & 0 & 0 & \cdots & 2 &-1 & 0 \\
    0 & 0 & 0 & \cdots &-1 & 2 &-2 \\
    0 & 0 & 0 & \cdots & 0 &-1 & 2
  \end{array} \right),
 \ee
\item $C(n)$ or $sp(n,\bf{C})$:
 \be
  K_{C(n)}=\left( \begin{array}{rrrcrrr}
    2 &-1 & 0 & \cdots & 0 & 0 & 0 \\
   -1 & 2 &-1 & \cdots & 0 & 0 & 0 \\
    0 &-1 & 2 & \cdots & 0 & 0 & 0 \\
    \vdots & \vdots & \vdots & \ddots & \vdots & \vdots & \vdots \\
    0 & 0 & 0 & \cdots & 2 &-1 & 0 \\
    0 & 0 & 0 & \cdots &-1 & 2 &-1 \\
    0 & 0 & 0 & \cdots & 0 &-2 & 2
  \end{array} \right),
 \ee
\item $D(n)$ or $so(2n)$:
 \be
  K_{D(n)}=\left( \begin{array}{rrrcrrr}
    2 &-1 & 0 & \cdots & 0 & 0 & 0 \\
   -1 & 2 &-1 & \cdots & 0 & 0 & 0 \\
    0 &-1 & 2 & \cdots & 0 & 0 & 0 \\
    \vdots & \vdots & \vdots & \ddots & \vdots & \vdots & \vdots \\
    0 & 0 & 0 & \cdots & 2 &-1 &-1 \\
    0 & 0 & 0 & \cdots &-1 & 2 & 0 \\
    0 & 0 & 0 & \cdots &-1 & 0 & 2
  \end{array} \right),
 \ee
\end{itemize}
\begin{itemize}
\item $E(6)$:
 \be
  K_{E(6)}=\left( \begin{array}{rrrrrr}
    2 &-1 & 0 & 0 & 0 & 0 \\
   -1 & 2 &-1 & 0 & 0 & 0 \\
    0 &-1 & 2 &-1 & 0 &-1 \\
    0 & 0 &-1 & 2 &-1 & 0 \\
    0 & 0 & 0 &-1 & 2 & 0 \\
    0 & 0 &-1 & 0 & 0 & 2
  \end{array} \right),
 \ee
\item $E(7)$:
 \be
  K_{E(7)}=\left( \begin{array}{rrrrrrr}
    2 &-1 & 0 & 0 & 0 & 0 & 0 \\
   -1 & 2 &-1 & 0 & 0 & 0 & 0 \\
    0 &-1 & 2 &-1 & 0 & 0 &-1 \\
    0 & 0 &-1 & 2 &-1 & 0 & 0 \\
    0 & 0 & 0 &-1 & 2 &-1 & 0 \\
    0 & 0 & 0 & 0 &-1 & 2 & 0 \\
    0 & 0 &-1 & 0 & 0 & 0 & 2
  \end{array} \right),
 \ee
\item $E(8)$:
 \be
  K_{E(8)}=\left( \begin{array}{rrrrrrrr}
    2 &-1 & 0 & 0 & 0 & 0 & 0 & 0 \\
   -1 & 2 &-1 & 0 & 0 & 0 & 0 & 0 \\
    0 &-1 & 2 &-1 & 0 & 0 & 0 &-1 \\
    0 & 0 &-1 & 2 &-1 & 0 & 0 & 0 \\
    0 & 0 & 0 &-1 & 2 &-1 & 0 & 0 \\
    0 & 0 & 0 & 0 &-1 & 2 &-1 & 0 \\
    0 & 0 & 0 & 0 & 0 &-1 & 2 & 0 \\
    0 & 0 &-1 & 0 & 0 & 0 & 0 & 2
  \end{array} \right),
 \ee
\item $F(4)$:
 \be
  K_{F(4)}=\left( \begin{array}{rrrr}
    2 &-1 & 0 & 0 \\
   -1 & 2 &-2 & 0 \\
    0 &-1 & 2 &-1 \\
    0 & 0 &-1 & 2
  \end{array} \right),
 \ee
\item $G(2)$:
 \be
  K_{G(2)}=\left( \begin{array}{rr}
    2 &-3 \\
   -1 & 2
  \end{array} \right).
 \ee
\end{itemize}

Turning back to the problem of intersecting branes we see that
after the substitution:
 \be
  x_i=-2 \ln( \frac{p_i \sqrt{\Delta_{ii}}}{2 \omega_i})
 \ee
into (\ref{Toda_1}) the equations take the form of (\ref{Toda_2})
where:
 \be
  K_{ij}=\frac{2\Delta_{ij}}{\Delta{jj}}.
  \label{K_Delta}
 \ee
But for arbitrary $\Delta$ the matrix $K$ does not to have to obey
the Cartan matrix conditions (\ref{Cartan_m_11}) --
(\ref{Cartan_m_13}). Such generalizations of the Toda systems are
called Toda-like systems.

The fact that the models of the intersecting branes in full
generality reduce to Toda-like but not exactly Toda systems causes
a problem since methods of integration are developed mainly for
the latter ones. Fortunately quite large set of reasonable brane
configurations can be described in terms of the standard Toda
systems.

It is important to note that while any given matrix $\Delta$ is
necessarily symmetric the corresponding matrix $K$ defined by the
identity (\ref{K_Delta}) may be not symmetric. So those of the
above algebras which have not symmetric Cartan matrices can also
be relevant for some configurations of the intersecting branes.
Another very useful fact which significantly enlarges set of known
solvable cases is that if any matrix $K$ can be written in a block
diagonal form:
 \be
  K = \left( \begin{array}{ll}
   K_1 & 0 \\
   0   & K_2
  \end{array} \right)
 \ee
then the problem of solving the corresponding Toda system
decomposes into two separate problems related with $K_1$ and $K_2$
respectively. On a ground of Lie algebras it means that from
solutions of the simple Lie algebras one can immediately construct
solutions with semisimple algebras being a sum of the simple ones.
In particular the case of diagonal matrix $K$ is related to
semi-simple $\oplus^{n} A(1)$ algebra and can be treated as $n$
independent equations. Each of the equations is the Liouville
equation:
 \be
  \ddot{q} = - e^{-2q},
  \label{Liouville_1}
 \ee
which can be solved:
 \be
  e^{q(\vartheta)} = \left\{ \begin{array}{ll}
   \frac{1}{\sqrt{ \kappa}}
   \sin (\sqrt{ \kappa}(\vartheta-\theta)), & \mbox{for   } \kappa>0, \\
   & \\
   \left(         \vartheta-\theta \right), & \mbox{for   } \kappa=0, \\
   & \\
   \frac{1}{\sqrt{-\kappa}}
   \sinh(\sqrt{-\kappa}(\vartheta-\theta)), & \mbox{for   } \kappa<0,
  \end{array} \right.
 \label{lio_sol}
 \ee
where $\theta$ is a real constant.

A very elegant method leading to a solution of the Toda system
related to $A(n)$ algebra was given in \cite{Anderson} and
furthermore extended to $B(n)$ and $C(n)$ and applied to the
problem of intersecting branes under assumption of the harmonic
gauge \cite{CGM}. But of course the method works also in a general
situation without the harmonic gauge.

The solution for $A(n)$ is then:
\be
 e^{-q_k(\vartheta)}=
 i^{k(n+1-k)} \sum_{j_1<\cdots <j_k}^{n+1} f_{j_1} \cdots f_{j_k}
 V^2(j_1,\ldots,j_k) e^{(\mu_{j_1}+\cdots+\mu_{j_k})\vartheta},
 \label{q_sol_2}
\ee
where $V$ is the Vandermonde determinant defined as:
\be
 V(j_1,\ldots,j_k)=\prod_{j_m<j_l}\left(\mu_{j_m}-\mu_{j_l}\right),
 \qquad
 V(\mu_j)=1,
\ee
the complex constants $f_j$, $\mu_j$ satisfy:
\be
 \prod_{j=1}^{n+1} f_j = V^{-2}(1,\ldots,n+1),
 \qquad
 \sum_{j=1}^{n+1} \mu_j=0
\ee
and additionally the constraint (\ref{eq_en}).

%%%%%%%%%%%%%%%%%%%%%%%%%%%%%%%%%%%%%%%%%%%%%%%%%%%%%%%%%%%%%%%%%
%%%%%%%%%%%%%%%%%%%%%%%%%%%%%%%%%%%%%%%%%%%%%%%%%%%%%%%%%%%%%%%%%
\subsection{Diagonal $\Delta$ solution.}
\label{diag_delta_sol}

\akapit Consider the case when matrix $\Delta$ is diagonal and nonsingular.
Then with use of (\ref{lio_sol}) the equation (\ref{Toda_1}) gives:
 \be
  \omega_i(\vartheta) = \left\{ \begin{array}{ll}
   \left| \frac{p_i\sqrt{\Delta_{ii}}}{2\sqrt{ \kappa_i}}
   \sin (\sqrt{ \kappa_i}(\vartheta-\theta_i))
   \right|,
  & \mbox{for   } \kappa_i>0, \\
    \\
   \left| \frac{p_i\sqrt{\Delta_{ii}}}{2}
  (\vartheta-\theta_i) \right|, & \mbox{for   } \kappa_i=0, \\
   \\
   \left| \frac{p_i\sqrt{\Delta_{ii}}}{2\sqrt{-\kappa_i}}
 \sinh(\sqrt{-\kappa_i}(\vartheta-\theta_i)) \right|,
  & \mbox{for   } \kappa_i<0,
  \end{array} \right.
 \label{om_sol_1}
 \ee
where real phases $\theta_i$ are independent but (\ref{eq_en})
gives a restriction on $\kappa_i$ which for $\td \neq 0$ reads:
 \be
  \sum_i \frac{\kappa_i}{\Delta_{ii}} =
  \frac12 (\Lambda_\chi + \Lambda_c).
  \label{kappa_constr}
 \ee
Substituting (\ref{om_sol_1}) into (\ref{sol_11}--\ref{sol_13})
and solving (\ref{sol_14}):
 \be
  e^{C_i(\vartheta)} = \left\{ \begin{array}{ll}
   E_i-\frac{4\sqrt{ \kappa_i}}{p_i\Delta_{ii}}
   \cot (\sqrt{ \kappa_i}(\vartheta-\theta_i)), & \mbox{for   } \kappa_i>0, \\
    \\
   E_i-\frac{4}{p_i\Delta_{ii}}
   (\vartheta-\theta_i)^{-1},  & \mbox{for   } \kappa_i=0, \\
    \\
   E_i-\frac{4\sqrt{-\kappa_i}}{p_i\Delta_{ii}}
   \coth(\sqrt{-\kappa_i}(\vartheta-\theta_i)), & \mbox{for   } \kappa_i<0,
  \end{array} \right.
  \label{ec_sol_1}
 \ee
where $E_i$ are integration constants. Therefore in this case we
have an explicit form of the solution.

It is not necessary to apply the absolute value to the $\omega_i$
in (\ref{om_sol_1}). However if any $\omega_i$ is a solution of
(\ref{Toda_1}) then $-\omega_i$ and more general any
$\tilde{\omega}_i(\vartheta) = j(\vartheta) \omega_i(\vartheta)$
such that $|j|=1$ is also a good solution of the system of
equations everywhere except points where $j$ is discontinuous. But
because the equation (\ref{Toda_1}) itself is not well defined for
$\omega_i=0$, what means $\vartheta=\theta$ there is no obstacle
to assume that $j$ can change sign there. It gives a possibility
to write the functions $\omega_i$ as everywhere nonnegative which
is in an agreement with the definition (\ref{omega_def}).

%%%%%%%%%%%%%%%%%%%%%%%%%%%%%%%%%%%%%%%%%%%%%%%%%%%%%%%%%%%%%%%%%
%%%%%%%%%%%%%%%%%%%%%%%%%%%%%%%%%%%%%%%%%%%%%%%%%%%%%%%%%%%%%%%%%
\subsection{Single component solution.}
\label{single_charge_noharm}

\akapit Let us restrict our attention to the simplest case -- the
single component solution which was already discussed in the
section \ref{single_charge_harm} with the harmonic gauge condition
imposed. Let us now consider the case when the condition is
dropped.

%%%%%%%%%%%%%%%%%%%%%%%%%%%%%%%%%%%%%%%%%%%%%%%%%%%%%%%%%%%%%%%%%
\subsubsection{Solution without dilaton for $\td \geq 1$ -- general remarks.}
\label{sol_nd_gr}

\akapit If $N_A=1$, $N_\phi=0$ and $\td\neq 0$ the above solution
can be written as:
 \bea
  e^A &=& \omega^{-\frac{1}{d}}, \label{s_nd_A_1} \\
  e^\Btheta &=& \left( \frac12 e^{\epsilon_\chi}\right)^{\frac{1}{\td}}
  \omega^{\frac{1}{\td}} \rho^{-1-\frac{1}{\td}},
  \label{s_nd_B_1}
 \eea
with $e^C = e^{C_{i=1}}$ given by (\ref{ec_sol_1}), $\omega =
\omega_{i=1}$ by (\ref{om_sol_1}) and $E_C=E_{i=1}$, $p=p_{i=1}$,
$\theta=\theta_{i=1}$, $\kappa = \kappa_{i=1}$, $\Delta =
\Delta_{11}$, where:
 \be
  \kappa = \frac{\Delta}{2} \Lambda_\chi.
 \ee
The parameter $E_C$ has no physical meaning because it describes
only freedom of shift of the antisymmetric potential: $e^C
\rightarrow E_C+e^C$ and values of $\theta$ and $\epsilon_\chi$
can be fixed if we assume that at a given point $\vartheta_0$ the
functions $A$ and $\Btheta$ (or B) take certain values.

For example if with $\vartheta \rightarrow \vartheta_0$ the
spacetime tends to flat one, it is natural to set $A(\vartheta_0)
= B(\vartheta_0)=0$ what equivalently means:
 \be
  \omega(\vartheta_0)=1, \qquad \chi(\vartheta_0)=0.
  \label{bound_cond}
 \ee
In such a situation calculating the energy density of the brane
from a formula derived form (\ref{energ_tot_3}):
 \be
  \calE = -\frac{e^{\epsilon_\chi}p^2}{2} \Omega_{(\td+1)} |V|
  \int \frac{d\vartheta}{\omega^2}
       \label{energ_tot_nd}
 \ee
we should take a boundary of the integration exactly at
$\vartheta_0$ what gives \cite{BMO}:
 \be
  \calE = \Omega_{\td+1} \left|V \right| \frac{2}{\sqrt{\Delta}}
   \sqrt{\lambda^2-\frac{1}{2} \Lambda_\chi }.
   \label{energ_nd}
 \ee
When the solution is supersymmetric ($R=0$ and consequently
$\Lambda_\chi=0$) it reproduces properly the BPS equality between
mass and charge densities (\ref{1_energ_3}). And if $0 < R <
\infty$ and $s_\chi = +1$ it gives $\calE > Q$ as it should for
nonsupersymmetric solutions. There is an exception for the
solution with $R=\infty$ (which evidently is not supersymmetric
because when applied to the electric $2$-brane solution in $D=11$
it does not satisfy (\ref{susy_unb1_11e})) -- we have then $\calE
= Q$. This leads to a contradiction with the rule that the BPS
states are supersymmetric. Moreover, for $s_\chi=-1$ we get even
$\calE < Q$.

This is only an apparent paradox, because we should remember that
the formula (\ref{energ_nd}) is well defined only when certain
conditions are satisfied. The conditions which assure that
performing the integration of (\ref{energ_tot_nd}) we keep all
energy inside the integrated area but do not encounter any naked
singularity. Additionally it may happen that it is impossible to
impose the boundary conditions (\ref{bound_cond}). So the validity
of the formula (\ref{energ_nd}) can be limited to only some
subspace of the whole parameter space. To describe the subspace we
need to examine more precisely some aspects of a geometry of
various variants of the solution and find where there are
singularities, horizons and boundary points at infinity.

\vskip 2em
{\bf Invariants of the metric.}
\vskip 1em

Let us calculate invariants constructed by coefficients of the
metric tensor:
 \bea
  R_{MN} R^{MN} &=& \frac{1}{4(D-2)^2}
  \left( \td^2(d+1) + d^2(\td+1) \right)
  e^{-4B} {S'}^4, \label{Rt_inv} \\
  R &=& \frac{d-\td}{2(D-2)} e^{-2B} {S'}^2, \\
  G_{MN} G^{MN} &=& \frac{D}{16} e^{-4B} {S'}^4
  \label{Gt_inv}
 \eea
and check where the quantities take infinite or zero values. Their
poles indicate the points which can be "suspected" to be
singularities and the zeros indicate the points where the geometry
tends to be flat. Except the situation when $d=\td$ and the Ricci
scalar identically vanishes all the quantities are proportional to
a positive power of the function $e^{-B}S'$, so they all have
zeros and poles at the same points. For $e^{-B}S'$ we have:
 \be
  e^{-B}S' = e^{-\Btheta} \dot{S} =
  \const \left( \frac{\rho}{\omega} \right)^{1+1/\td}\,.
 \ee
Therefore to determine poles and zeros it is enough to study
properties of the functions $\omega$ and $\rho$. It is interesting
that if $\td=-1$ then all the invariants are constant -- this case
should be considered separately and we assume that $\td \geq 1$
from now on. We know that the zeros of $\omega$ are given by
$\vartheta = \theta$, so the points are "suspected" to describe
localization of the singularities. But to verify real nature of
the points it is very helpful to examine behavior of test
particles freely falling onto them.

\vskip 2em
{\bf The proper and the coordinate time.}
\vskip 1em

Because of the spherical symmetry of the model it is enough to
look at a test particle moving only in the radial direction i.e.
consider only time $t$ and radial coordinate $\vartheta$. The
equation for geodesic in coordinates $X^M$ reads:
 \be
  \frac{d^2}{d\tau^2} {X}^M+
  \Gamma^M_{NR} \frac{d}{d\tau} X^N \frac{d}{d\tau} X^R =0,
 \ee
where $\tau$ describes an affine parameter and the Christoffel
symbols are:
 \be
  \left. \begin{array}{lll}
   \Gamma^t_{tt}=0,       & \quad & \Gamma^\vartheta_{tt}=
   \dot{A} e^{2(A-\Btheta)}, \\
   \Gamma^t_{\vartheta t}=\dot{A}, & \quad &
   \Gamma^\vartheta_{\vartheta t}=0, \\
   \Gamma^t_{\vartheta\vartheta}=0 ,      &
   \quad & \Gamma^\vartheta_{\vartheta\vartheta}=\dot{\Btheta}.
  \end{array} \right.
 \ee
Therefore equations for the trajectory of the test particle read:
 \bea
  \frac{d^2}{d\tau^2} t + 2\dot{A} \frac{d}{d\tau}t
  \frac{d}{d\tau}\vartheta &=& 0, \\
  \frac{d^2}{d\tau^2} \vartheta + \dot{\Btheta}
  \left(\frac{d}{d\tau} \vartheta \right)^2
    + \dot{A} e^{2(A-\Btheta)} \left( \frac{d}{d\tau} t \right)^2 &=&
    0.
 \eea
The first equation leads to:
 \be
  E = e^{2A} \frac{d}{d\tau} t = \const,
 \ee
where $E$ has interpretation of test particle energy and then the
second equation can be rewritten as:
 \be
  \frac{d^2}{d\tau^2} r+ \dot{\Btheta}
  \left(\frac{d}{d\tau} r \right)^2 + \dot{A} e^{-2(A+\Btheta)}E^2 = 0.
 \ee
Solving this equation one finally obtains:
 \be
  \frac{d}{d\tau} r = e^{-\Btheta} \sqrt{C+E^2e^{-2A}},
 \ee
where $C$ is another integration constant. For radially
propagating test particle one has:
 \be
  ds^2=-e^{2A} dt^2 + e^{2\Btheta} d\vartheta^2 = C d\tau^2.
 \ee
Since any massive particle must have $ds^2<0$ it is convenient to
set in that case $C=-1$, what defines units of the proper time.
For photons and other massless particles we must set $C=0$.

We should calculate the coordinate time $\delta t$ and proper time
(or affine parameter if the particle is massless) $\delta \tau$
that passes when the test particle falls from an arbitrary point
$\vartheta_1$ to the examined one $\vartheta_0$. For our needs it
is enough to check if the values are finite or infinite for
$\vartheta_1$ arbitrarily close to $\vartheta_0$. These times, for
the massive test particle, are:
 \bea
  \delta \tau &=& \int_{\vartheta_1}^{\vartheta_0}
  \frac{e^\Btheta d\vartheta}{\sqrt{E^2e^{-2A}-1}},
  \label{tau_massive_1} \\
  \delta t    &=& \int_{\vartheta_1}^{\vartheta_0}
  \frac{Ee^{\Btheta-2A} d\vartheta}{\sqrt{E^2e^{-2A}-1}}
  \label{t_massive_1}
 \eea
and for the massless test particle:
 \bea
  \delta \tau &=& \int_{\vartheta_1}^{\vartheta_0}
  \frac{1}{E} e^{A+\Btheta} d\vartheta,
  \label{tau_massless_1} \\
  \delta t    &=& \int_{\vartheta_1}^{\vartheta_0}
  e^{\Btheta-A} d\vartheta.
  \label{t_massless_1}
 \eea
For the discussed model the last integrals can be rewritten as:
 \bea
  \delta \tau &=& \int_{\vartheta_1}^{\vartheta_0} \frac{1}{E}
           \omega^{-\frac{1}{d}+\frac{1}{\td}}
           \rho^{-1-\frac{1}{\td}} d\vartheta,
           \label{tau_massless_2} \\
  \delta t    &=& \int_{\vartheta_1}^{\vartheta_0}
           \omega^{\frac{1}{d}+\frac{1}{\td}}
           \rho^{-1-\frac{1}{\td}} d\vartheta,
           \label{t_massless_2}
 \eea
so everything depends on the functions $\omega$ and $\rho$.

%%%%%%%%%%%%%%%%%%%%%%%%%%%%%%%%%%%%%%%%%%%%%%%%%%%%%%%%%%%%%%%%%
\subsubsection{Solution without dilaton for $\td \geq 1$ -- variants review.}
\label{sol_nd_vr}

\akapit
We decompose the general solution into several variants with
respect to values taken by parameters $R, s_\chi$ and $\theta$.

We distinguish:
 \begin{itemize}
 \item Variant I where $R=0$, what immediately gives $\kappa=0$,
 \item Variant II where $R\in (0,\infty)$ and $s_\chi=+1$, so $\kappa<0$,
 \item Variant III where $R=\infty$, so $\kappa=0$,
 \item Variant IV where $R\in (0,\infty)$ and $s_\chi=-1$, so $\kappa>0$
 \end{itemize}
 and
 \begin{itemize}
 \item Variant A when $\vartheta=\theta$ does not belong to
       an area described by positive $r$,
 \item Variant B when $\vartheta=\theta$ is on an edge of the area
      with positive $r$,
 \item Variant C when $\vartheta=\theta$ belongs to the area of positive $r$.
 \end{itemize}

\vskip 1em
{\bf Variant IA} where $R = 0$, $\kappa=0$ and $\theta < -\frac{1}{2\td}$.
Then:
 \bea
  \rho &=& \left| \td\vartheta + \frac{1}{2} \right|,
  \label{rho_IA} \\
  \omega &=& \frac{|p|\sqrt{\Delta}}{2} \left| \vartheta-\theta \right|,
  \label{om_IA}
 \eea

\begin{table}[htbp]
\begin{tabular}{|c||c|c|c|c|c|}
\hline
 $\vartheta$    & $-\infty$ & $\theta$ & $-\frac{1}{2\td}$ & $0$ & $+\infty$ \\
\hline
 $r$     & undefined   & undefined  & $\infty$  & $1$       & $0$       \\
\hline
\hline
 $\dot{S}e^{-\Btheta}$  & finite & $\infty$   & $0$    & finite    & finite    \\
\hline
 $\delta t$ & $\infty$    & finite     & $\infty$    & finite    & $\infty$  \\
\hline
 $\delta \tau$      & finite & finite     & $\infty$ & finite    & finite    \\
\hline
\hline
 $e^A$          & $0$ & $\infty$   & finite            & finite    & $0$  \\
\hline
 $e^\Btheta$   & $0$   & $0$        & $\infty$          & finite    & $0$  \\
\hline
 $e^B$   & undefined   & undefined  & finite            & finite    & $\infty$ \\
\hline
\end{tabular}
\caption{Variant IA.}
\label{v_IA}
\end{table}

This gives exactly the supersymmetric single brane solution with a
naked singularity hidden behind a horizon which was discussed in
the section \ref{single_charge_harm}. The point
$\vartheta=-\frac{1}{2\td}$ (or equivalently $r=\infty$) describes
boundary of the spacetime at infinity where the geometry tends to
the flat one. Because $e^A$ and $e^B$ are finite there it is
possible to normalize them to one with the conditions
(\ref{bound_cond}) and get:
 \bea
  \left|\theta+\frac{1}{2\td}\right| &=& \frac{2}{|p|\sqrt{\Delta}}, \\
  \epsilon_\chi &=& 0.
 \eea
The area where $\vartheta > -\frac{1}{2\td}$ is related to
positive values of the isotropic radial coordinate $r$ and at
$\vartheta=\infty$ (or $r=0$) where metric tensor coefficients
$e^A$ and $e^\Btheta$ become singular but the invariants
(\ref{Rt_inv} -- \ref{Gt_inv}) are finite the horizon is located.
The solution can be extended beyond the horizon by an
identification of $\vartheta=+\infty$ with $\vartheta=-\infty$ and
continued to $\vartheta = \theta$ where the naked singularity is
settled. The singularity means a point which is reachable with
finite $\delta \tau$ and where the invariants of the metric grows
to infinity. An observer located at positive $r$ can see only a
region of the spacetime described by $\vartheta \in
\left(-\frac{1}{2\td},\infty \right)$ (or equivalently $r \in
(\infty,0)$), because any piece of information sent to the
observer from other points needs to travel infinite amount of
coordinate time. So counting the total energy of the brane we have
to integrate (\ref{energ_tot_nd}) over the area and get:
 \be
  \calE = -\frac{e^{\epsilon_\chi}p^2}{2} \Omega_{(\td+1)} |V|
       \int^{\vartheta=-\frac{1}{2\td}}_{\vartheta=\infty}
       \frac{d\vartheta}{\omega^2(\vartheta)}
     = \Omega_{(\td+1)}|V|\frac{2}{\sqrt{\Delta}}|\lambda|.
       \label{energ_tot_IA}
 \ee

\vskip 1em
{\bf Variant IB} where $R = 0$, $\kappa=0$ and $\theta = -\frac{1}{2\td}$, so:
 \bea
 \rho &=& \left| \vartheta + \frac{1}{2\td} \right|, \label{rho_IB} \\
  \omega &=& \frac{|p|\sqrt{\Delta}}{2} \left| \vartheta +
  \frac{1}{2\td} \right|. \label{om_IB}
 \eea

\begin{table}[htbp]
\begin{tabular}{|c||c|c|c|c|}
\hline
 $\vartheta$  & $-\infty$   & $\theta=-\frac{1}{2\td}$    & $0$  & $+\infty$ \\
\hline
 $r$      & undefined   & $\infty$            & $1$       & $0$       \\
\hline
\hline
 $\dot{S}e^{-\Btheta}$  & finite      & finite      & finite    & finite    \\
\hline
 $\delta t$         & $\infty$    & finite         & finite    & $\infty$  \\
\hline
 $\delta \tau$      & finite      & $\infty$        & finite    & finite    \\
\hline
\hline
 $e^A$          & $0$         & $\infty$            & finite    & $0$  \\
\hline
 $e^\Btheta$        & $0$         & $\infty$         & finite    & $0$  \\
\hline
 $e^B$          & undefined   & $0$               & finite    & $\infty$    \\
\hline
\end{tabular}
\caption{Variant IB.}
\label{v_IB}
\end{table}

Then $\dot{S}e^{-\Btheta}$ is everywhere constant, non zero and
finite what means that there is no singularity nor flat places at
all. The point $\vartheta=-\frac{1}{2\td}$ is characterized by
$\delta \tau = \infty$, so we can interpret it as the boundary at
infinity. But a geometry at the infinity is not flat and the metric
tensor coefficients are singular so one should expect an external
energy source attached at the point sustaining the nonzero
curvature -- the brane. Note also that a travel to the brane takes
a finite amount of coordinate time $t$, so an observer located at
positive $r$ can "feel" a presence of the object. Integrating
(\ref{energ_tot_nd}) from $\vartheta = \infty$ where a horizon is
localised to $\vartheta=-\frac{1}{2\td}$ and calculating in this
way the total energy of the brane one gets an infinite value. It
suggests that to evaluate the energy properly one should take into
account also an external contribution originating from the point
$\vartheta=\theta$.
%Additional problem is caused by the fact that
%(\ref{bound_cond}) cannot be imposed in the variant. While we have
%a priori $\theta = -\frac{1}{2\td}$ there is no similar condition
%determining $\epsilon_\chi$.
The solution can be continued beyond the horizon at
$\vartheta=+\infty$ when it is identified with $\vartheta =
-\infty$. At $\vartheta=-\frac{1}{2\td}$ reached now from the left
side one again encounters the external energy source. This
situation (and all other type B variants) are quite similar to a
solution describing spacetime enclosed between two domain walls.
But in this case we have not one but three or more direction
transversal to the branes.

\vskip 1em {\bf Variant IC} where $R = 0$, $\kappa=0$ and $\theta
> -\frac{1}{2\td}$. It is very similar to the variant IA but now
the singularity appears in the area corresponding to positive $r$.

\vskip 1em
{\bf Variant IIC/A} where $R\in (0,\infty)$, $s_\chi=+1$,
$\kappa = \frac{\Delta}{2} \Lambda_\chi <0$ and
 $\theta < \vartheta_0 = -\frac{1}{2\td}
 \left|(1/R)^{\td}-R^{\td}\right| \Arth R^\td$ then:
  \bea
   \rho &=& \frac{1}{4} \left|(1/R)^{\td}-R^{\td}\right|
    \left|\sinh\left(\frac{4\td\vartheta}{\left|(1/R)^{\td}-R^{\td}\right|}+
    2\Arth R^\td \right) \right|, \\
   \omega &=& \frac{p\sqrt{\Delta}}{2\sqrt{-\kappa}}
   \left| \sinh(\sqrt{-\kappa}(\vartheta-\theta)) \right|.
  \eea

\begin{table}[htbp]
\begin{tabular}{|c||c|c|c|c|c|}
\hline
 $\vartheta$ & $-\infty$ & $\theta$  & $\vartheta_0$     & $0$   & $+\infty$ \\
\hline
 $r$     & $R$       & finite    & $0|\infty$        & $1$       & $R$       \\
\hline
\hline
 $\dot{S}e^{-\Btheta}$  & $0$       & $\infty$  & $0$   & finite    & $0$    \\
\hline
 $\delta t$         & $\infty$  & finite    & $\infty$ & finite    & $\infty$  \\
\hline
 $\delta \tau$      & finite    & finite    & $\infty$ & finite    & finite    \\
\hline
\hline
 $e^A$          & $0$       & $\infty$  & finite      & finite    & $0$  \\
\hline
 $e^\Btheta$        & $0$       & $0$       & $\infty$   & finite    & $0$  \\
\hline
 $e^B$ & $\infty$  & $0$  & $\infty|\mbox{finite}$    & finite    & $\infty$    \\
\hline
\end{tabular}
\caption{Variant IIC/A.}
\label{v_IICA}
\end{table}

In the variant both coordinates $\vartheta \in [-\infty,+\infty]$
and $r\in [0,\infty]$ cover the same area of the spacetime. When
$\vartheta\rightarrow\vartheta_0$ (or equivalently
$r\rightarrow\infty$) a geometry of the spacetime tends to
Minkowski spacetime and we can normalize $e^A$ and $e^B$ with
(\ref{bound_cond}). Going in the opposite direction at
$\vartheta=\infty$ ($r=R$) one encounters a horizon. So we can
give now a physical interpretation to the parameter $R$ as a
length where in the isotropic coordinate system the horizon is
situated. When $R\rightarrow 0$ the localization goes to $r=0$ and
in this limit we recover the supersymmetric solution described by
the variant IA. But let us turn back to the variant IIC/A. Near
the horizon the quantity $\dot{S}e^{-\Btheta}$ behaves as:
 \be
  \dot{S}e^{-\Btheta} =
   \exp \left( \frac{4(\td+1)
   \left(1-\sqrt{\frac{d\td+d}{d+\td}}\right)}{\left|1/R^\td-R^\td \right|}
   \vartheta \right)
   \stackrel{\vartheta \rightarrow \infty}{\longrightarrow} 0,
 \ee
so it vanishes at $r=R$. Behind the horizon we can continue the
solution with negative $\vartheta$ ($r<R$) and at
$\vartheta=\theta$ we find a naked singularity. Since the
singularity is placed in the area covered by $r$ we should
classify the variant as type C. But if we restrict our
consideration to only $r>R$ we get a situation of type A, and this
is an explanation why the variant is called IIC/A.

An observer living in the part of the spacetime where there is no
singularity, it means at $r \in (R,\infty)$ sees only the area
between the flat infinity at $r=\infty$ and the horizon $r=R$. So
counting the energy density from (\ref{energ_tot_nd}) one should
take exactly such limits of the integration. And this gives:
 \be
  \calE = \Omega_{\td+1} \left|V \right| \frac{2}{\sqrt{\Delta}}
   \sqrt{\lambda^2+\frac{8\td(\td+1)}{\left|1/R^\td-R^\td \right|^2}},
   \label{energ_tot_IIA}
 \ee
where we used the definition (\ref{kosmo_1}) of $\Lambda_\chi$
with $s_\chi=+1$. So the energy is always bigger than the charge
of the brane as it should be expected for a nonsupersymmetric
solution.

\vskip 1em
{\bf Variant IIB} where $R\in (0,\infty)$, $s_\chi=+1$, $\kappa<0$ and
 $\theta = -\frac{1}{2\td} \left|(1/R)^{\td}-R^{\td}\right| \Arth R^\td $, so:
  \bea
   \rho &=& \frac{1}{4} \left|(1/R)^{\td}-R^{\td}\right|
    \left|\sinh\left(\frac{4\td\vartheta}{\left|(1/R)^{\td}-R^{\td}\right|}+
    2\Arth R^\td \right) \right|, \\
   \omega &=& \frac{p\sqrt{\Delta}}{2\sqrt{-\kappa}} \left|
     \sinh\left(\sqrt{-\kappa}\left(\vartheta+\frac{1}{2\td}
     \left|(1/R)^{\td}-R^{\td}\right| \Arth R^\td
    \right)\right) \right|.
  \eea

\begin{table}[htbp]
\begin{tabular}{|c||c|c|c|c|}
\hline
 $\vartheta$            & $-\infty$ & $\theta$  & $0$       & $+\infty$ \\
\hline
 $r$                    & R     & $\infty|0$& $1$       & $R$       \\
\hline
\hline
 $\dot{S}e^{-\Btheta}$  & $0$       & finite    & finite    & $0$    \\
\hline
 $\delta t$         & $\infty$  & finite    & finite    & $\infty$  \\
\hline
 $\delta \tau$      & finite    & $\infty$  & finite    & finite    \\
\hline
\hline
 $e^A$          & $0$       & $\infty$  & finite    & $0$  \\
\hline
 $e^\Btheta$        & $0$       & $\infty$  & finite    & $0$  \\
\hline
 $e^B$          & $\infty$  & $\infty|0$& finite    & $\infty$    \\
\hline
\end{tabular}
\caption{Variant IIB.}
\label{v_IIB}
\end{table}

The variant is quite similar to the IB, but now the isotropic
coordinates cover the whole spacetime. Again we find the spacetime
to be free of singularities and stretched between two points
reachable in finite coordinate time where the curvature is finite
but non zero, so we should expect the points to be localizations
of the branes. Both the points are described in the $\vartheta$
coordinate system  as $\vartheta=\theta$ (but as $r=0$ and
$r=\infty$) and they are separated by a horizon situated at $r=R$.

\vskip 1em

{\bf Variant IIA/C} where $R\in (0,\infty)$, $s_\chi=+1$, $\kappa
= \frac{\Delta}{2} \Lambda_\chi <0$ and $\theta > \vartheta_0 =
-\frac{1}{2\td} \left|(1/R)^{\td}-R^{\td}\right| \Arth R^\td$. It
can be obtained form the variant IIC/A by shifting the point
$\vartheta=\theta$ with the naked singularity to the area where
$r>R$. So, the region "behind the horizon" described by $r<R$ is
nonsingular now. We can count the energy density and find again
(\ref{energ_tot_IIA}). But to achieve it we should not
determine $\epsilon_\chi$ by the boundary conditions for $e^B$
$r=\infty$, what is a point separated from the discussed area by
the horizon and the singularity. We should rather make a coordinate
change to introduce $\bar{r}=1/r$ and normalize finite value of
$e^{\bar{B}}$ to one at $\bar{r}=\infty$ what means $r=0$. However,
both the methods gives the same result $\epsilon_\chi=0$.

\vskip 1em
{\bf Variant IIIA} where $R =\infty$, $\kappa=0$ and $\theta > \frac{1}{2\td}$.
 Then the functions $\rho$ and $\omega$ satisfy:
 \bea
  \rho &=& \left| \td \vartheta - \frac{1}{2} \right|, \label{rho_IIIA} \\
  \omega &=& \frac{|p|\sqrt{\Delta}}{2} \left| \vartheta-\theta \right|.
  \label{om_IIIA}
 \eea

\begin{table}[htbp]
\begin{tabular}{|c||c|c|c|c|c|}
\hline
 $\vartheta$   & $-\infty$ & $0$   & $\frac{1}{2\td}$& $\theta$  & $+\infty$ \\
\hline
 $r$        & $\infty$  & $1$   & $0$           & undefined & undefined       \\
\hline
\hline
 $\dot{S}e^{-\Btheta}$  & finite    & finite    & $0$  & $\infty$  & finite    \\
\hline
 $\delta t$         & $\infty$  & finite    & $\infty$  & finite    & $\infty$  \\
\hline
 $\delta \tau$      & finite    & finite    & $\infty$ & finite    & finite    \\
\hline
\hline
 $e^A$          & $0$       & finite    & finite      & $\infty$  & $0$  \\
\hline
 $e^\Btheta$        & $0$       & finite    & $\infty$   & $0$       & $0$  \\
\hline
 $e^B$     & $O(r^{\td/d-2})$  & finite& $\infty$  & undefined & undefined   \\
\hline
\end{tabular}
\caption{Variant IIIA.}
\label{v_IIIA}
\end{table}

While the variant IA can be understood as a some kind of a limit
of the variant IIIC/A when $R$ converges to $0$, the variant IIIA
is a similar limit of IIIA/C when $R\rightarrow \infty$. We have
now a singularity at $\vartheta=\theta$ hidden behind a horizon
located at $\vartheta=\pm \infty$ (or $r=\infty$). But for
$\vartheta \in (-\infty, \frac{1}{2\td})$ which corresponds to
positive $r$ the solution is regular. Finally at $\vartheta =
\frac{1}{2\td}$ ($r=0$) a boundary at infinity is placed where the
spacetime tends to the flat one. The last statement can be easily
seen when the coordinate $r$ is replaced by $\bar{r}=1/r$. Then
both functions $e^A$ and $e^{\bar{B}}$ tend asymptotically to
finite values with $\bar{r} \rightarrow \infty$. Taking the values
as equal to $1$ we get:
 \bea
  \left|\theta-\frac{1}{2\td}\right| &=& \frac{2}{|p|\sqrt{\Delta}}, \\
  \epsilon_\chi &=& 0,
 \eea
what furthermore gives:
 \be
  \calE = \Omega_{(\td+1)}|V|\frac{2}{\sqrt{\Delta}}|\lambda|.
       \label{energ_tot_IIIA}
 \ee
To achieve the positive sign in the above formula the integration
leading to it has to be conducted from $\bar{r}=0$ ($r=\infty$) to
$\bar{r}=\infty$ ($r=0$) instead of from $r=0$ to $r=\infty$. But
the obtained result is strange at a first sight because it says
that this variant saturates BPS bound while being
nonsupersymmetric (since the function $\chi$ does not vanish in
this case). An explanation of the fact is given below in the
paragraph \ref{susy_nd}.

\vskip 1em

{\bf Variant IIIB} where $R =\infty$, $\kappa=0$ and $\theta =
\frac{1}{2\td}$ analogously as the variant IB describes a
spacetime stretched between two boundaries where the branes
sustain nonzero curvature.

\begin{table}[htbp]
\begin{tabular}{|c||c|c|c|c|}
\hline
 $\vartheta$      & $-\infty$ & $0$  & $\theta=\frac{1}{2\td}$   & $+\infty$ \\
\hline
 $r$        & $0$       & $1$       & $\infty$             & undefined       \\
\hline
\hline
 $\dot{S}e^{-\Btheta}$  & finite    & finite    & finite      & finite    \\
\hline
 $\delta t$         & $\infty$  & finite    & finite           & $\infty$  \\
\hline
 $\delta \tau$      & finite    & finite    & $\infty$          & finite    \\
\hline
\hline
 $e^A$          & $0$       & finite    & $\infty$           & $0$  \\
\hline
 $e^\Btheta$        & $0$       & finite    & $\infty$          & $0$  \\
\hline
 $e^B$     & $O(r^{\td/d-2})$  & finite & $O(r^{\td/d-2})$   & undefined    \\
\hline
\end{tabular}
\caption{Variant IIIB.}
\label{v_IIIB}
\end{table}

\vskip 1em

{\bf Variant IIIC} where $R =\infty$, $\kappa=0$ and $\theta <
\frac{1}{2\td}$ is similar to the variant IIIA but now the naked
singularity emerges in the area covered by isotropic coordinates.

\vskip 1em

{\bf Variant IVC} where $R\in (0,\infty)$, $s_\chi=-1$, $\kappa>0$
and $\theta \neq \vartheta_0$, where it is defined
$\vartheta_0 = -\frac{(1/R)^{\td}+R^{\td}}{2\td} \arctan R^\td$, then:
  \bea
   \rho &=& \frac{1}{4} \left((1/R)^{\td}+R^{\td}\right)
    \left| \sin\left(\frac{4\td\vartheta}{(1/R)^{\td}+R^{\td}}+
    2\arctan R^\td\right) \right|, \\
   \omega &=& \frac{|p|\sqrt{\Delta}}{2\sqrt{\kappa}}
   \left| \sin(\sqrt{\kappa}(\vartheta-\theta)) \right|.
  \eea

\begin{table}[htbp]
\begin{tabular}{|c||c|c|c|c|}
\hline
 $\vartheta$ & $\vartheta_0$ & $0$ & $\theta$  &
                    $\vartheta_0+\frac{\pi((1/R)^{\td}+R^{\td})}{4\td}$ \\
\hline
 $r$          & $0|\infty$    & $1$       & finite    & $0|\infty$       \\
\hline
\hline
 $\dot{S}e^{-\Btheta}$  & $0$           & finite    & $\infty$  & $0$    \\
\hline
 $\delta t$         & $\infty$      & finite    & finite    & $\infty$  \\
\hline
 $\delta \tau$      & $\infty$      & finite    & finite    & $\infty$  \\
\hline
\hline
 $e^A$          & finite        & finite    & $\infty$  & finite  \\
\hline
 $e^\Btheta$        & $\infty$      & finite    & $0$   & $\infty$  \\
\hline
 $e^B$          & finite        & finite    & $0$   & finite  \\
\hline
\end{tabular}
\caption{Variant IVC.}
\label{v_IVC}
\end{table}

In this variant the functions $\omega$ and $\rho$ are periodic and
a interval of $\vartheta$ closed between two adjacent zeros of
$\rho$ corresponds to $r\in(0,\infty)$. At the both ends given by
$\vartheta=\vartheta_0$ ($r=\infty$) and
$\vartheta=\vartheta_0+\frac{\pi((1/R)^{\td}+R^{\td})}{4\td}$
($r=0$) the spacetime tends to flat Minkowski spacetime. But
because the period of $\omega$ is shorter than the period of
$\rho$ there is always at least one naked singularity between the
infinities. Localization of the singularities is given by
$\vartheta=\theta + \frac{\pi}{\sqrt{\kappa}}$.

\vskip 2em
{\bf Variant IVC/B} where $R\in (0,\infty)$, $s_\chi=-1$, $\kappa>0$ and
 $\theta = -\frac{(1/R)^{\td}+R^{\td}}{2\td} \arctan R^\td$.

\begin{table}[htbp]
\begin{tabular}{|c||c|c|c|c|}
\hline
 $\vartheta$   & $\theta=\vartheta_0$  & $0$       &
            $\theta+\frac{\pi}{\sqrt{\kappa}}$
                 & $\vartheta_0+\frac{\pi((1/R)^{\td}+R^{\td})}{4\td}$ \\
\hline
 $r$      & $0|\infty$        & $1$       & finite        & $0|\infty$       \\
\hline
\hline
 $\dot{S}e^{-\Btheta}$  & finite     & finite    & $\infty$      & $0$    \\
\hline
 $\delta t$         & finite       & finite    & finite        & $\infty$  \\
\hline
 $\delta \tau$      & $\infty$      & finite    & finite        & $\infty$  \\
\hline
\hline
 $e^A$          & $\infty$          & finite    & $\infty$      & finite  \\
\hline
 $e^\Btheta$        & $\infty$          & finite    & $0$       & $\infty$  \\
\hline
 $e^B$          & $0$               & finite    & $0$       & finite  \\
\hline
\end{tabular}
\caption{Variant IVC/B.}
\label{v_IVCB}
\end{table}
This variant is a variation of the previous one and describes a
situation when $\theta$ coincides with $\vartheta_0$. Then at the
point $\vartheta = \theta = \vartheta_0$ curvature of the
spacetime is finite an nonzero due to presence of the brane
attached at this point. But there is necessarily another brane at
$\vartheta=\theta + \frac{\pi}{\sqrt{\kappa}}$ which produces a
naked singularity.

\vskip 1em
Summarizing, we see that the parameter $R$ gives a
position of a horizon expressed with the radial isotropic
coordinate $r$. Similarly a point described in $\vartheta$
coordinate as $\vartheta=\theta$ can be interpreted as a
localization of the brane. In the variants of type A and type C
the points $\vartheta=\theta$ are singular.

%%%%%%%%%%%%%%%%%%%%%%%%%%%%%%%%%%%%%%%%%%%%%%%%%%%%%%%%%%%%%%%%%
\subsubsection{Supersymmetry.}
\label{susy_nd}

\akapit When $D=11$ and $d=3$ the variants reviewed in the
previous paragraph can be naturally interpreted as describing the
electric $2$-brane in the eleven dimensional supergravity. This
gives us a possibility to test supersymmetric properties of the
variants directly i.e. by examining the formulae
(\ref{susy_el_psi1}) and (\ref{susy_el_psi2}). From those we have
the following rules:

\begin{itemize}
\item $\delta_\eta \Psi_\mu = 0$ only if:
 \be
  e^{N_+}\frac{d}{dr}\left(e^{3A}+e^C\right)\eta_{0+} +
  e^{N_-}\frac{d}{dr}\left(e^{3A}-e^C\right)\eta_{0-} = 0,
  \label{susy_nd_1}
 \ee
\item $\delta_\eta \Psi_m = 0$ only if:
 \bea
  \left(N'_+ + \frac{1}{6} e^{C-3A} C' \right) e^{N_+}\eta_{0+} +
  \left(N'_- - \frac{1}{6} e^{C-3A} C' \right) e^{N_-}\eta_{0-}
  &=& 0 , \label{susy_nd_2} \\
  \left(B' - \frac{1}{6} e^{C-3A} C' \right) e^{N_+}\eta_{0+} +
  \left(B' + \frac{1}{6} e^{C-3A} C' \right) e^{N_-}\eta_{0-}
  &=& 0, \label{susy_nd_3}
 \eea
\end{itemize}
where we decomposed the parameter $\eta$ with respect to eigenstates
of the operator $\Gamma_V$ (\ref{gamma_v}):
 \be
  \eta=e^{N_+}\eta_{0+} + e^{N_-} \eta_{0-} ,
 \ee
 where
 \be
  \Gamma_V \eta_{0\pm} = \pm \sigma \eta_{0\pm}
 \ee
and $\eta_{0\pm}$ are constant spinors and $N_\pm$ are functions
dependent on the radial coordinate.

We can always choose such $N_+$ and $N_-$ which satisfy
(\ref{susy_nd_2}) identically for arbitrary $\eta_{0+}$ and
$\eta_{0-}$. In the group of supersymmetry parameters $\eta$ these
$N_+$ and $N_-$ define a subgroup which is possibly preserved by
the solution. So the preserved supersymmetry is necessary rigid
with respect to the coordinate $r$. Examining the other conditions
we verify if the supersymmetry is really preserved.

For the variant I we find that:
\bea
 \delta_\eta \Psi_\mu &=& \frac{1}{3} e^{N_+}
 \Gamma_\mu \Gamma^m \frac{y_m}{r}
 \left( e^{3A} \right) ' \eta_{0+},
 \label{susy_nd_1_I}\\
 \delta_\eta \Psi_m &=& \frac{1}{4} \Gamma_m{}^n \frac{y_n}{r}
 \left( e^{-\frac{1}{2}A} \right) ' \eta_{0+}
 \label{susy_nd_2_I},
\eea
so this part of supersymmetry which is described by $\eta_{0-}$ is
always preserved. Similarly for the variant III we have:
\bea
 \delta_\eta \Psi_\mu &=& \frac{1}{3} e^{N_+}
 \Gamma_\mu \Gamma^m \frac{y_m}{r}
 \left( e^{3A} \right) ' \eta_{0+},
 \label{susy_nd_1_III} \\
 \delta_\eta \Psi_m &=& \frac{1}{4} \Gamma_m{}^n \frac{y_n}{r}
 \left( e^{+\frac{1}{2}A} \right) ' \eta_{0-},
 \label{susy_nd_2_III}
\eea
The first condition is the same as for the variant I, but the second
is changed replacing $\eta_{0+}$ with $\eta_{0-}$. Consequently
no supersymmetry can be preserved.

It is a very important fact that the supersymmetry transformations
of the gravitino components with the vector index corresponding to
the directions tangent to the brane (\ref{susy_nd_1_III}) break
other part of supersymmetry than the components which have the
index related to the transversal directions (\ref{susy_nd_2_III}).
So, considering a model dimensionally reduced only to the subspace
parallel to the brane or only to the subspace orthogonal to the
brane we can see only a term breaking half of supersymmetry. This
explains why for the nonsupersymmetric variant III we have the
same relation between the charge and the energy density as for the
supersymmetric variant I. Calculating the densities we conduct
only an integration over the transversal directions, so we in fact
work only with the subspace orthogonal to the brane. But in the
subspace the $\eta_{+0}$ part of supersymmetry is still preserved
and the BPS inequality has to be saturated.

For variants II and IV analogs of the both conditions
(\ref{susy_nd_1_I}--\ref{susy_nd_2_I}) possess terms proportional
to $\eta_{0+}$ and $\eta_{0-}$, so supersymmetry is broken in both
tangent and transversal subspaces separately.

%%%%%%%%%%%%%%%%%%%%%%%%%%%%%%%%%%%%%%%%%%%%%%%%%%%%%%%%%%%%%%%%%
\subsubsection{Solution without dilaton for $\td = -1$.}

\akapit It was already noted that the case of $\td=-1$ is specific
because then all the invariants (\ref{Rt_inv} -- \ref{Gt_inv}) of
the metric tensor are constant finite and nonzero numbers.
However, we shoul note that the conditions defining the considered
case are rather unphysical, because there is no known supergravity
theory without any dilaton but with an antisymmetric field
supporting a $(D-2)$-brane. So we can treat the case considered
here only as a toy model helping to understand properties of some
more realistic brane configurations with $\td=-1$. See for example
the one studied in the paragraph \ref{3m_sol_11sugra}.

As was already mentioned because of the assumptions made we should
not expect any naked singularity in the solution. See also that
(\ref{s_nd_A_1}) and (\ref{s_nd_B_1}) take forms:
 \bea
  e^A &=& \omega^{-\frac{1}{d}} \label{s_nd_A_-1_1}, \\
  e^\Btheta &=& 2 e^{-\epsilon_\chi} \omega^{-1},
  \label{s_nd_B_-1_1}
 \eea
so the solution expressed in terms of $\vartheta$ does not have
any dependence on the function $\rho$. Consequently
(\ref{tau_massless_2} -- \ref{t_massless_2}) can be rewritten as:
 \bea
  \delta \tau &=& \int_{\vartheta_1}^{\vartheta_0} \frac{1}{E}
           \omega^{-1-\frac{1}{d}} d\vartheta,
           \label{tau_massless_3} \\
  \delta t    &=& \int_{\vartheta_1}^{\vartheta_0}
           \omega^{\frac{1}{d}-1} d\vartheta.
           \label{t_massless_3}
 \eea
Moreover if $\td=-1$ then not only the constant $\Lambda_c$
vanishes (as it is for $\td>0$) but we have also $\Lambda_\chi=0$
and the identity holds for all values of $R$. It suggests that for
the solution we have $\calE = |V| \frac{2}{\sqrt{\Delta}}
\lambda$. But further analysis shows that it is not true, because
to calculate properly the energy density we should add an external
contribution originating from the brane in each variant.

It can be proved that for the variants I, II and III there is a
coordinate system singularity and an event horizon at
$\vartheta=\pm \infty$ where the functions $e^A$ and $e^\Btheta$
vanish. It is possible to identify the points $\vartheta=-\infty$
and $\vartheta=+\infty$, so the areas of large positive and
negative values of $\vartheta$ are related to spacetime at
opposite sides of the horizon. A point where $\vartheta = \theta$
can be reached from the left and the right side by a massless
particle in infinite affine parameter but a finite amount of the
coordinate time, so it describes boundaries of the spacetime where
domain walls -- the branes -- are localized. The variant IV is
different, because then we have no horizon and a series of
"infinite" points $\vartheta = \theta+ n
\frac{\pi}{\sqrt{\kappa}}$. So any interval $\vartheta \in
(\theta+\frac{n\pi}{\sqrt{\kappa}},\theta+
\frac{(n+1)\pi}{\sqrt{\kappa}})$ describes a geodesically complete
universe between two domain walls.

%%%%%%%%%%%%%%%%%%%%%%%%%%%%%%%%%%%%%%%%%%%%%%%%%%%%%%%%%%%%%%%%%
\subsubsection{Solution with dilaton for $\td \geq 1$.}
\label{sol_dil}

\akapit Imposing in the diagonal $\Delta$ solution constraints
$N_A=1$ and $\td \geq 1$ we recover the single brane solution
found in \cite{ZZ1} and discussed in \cite{BMO}:
 \bea
  e^A &=& E_A \omega^{-\frac{2\td}{(D-2)\Delta}} e^{c_A\vartheta},
  \label{s_d_A_1} \\
  e^\Btheta &=& \left( \frac12 E_A^{-d} e^{\epsilon_\chi}\right)^{1/\td}
   \omega^{\frac{2d}{(D-2)\Delta}} e^{-\frac{d}{\td} c_A\vartheta}
   \rho^{-1-\frac{1}{\td}},
  \label{s_d_B_1} \\
  e^\phi &=& E_A^{2\varsigma d/a} \omega^{2a/\Delta}
  e^{-\frac{\varsigma 2d}{a} c_A\vartheta}, \\
  e^C &=& \left\{ \begin{array}{ll}
   E_C-\frac{4\sqrt{ \kappa}}{p\Delta}
   \cot (\sqrt{ \kappa}(\vartheta-\theta)), & \mbox{for   } \kappa>0, \\
    \\
   E_C-\frac{4}{p\Delta}
   (\vartheta-\theta)^{-1},  & \mbox{for   } \kappa=0, \\
    \\
   E_C-\frac{4\sqrt{-\kappa}}{p\Delta}
   \coth(\sqrt{-\kappa}(\vartheta-\theta)), & \mbox{for   } \kappa<0,
  \end{array} \right.
  \label{ec_sol_dil}
 \eea
 where:
 \bea
  \kappa &=& \frac{\Delta}{2} \left( \Lambda_\chi+\Lambda_c \right), \\
  \Lambda_c &=& \frac{c_A^2 d}{a^2 \td} (D-2) \Delta.
 \eea
It seems that the solution depends on seven parameters: $E_A,C_A,
\epsilon_\chi, E_C, \theta, R$ and $p$. But $E_C$ has no physical
meaning and describes gauge freedom of the antisymmetric
potential. Other three can be determined if we assume that at a
given point $\vartheta=\vartheta_0$ the functions $A=A_{I=\{1\}}$,
$B$ and $\phi=\phi_{i=1}$ have a certain value. It is convenient
to choose $\vartheta_0$ at a point which is reachable only in
infinite affine parameter and where the spacetime is flat and
demand:
 \bea
  \chi(\vartheta_0) &=& 0, \label{constr_21} \\
  \omega(\vartheta_0) &=& 1, \label{constr_22} \\
  E_A &=& e^{-c_A \vartheta_0}. \label{constr_23}
 \eea
Then $e^A$, $e^B$ and $e^\phi$ are normalized to one at this
point. The constraints (\ref{constr_21} -- \ref{constr_23}) allow
then to express $e_\chi$, $\theta$ and $E_A$ by the remaining
parameters. The remaining three degrees of freedom should then be
related to physical quantities: charge, mass and the tachyon
condensate \cite{BMO}.

Calculating energy density of the brane, we obtain:
 \be
  \calE(A) = \Omega_{\td+1} \left|V \right| \frac{2}{\sqrt{\Delta}}
   \sqrt{\lambda^2-\frac{1}{2} \left( \Lambda_\chi + \Lambda_c \right) }.
 \ee
But analogously as it was discussed in the paragraphs
\ref{sol_nd_gr} and \ref{sol_nd_vr} it is valid only when certain
conditions are fulfilled assuring that the integration has a well
defined physical meaning. Comparing to (\ref{energ_nd}) we see
that an additional term $-\frac{1}{2}\Lambda_c$ appeared. The term
contributes with a nonpositive value (if $\td\geq 0$) and depends
only on the parameter $c_A$. This is the reason why we can
identify $c_A$ with the tachyon condensate.

Note that nonzero $c_A$ and thus $\Lambda_c$ is not a direct
consequence of a presence of the dilaton in the considered model.
See that a total number of constants $c_I, c_\alpha$ and $c_B$ is
equal to $N_g+N_\phi$ and there are $N_A+1$ constraints imposed on
them (\ref{sol_const_1}) and (\ref{sol_const_3}). So we should
expect similar effect in any situation when a total number of
functions $A_I$ and $\phi_\alpha$ in the model exceeds the number
of branes $N_A$.

%%%%%%%%%%%%%%%%%%%%%%%%%%%%%%%%%%%%%%%%%%%%%%%%%%%%%%%%%%%%%%%%%
\subsubsection{$r\rightarrow 1/r$ duality.}

\akapit
Analyzing the formula (\ref{th_1}) giving the definition of the
coordinate $\vartheta$ we can see that it exhibits the duality:
 \be
  \vartheta(r;R,s_\chi)=-\vartheta(1/r;1/R,s_\chi),
  \label{sym_1}
 \ee
what establishes a relation between $\vartheta$ defined for different
values of the parameter $R$. The same transformation applied to the
function $\rho$ (\ref{rho_1}) leads to:
 \be
  \rho(\vartheta;R,s_\chi)=\rho(-\vartheta;1/R,s_\chi)
  \label{sym_2}.
 \ee
Further, having a particular solution, for example
(\ref{om_sol_1}) one can extend the duality to all parameters
appearing in the solution:
 \be
  \left( \begin{array}{c}
   \vartheta; \theta_i, \kappa_i \\
   R, s_\chi, \epsilon_\chi \\
   p_i, \sigma_i \\
   c_I, c_B, c_\alpha \\
   E_I, E_B, E_\alpha
  \end{array} \right)
  \rightarrow
  \left( \begin{array}{c}
   -\vartheta; -\theta_i, \kappa_i \\
   1/R, s_\chi, \epsilon_\chi \\
   -p_i, -\sigma_i \\
   -c_I, -c_B, -c_\alpha \\
   E_I, E_B, E_\alpha
  \end{array} \right)
  \label{sym_3}
 \ee
and check that the transformation is a symmetry of the fields
$e^{A_I}, e^{B_\vartheta}, e^{\phi_\alpha}$ and $e^{C_i}$
(\ref{sol_11}--\ref{sol_13}, \ref{ec_sol_1}). The function $e^B$ is
not invariant under (\ref{sym_3}), but $e^B dr$ is.

One can be afraid that the duality is only a mathematical trick
and both the solutions tied by the duality are physically
equivalent up to reversing sing of the brane charge. However, it
can be checked that it is not true. To see that it is enough to
look at the function $\chi$. Under (\ref{sym_3}) one has:
 \be
 \chi(1/r;1/R,s_\chi,\epsilon_\chi)
  = \chi(r;R,s_\chi,\epsilon_\chi) + 2\td\ln r.
 \ee
This gives in particular that the supersymmetric solution given by
$R=0$ (variant I) has a non-supersymmetric partner which can be
described by $R=\infty$ (variant III).

{}From (\ref{sym_3}) it follows that while the metric tensor and
the dilaton are unchanged under the duality, the antisymmetric
tensor changes its sign. However our knowledge about the
antisymmetric tensor is limited to only those components of the
field which spun the brane. Even less we know about fermions.
Because the considered model is restricted only to the bosonic
truncation of some supergravity theory it is difficult to guess
how the spinor fields appearing in the original theory behave
under the transformation. But we can collect some information by
comparing properties of the supersymmetry transformations
(\ref{susy_nd_1_I}) and (\ref{susy_nd_2_I}) with
(\ref{susy_nd_1_III}) and (\ref{susy_nd_2_III}). This suggests
that the duality (\ref{sym_3}) has a different effect when acting
on the components $\Psi_\mu$ and $\Psi_m$.

%%%%%%%%%%%%%%%%%%%%%%%%%%%%%%%%%%%%%%%%%%%%%%%%%%%%%%%%%%%%%%%%%
%%%%%%%%%%%%%%%%%%%%%%%%%%%%%%%%%%%%%%%%%%%%%%%%%%%%%%%%%%%%%%%%%
\subsection{Composite branes.}
\label{compo_brane}

\akapit Let us assume now that all except one condition imposed to
construction of the model discussed in the paragraph \ref{model}
are still valid. Therefore we allow branes to be composite now. We
should recall that validity of the intersecting branes solution
derived in the previous model (where all branes are supported by
different fields) can be extended to the case of composite branes
if the nondiagonal elements of the stress-energy tensor
$T(A)_{MN}$ (\ref{1_s_en_2}) for each antisymmetric field vanish
i.e. identity (\ref{s_en_diag}) is satisfied. With results
developed in \cite{IM1} we see that the tensor $T(A)_{MN}$ can
(but does not have to) be nondiagonal only if at least one of the
following situations occur:
\begin{enumerate}

\item There are two electric branes with worldvolumes $V_1$ and $V_2$ such that:
 \be
  \dim \left( V_1 \cap V_2 \right) = \dim V_1 - 1. \label{compo_1}
 \ee
 Of course $\dim V_1 = \dim V_2$.

\item There are two magnetic branes with worldvolumes $V_1$ and $V_2$ such that:
 \be
  \dim \left( V_1 \cap V_2 \right) = \dim V_1 - 1. \label{compo_2}
 \ee
 In this case also $\dim V_1 = \dim V_2$.

\item There is an electric brane with worldvolume $V_e$ and a magnetic
      with $V_m$ such that:
 \be
  \dim \left( V_e \cap V_m \right) = 0 \quad \mbox{and} \quad
  \td=-1. \label{compo_3}
 \ee

\item There is an electric brane with worldvolume $V_e$ and a magnetic
       with $V_m$ such that:
 \be
  \dim \left( V_e \cap V_m \right) = 0 \quad \mbox{and} \quad
  \td=0. \label{compo_4}
 \ee

\item There is an electric brane with worldvolume $V_e$ and a magnetic
       with $V_m$ such that:
 \be
  \dim \left( V_e \cap V_m \right) = 1 \quad \mbox{and} \quad
  \td=0. \label{compo_5}
 \ee

\item There is an electric brane with worldvolume $V_e$ and a magnetic
        with $V_m$ such that:
 \be
  \dim \left( V_e \cap V_m \right) = 1 \quad \mbox{and} \quad
  \td=1. \label{compo_6}
 \ee
\end{enumerate}

%%%%%%%%%%%%%%%%%%%%%%%%%%%%%%%%%%%%%%%%%%%%%%%%%%%%%%%%%%%%%%%%%
\subsubsection{Intersecting composite branes of $D=11$ supergravity.}

\akapit Let us focus on the eleven dimensional supergravity theory
now. The theory contains only one antisymmetric field, so if one
wants to consider commonly orthogonally intersecting branes in the
theory they have to be composite. The previously developed model
can be employed to give a description of the configuration only if
the branes do not fall into classes characterized by
(\ref{compo_1} -- \ref{compo_6}) and satisfy all other conditions
presented in paragraph \ref{model}. But even that it is not
sufficient yet.

The $D=11$ supergravity possesses the Chern-Simons term $F\wedge F
\wedge A$ (\ref{D11sugra}) which was neglected in our model, so to
be precise instead of $a=0$ version of the equation
(\ref{eqm_12}):
 \be
  \nabla_M F^{MN_1N_2N_3} = 0,
 \ee
we should discuss in the theory:
 \be
  \nabla_M F^{MN_1N_2N_3}+ \frac{1}{2(4!)^2}
  \epsilon^{N_1 \ldots N_{11}} F_{N_4\ldots N_7}F_{N_8\ldots N_{11}}=0.
 \ee
Fortunately it can be checked that if the branes are commonly
orthogonally intersecting, delocalized (i.e. with the solution
depending only on a radial coordinate in the overall transverse
space) and with timelike worldvolumes, a set of configurations
with nonzero contribution from the Chern-Simons term is very
limited. We can describe the configurations as containing at least
one electric brane and at least two magnetic branes such that:
 \be
  V_e = V_{m,1} \cap V_{m,2} \quad \mbox{and} \quad \td=0,
 \ee
where $V_e$ denotes the electric brane worldvolume and $V_{m,1}$,
$V_{m,2}$ worldvolumes of the magnetic branes.

\begin{table}[htbp]
 \begin{tabular}{|r||c|c||c|c|c|c||c|c|}
  \hline
  no. & $D_1$ & $D_2$ & $d_{\{1,2\}}$ & $d_{\{1\}}$ &
  $d_{\{2\}}$ & $\dim V_{\emptyset}$ & $\td$ & diag T \\
  \hline
  \hline
  \multicolumn{9}{|c|}{two magnetic $5$-branes}  \\
  \hline
  1. & 6 & 6 & 5 & 1 & 1 & 4 & 2 & 2. \\
  2. & 6 & 6 & 4 & 2 & 2 & 3 & 1 & y  \\
  3. & 6 & 6 & 3 & 3 & 3 & 2 & 0 & y  \\
  4. & 6 & 6 & 2 & 4 & 4 & 1 &-1 & y  \\
  \hline
  \multicolumn{9}{|c|}{a magnetic $5$-brane and an electric $2$-brane}  \\
  \hline
  5. & 6 & 3 & 3 & 3 & 0 & 5 & 3 & y  \\
  6. & 6 & 3 & 2 & 4 & 1 & 4 & 2 & y  \\
  7. & 6 & 3 & 1 & 5 & 2 & 3 & 1 & 6. \\
  \hline
  \multicolumn{9}{|c|}{two electric $2$-branes}  \\
  \hline
  8. & 3 & 3 & 2 & 1 & 1 & 7 & 5 & 1. \\
  9. & 3 & 3 & 1 & 2 & 2 & 6 & 4 & y  \\
  \hline
 \end{tabular}
 \caption{Two intersecting branes.}
 \label{2_intersect_branes}
\end{table}

Let us make a brief scan of possible two-brane and three-brane
intersections in the eleven dimensional supergravity. There are
nine inequivalent two-brane configurations which fall into three
categories:
\begin{itemize}
\item two electric branes, there are two configurations of this kind,
     one with the tensor $T(A)_{MN}$ diagonal,
\item electric and magnetic brane -- three cases, two with diagonal $T(A)_{MN}$.
\item two magnetic branes -- four configurations including three
       described by diagonal $T(A)_{MN}$.
\end{itemize}
All of them are shown in the table \ref{2_intersect_branes}. The
letter "y" in the last column of the table tells that the tensor
$T(A)_{MN}$ is diagonal. And the numbers in the column indicate
which of the reason given in at the beginning of the section makes
the tensor possibly nondiagonal. The same convention is used in
the tables \ref{3m_intersect_branes}, \ref{2me_intersect_branes},
\ref{m2e_intersect_branes} and \ref{3e_intersect_branes}.

\begin{table}[htbp]
 \begin{tabular}{|r||c|c|c|c|c|c|c|c||c|c|}
  \hline
  no. & $d_{\{1,2,3\}}$ &
  $d_{\{1,2\}}$ & $d_{\{2,3\}}$ & $d_{\{1,3\}}$ &
  $d_{\{1\}}$ & $d_{\{2\}}$ & $d_{\{3\}}$ &
  $\dim V_{\emptyset}$ & $\td$ & diag T \\
  \hline
  1. & 5 & 0 & 0 & 0 & 1 & 1 & 1 & 3 & 1 & 2. \\
  \hline
  2. & 4 & 1 & 1 & 1 & 0 & 0 & 0 & 4 & 2 & 2. \\
  3. & 4 & 1 & 1 & 0 & 1 & 0 & 1 & 3 & 1 & 2. \\
  4. & 4 & 1 & 0 & 0 & 1 & 1 & 2 & 2 & 0 & 2. \\
  5. & 4 & 0 & 0 & 0 & 2 & 2 & 2 & 1 & -1& y  \\
  \hline
  6. & 3 & 2 & 1 & 1 & 0 & 0 & 1 & 3 & 1 & 2. \\
  7. & 3 & 2 & 1 & 0 & 1 & 0 & 2 & 2 & 0 & 2. \\
  8. & 3 & 1 & 1 & 1 & 1 & 1 & 1 & 2 & 0 & y  \\
  9. & 3 & 2 & 0 & 0 & 1 & 1 & 3 & 1 & -1& 2. \\
 10. & 3 & 1 & 1 & 0 & 2 & 1 & 2 & 1 & -1& y  \\
  \hline
 11. & 2 & 2 & 2 & 2 & 0 & 0 & 0 & 3 & 1 & y  \\
 12. & 2 & 3 & 1 & 1 & 0 & 0 & 2 & 2 & 0 & 2. \\
 13. & 2 & 2 & 2 & 1 & 1 & 0 & 1 & 2 & 0 & y  \\
 14. & 2 & 3 & 1 & 0 & 1 & 0 & 3 & 1 & -1& 2. \\
 15. & 2 & 2 & 2 & 0 & 2 & 0 & 2 & 1 & -1& y  \\
 16. & 2 & 2 & 1 & 1 & 1 & 1 & 2 & 1 & -1& y  \\
  \hline
 17. & 1 & 3 & 2 & 2 & 0 & 0 & 1 & 2 & 0 & y  \\
 18. & 1 & 4 & 1 & 1 & 0 & 0 & 3 & 1 & -1& 2. \\
 19. & 1 & 3 & 2 & 1 & 1 & 0 & 2 & 1 & -1& y  \\
 20. & 1 & 2 & 2 & 2 & 1 & 1 & 1 & 1 & -1& y  \\
  \hline
 \end{tabular}
 \caption{Three intersecting magnetic $5$-branes.
  $D_1=D_2=D_3=6$.}
 \label{3m_intersect_branes}
\end{table}

The table \ref{3m_intersect_branes} gives twenty inequivalent
configurations which can be constructed of three magnetic
$5$-branes. Ten of the configurations are described by diagonal
$T(A)_{MN}$, so the method developed in the previous sections can
by applied.

\begin{table}[htbp]
 \begin{tabular}{|r||c|c|c|c|c|c|c|c||c|c|}
  \hline
  no. & $d_{\{1,2,3\}}$ &
  $d_{\{1,2\}}$ & $d_{\{2,3\}}$ & $d_{\{1,3\}}$ &
  $d_{\{1\}}$ & $d_{\{2\}}$ & $d_{\{3\}}$ &
  $\dim V_{\emptyset}$ & $\td$ & diag T \\
  \hline
  1. & 3 & 2 & 0 & 0 & 1 & 1 & 0 & 4 & 2 & 2. \\
  2. & 3 & 1 & 0 & 0 & 2 & 2 & 0 & 3 & 1 & y \\
  3. & 3 & 0 & 0 & 0 & 3 & 3 & 0 & 2 & 0 & y \\
  \hline
  4. & 2 & 3 & 1 & 0 & 1 & 0 & 0 & 4 & 2 & 2. \\
  5. & 2 & 3 & 0 & 0 & 1 & 1 & 1 & 3 & 1 & 2. \\
  6. & 2 & 2 & 1 & 0 & 2 & 1 & 0 & 3 & 1 & y \\
  7. & 2 & 2 & 0 & 0 & 2 & 2 & 1 & 2 & 0 & y \\
  8. & 2 & 1 & 1 & 0 & 3 & 2 & 0 & 2 & 0 & y \\
  9. & 2 & 1 & 0 & 0 & 3 & 3 & 1 & 1 & -1& y \\
 10. & 2 & 0 & 1 & 0 & 4 & 3 & 0 & 1 & -1& y \\
  \hline
 11. & 1 & 4 & 1 & 1 & 0 & 0 & 0 & 4 & 2 & 2. \\
 12. & 1 & 4 & 1 & 0 & 1 & 0 & 1 & 3 & 1 & 2. 6. \\
 13. & 1 & 3 & 2 & 0 & 2 & 0 & 0 & 3 & 1 & 6. \\
 14. & 1 & 3 & 1 & 1 & 1 & 1 & 0 & 3 & 1 & y \\
 15. & 1 & 4 & 0 & 0 & 1 & 1 & 2 & 2 & 0 & 2. 5.\\
 16. & 1 & 3 & 1 & 0 & 2 & 1 & 1 & 2 & 0 & 5. \\
 17. & 1 & 2 & 2 & 0 & 3 & 1 & 0 & 2 & 0 & 5. \\
 18. & 1 & 2 & 1 & 1 & 2 & 2 & 0 & 2 & 0 & y \\
 19. & 1 & 3 & 0 & 0 & 2 & 2 & 2 & 1 & -1& y \\
 20. & 1 & 2 & 1 & 0 & 3 & 2 & 1 & 1 & -1& y \\
 21. & 1 & 1 & 2 & 0 & 4 & 2 & 0 & 1 & -1& y \\
 22. & 1 & 1 & 1 & 1 & 3 & 3 & 0 & 1 & -1& y \\
  \hline
 \end{tabular}
\caption{Intersecting two magnetic $5$-branes and an electric
$2$-brane. $D_1=D_2=6$, $D_3=3$.}
 \label{2me_intersect_branes}
\end{table}

Similarly in the table \ref{2me_intersect_branes} we have listed
all 22 examples of brane configurations containing two magnetic
and one electric brane. Again for ten of them the tensor
$T(A)_{MN}$ is diagonal, but in the case number 3. we encounter
nonvanishing Chern-Simons term so it has to excluded from the set
of configurations relevant for the model discussed before.

\begin{table}[htbp]
 \begin{tabular}{|r||c|c|c|c|c|c|c|c||c|c|}
  \hline
  no. & $d_{\{1,2,3\}}$ &
  $d_{\{1,2\}}$ & $d_{\{2,3\}}$ & $d_{\{1,3\}}$ &
  $d_{\{1\}}$ & $d_{\{2\}}$ & $d_{\{3\}}$ &
  $\dim V_{\emptyset}$ & $\td$ & diag T \\
  \hline
  1. & 2 & 1 & 0 & 1 & 2 & 0 & 0 & 5 & 3 & 1. \\
  2. & 2 & 1 & 0 & 0 & 3 & 0 & 1 & 4 & 2 & 1. \\
  3. & 2 & 0 & 0 & 0 & 4 & 1 & 1 & 3 & 1 & 1. \\
  \hline
  4. & 1 & 2 & 0 & 2 & 1 & 0 & 0 & 5 & 3 & y \\
  5. & 1 & 2 & 0 & 1 & 2 & 0 & 1 & 4 & 2 & y \\
  6. & 1 & 1 & 1 & 1 & 3 & 0 & 0 & 4 & 2 & 1. \\
  7. & 1 & 2 & 0 & 0 & 3 & 0 & 2 & 3 & 1 & 6. \\
  8. & 1 & 1 & 1 & 0 & 4 & 0 & 1 & 3 & 1 & 1. 6. \\
  9. & 1 & 1 & 0 & 1 & 3 & 1 & 1 & 3 & 1 & y \\
 10. & 1 & 1 & 0 & 0 & 4 & 1 & 2 & 2 & 0 & y \\
 11. & 1 & 0 & 1 & 0 & 5 & 1 & 1 & 2 & 0 & 1. 5. \\
 12. & 1 & 0 & 0 & 0 & 5 & 2 & 2 & 1 & -1& y \\
  \hline
 \end{tabular}
\caption{Intersecting a magnetic $5$-brane and two electric
$2$-branes. $D_1=6$, $D_2=D_3=3$.}
 \label{m2e_intersect_branes}
\end{table}

The table \ref{m2e_intersect_branes} shows 12 configurations with
two electric and one magnetic brane, five of them are
characterized by diagonal stress-energy tensor. And in the table
\ref{3e_intersect_branes} there are contained five three-brane
configurations made of electric branes exclusively, but only one
has diagonal $T(A)_{MN}$.

\begin{table}[htbp]
 \begin{tabular}{|r||c|c|c|c|c|c|c|c||c|c|}
  \hline
  no. & $d_{\{1,2,3\}}$ &
  $d_{\{1,2\}}$ & $d_{\{2,3\}}$ & $d_{\{1,3\}}$ &
  $d_{\{1\}}$ & $d_{\{2\}}$ & $d_{\{3\}}$ &
  $\dim V_{\emptyset}$ & $\td$ & diag T \\
  \hline
  1. & 2 & 0 & 0 & 0 & 1 & 1 & 1 & 6 & 4 & 1. \\
  \hline
  2. & 1 & 1 & 1 & 1 & 0 & 0 & 0 & 7 & 5 & 1. \\
  3. & 1 & 1 & 1 & 0 & 1 & 0 & 1 & 6 & 4 & 1. \\
  4. & 1 & 1 & 0 & 0 & 1 & 1 & 2 & 5 & 3 & 1. \\
  5. & 1 & 0 & 0 & 0 & 2 & 2 & 2 & 4 & 2 & y  \\
  \hline
 \end{tabular}
 \caption{Three intersecting electric $2$-branes. $D_1=D_2=D_3=3$.}
 \label{3e_intersect_branes}
\end{table}

Note, that for the $D=11$ supergravity the formula
(\ref{delta_m_1}) for the components of the matrix $\Delta$ can be
written in a simplified form:
 \be
 \Delta_{ij} = 4 - 2 \left( \min(D_i,D_j) - D_{ij} \right),
 \ee
where
 \be
 D_{ij} = \dim (V_i \cap V_j) = \sum_{I:i,j\in I} d_I.
 \ee

%%%%%%%%%%%%%%%%%%%%%%%%%%%%%%%%%%%%%%%%%%%%%%%%%%%%%%%%%%%%%%%%%
\subsubsection{Three magnetic brane solution in $D=11$ supergravity.}
\label{3m_sol_11sugra}

\akapit One of the brane intersections in the eleven dimensional
supergravity for which we can formulate and next solve the Toda-like
equation is the fifth case in the table \ref{3m_intersect_branes}.
The configuration consists of three magnetic branes having the
four dimensional common intersection and the one dimensional
overall transverse space. The case is very interesting because it
can serve as a simplified model of the most promising
compactification scheme where the $D=11$ spacetime decomposes like:
\be
 \calM_{11} \rightarrow \calM_4 \times \calK_6 \times \calI,
 \label{M_decomp}
\ee
where $\calK$ is a Calabi-Yau manifold and $\calI$ is a real interval.

In our model we have $\calM_4 = V_{\{1,2,3\}}$, $\calK_6 =
V_{\{1\}} \times V_{\{2\}} \times V_{\{3\}}$ and $\calI =
V_\emptyset$ what means that $d_{\{1,2,3\}}=4$, $d_{\{i\}}=2$ for
$i=1,2,3$ and $\td=-1$. This gives a diagonal form for the matrix
$\Delta$:
 \be
  \Delta = \left( \begin{array}{ccc}
   4 & 0 & 0 \\
   0 & 4 & 0 \\
   0 & 0 & 4
  \end{array} \right).
 \ee
Finally the appropriate solution can be written as:
 \bea
  e^A &=& E \omega_1^{-1/6} \omega_2^{-1/6}
  \omega_3^{-1/6} e^{c\vartheta}, \\
  e^{A_{\{1\}}} &=& E^{-2} \omega_1^{-1/6}
  \omega_2^{1/3} \omega_3^{1/3} e^{-2c\vartheta}, \\
  e^{A_{\{2\}}} &=& E^{-2} \omega_1^{1/3}
  \omega_2^{-1/6} \omega_3^{1/3} e^{-2c\vartheta}, \\
  e^{A_{\{3\}}} &=& E^{-2} \omega_1^{1/3}
  \omega_2^{1/3} \omega_3^{-1/6} e^{-2c\vartheta}, \\
  e^{\Btheta} &=& 2 e^{\epsilon_\chi} E^{-8}
  \omega_1^{1/3} \omega_2^{1/3} \omega_3^{1/3}
   e^{-8c\vartheta},
 \eea
where for simplicity $A$ is written instead of $A_{\{1,2,3\}}$ and
where $\omega_i$'s are given by (\ref{om_sol_1}). Since
 \be
  \Lambda_\chi=0, \qquad \Lambda_c = -36c^2,
 \ee
therefore for the constants $\kappa_i$ we have:
 \be
  \kappa_1+\kappa_2+\kappa_3 = -72 c^2.
 \ee
The relation between the energy density and the charges densities
is (in those cases when it is well defined):
 \be
  \calE = |V| \sum_i \sqrt{\lambda_i^2 - \frac{1}{4}\kappa_i}.
 \ee

This relation suggests that a necessary condition for preserving
supersymmetry is $\kappa_i=0$ and $c=0$. To verify this it is
instructive to consider again when $\delta_\eta \Psi_M=0$. To
discuss the condition we need first to check how the Dirac
matrices decompose under (\ref{M_decomp}). We have:
 \be
  \Gamma_M \rightarrow \left(
   e^A \Gamma_\mu,
   e^{A_{\{j\}}} \Gamma_{\mu^{\{j\}}},
   e^B \Gamma_{11} \right), \qquad \mbox{where} \quad j=1,2,3,
 \ee
 and:
 \be
  \begin{array}{ccccccccc}
  \Gamma_\mu
   &=& \gamma_\mu &\otimes &\gamma_{V_1}
    &\otimes &\gamma_{V_2}        &\otimes & \gamma_{V_3}, \\
  \Gamma_{\mu^{\{1\}}}
   &=& Id         &\otimes &\gamma_{\mu^{\{1\}}}&\otimes &\gamma_{V_2}
          &\otimes & \gamma_{V_3}, \\
  \Gamma_{\mu^{\{2\}}}
   &=& Id         &\otimes & Id
                  &\otimes &\gamma_{\mu^{\{2\}}}&\otimes & \gamma_{V_3}, \\
  \Gamma_{\mu^{\{3\}}}
   &=& Id         &\otimes & Id
                 &\otimes & Id
                  &\otimes & \gamma_{\mu^{\{3\}}}, \\
  \Gamma_{11}
   &=& \gamma_V   &\otimes &\gamma_{V_1}
     &\otimes &\gamma_{V_2}        &\otimes & \gamma_{V_3}.
  \end{array}
 \ee
The matrices $\gamma_\mu$ are Dirac matrices and
$\gamma_V=-i\gamma_1 \gamma_2 \gamma_3 \gamma_4$ the chiral
operator in the four dimensional subspacetime $V_{\{1,2,3\}}$.
Similarly $\gamma_{\mu^{\{j\}}}$ and $\gamma_{V_j}$ for $j=1,2,3$
are respectively Dirac matrices and chiral operator on the two
dimensional spaces $V_{\{j\}}$. So we can define:
 \be
  \begin{array}{cccccccccc}
  \Gamma_{V_1}
   &=& i & Id &\otimes &\gamma_{V_1} &\otimes & Id          &\otimes & Id, \\
  \Gamma_{V_2}
   &=& i & Id &\otimes & Id          &\otimes &\gamma_{V_2} &\otimes & Id, \\
  \Gamma_{V_3}
   &=& i & Id &\otimes & Id          &\otimes & Id          &\otimes &
   \gamma_{V_3}, \\
 \end{array}
 \ee
 which have to satisfy:
 \be
  \Gamma_{V_j} = \frac{1}{2} \epsilon_{\mu^{\{j\}}\nu^{\{j\}}}
  \Gamma^{\mu^{\{j\}}\nu^{\{j\}}}.
 \ee

Finally from the supersymmetry preserving condition $\delta_\eta
\Psi_M = 0$ it follows:
 \bea
  \delta_\eta \Psi_\mu &=&  \Gamma_\mu \Gamma^{11} e^{A-B}
   \left( \frac{1}{2} A' - \frac{1}{12} \sum_i P_i \right) \eta,
   \label{susy_3m_psi1} \\
  \delta_\eta \Psi_{\mu^{\{k\}}} &=&  \Gamma_{\mu^{\{k\}}}
  \Gamma^{11} e^{A_{\{k\}}-B}
   \left( \frac{1}{2} A'_{\{k\}} + \frac{1}{6} \sum_j P_j -
   \frac{1}{4} P_k \right)
   \eta, \qquad \mbox{for} \quad k=1,2,3,
   \label{susy_3m_psi2} \\
  \delta_\eta \Psi_{11} &=& \left( N' - \frac{1}{12} \sum_j P_j \right)
   \eta, \label{susy_3m_psi3}
 \eea
where we defined projection operators:
 \bea
  P_1 &=& e^{B-2A_{\{2\}}-2A_{\{3\}}} \Gamma^{11} \Gamma^{V_2}
  \Gamma^{V_3}\lambda_1, \\
  P_2 &=& e^{B-2A_{\{3\}}-2A_{\{1\}}} \Gamma^{11} \Gamma^{V_3}
  \Gamma^{V_1}\lambda_2, \\
  P_3 &=& e^{B-2A_{\{1\}}-2A_{\{2\}}} \Gamma^{11} \Gamma^{V_1}
  \Gamma^{V_2}\lambda_3
 \eea
and for the parameter $\eta$ we introduced $\eta(r)=e^{N(r)}
\eta_0$ with $\eta_0$ -- a constant spinor.

{}From the equation (\ref{susy_3m_psi3}) we obtain:
 \be
  N' = 2 A'
 \ee
an analog of (\ref{susy_unb3_11e}). Next the equations
(\ref{susy_3m_psi1} -- \ref{susy_3m_psi2}) lead to:
 \be
  c=0, \qquad \kappa_j = 0,
 \ee
for $j=1,2,3$. However, instead of the harmonic gauge condition
$\chi'=0$ we get only:
 \bea
  6 A' + 2 \left( A_{\{1\}} + A_{\{2\}} + A_{\{3\}}\right)' &=& 0, \\
  -2 A' - B' &=& \chi',
 \eea
where $\chi$ preserves its general form. This feature is peculiar
for all models where the overall transverse space is one
dimensional. In such situation all operators $\Gamma_{mn}$ have to
vanish so we do not have any analog of the term $\Gamma_m{}^n
\left(\de_n B - \frac{1}{6} e^{C-3A}  \de_n C \Gamma_V \right)$
appearing in (\ref{susy_el_psi2}). Consequently it is not possible
from the requirement of supersymmetry to derive  any equation
involving $B'$ and restore the harmonic gauge condition.

%%%%%%%%%%%%%%%%%%%%%%%%%%%%%%%%%%%%%%%%%%%%%%%%%%%%%%%%%%%%%%%%%
%%%%%%%%%%%%%%%%%%%%%%%%%%%%%%%%%%%%%%%%%%%%%%%%%%%%%%%%%%%%%%%%%
%%%%%%%%%%%%%%%%%%%%%%%%%%%%%%%%%%%%%%%%%%%%%%%%%%%%%%%%%%%%%%%%%
\newpage
\section{Conclusions}

\akapit The main purpose of this paper was to provide an overview
of the way that leads to the notion of branes, to discuss the
known supersymmetric brane solutions and to describe new
nonsupersymmetric brane solutions. It is generally accepted that
the route from the Standard Model through supersymmetric
extensions, Grand Unified theories, supergravities to superstrings
and M theory is the most promising (albeit extremely difficult in
practice) attempt to unify Quantum Field Theory and the General
Relativity i.e. two main pillars of contemporary physics. It is a
relatively recent result that the theory of strings has much
richer spectrum than just perturbative string excitations. The new
objects were known for many years in supergravity under the name
of branes and (in analogy to instantons in nonperturbative quantum
field theory) were solutions of the classical equations of motion.
Although we do not have yet a quantum theory of branes we suspect
that the elusive 11-dimensional M theory is just a quantum theory
of 2-branes (as string theory is perturbatively a quantum theory
of 1-branes i.e. strings). Therefore it is important to find and
classify as wide class of brane solutions as possible. A special
role is played by the presence (or the absence) of supersymmetry.
All known string theories are supersymmetric but it does not
exclude a possibility that some solutions break (spontaneously)
supersymmetry in a similar way as gauge symmetry can be
spontaneously broken by the Higgs mechanism. Since supersymmetry
is certainly broken at low energy scales it is interesting too
search for brane solutions that are nonsupersymmetric to gain some
insight into possible ways of supersymmetry breaking in
supergravity (and indirectly in string theory).

The branes are multidimensional objects possible to define in many
various environments (in the most general situation -- on a ground
of an arbitrary theory with antisymmetric tensor fields) and
provide a very useful method which allows to understand some
theories as theories living on brane worldvolumes (or intersection
of the worldvolumes) immersed in some other theory. In this scheme
we can for example interpret superstring theories as defined on
domain walls of the M-theory. Similarly various super-Yang-Mills
or nonsupersymmetric Yang-Mills theories can be regarded as
superstring theories on respectively BPS or non-BPS configurations
made of D- and NS-branes. Finally the Standard Model should appear
as a low energy limit of a theory of this kind related to a four
dimensional intersection of the brane worldvolumes which break
supersymmetry. A verification if it is really possible to derive
the Standard Model directly from brane solutions of string theory
would give us new insight alternative to the usual route string
theory - supergravity - rigid supersymmetry - Standard Model.

The branes appearing in superstrings and M-theory can be described
also from supergravity point of view as a special class of
solution of equations of motion. Because supergravities are low
energy limits of superstring and M-theory and are consistent only
at the classical level, in such approach some information
especially involved with quantum properties of the branes is lost.
But on the other hand branes in supergravities can be studied with
the use of exact methods while superstrings provide only
perturbative techniques. Therefore results obtained in both the
ways are in many cases complementary.

The main part of this work was dedicated to give an exact
descriptions of possibly wide class of configurations of the
branes in the framework of supergravity. It was shown that for
such configuration which can be described as commonly orthogonally
intersecting delocalized branes the respective equations of
motions reduce to the known Toda-like system i.e. a generalization
of the Liouville equation. The reduction to the Toda-like system
works both for supersymmetric and nonsupersymmetric cases since it
does not depend on imposing on the model a harmonic gauge $\chi=0$
(a necessary condition for preserving supersymmetry). The key of
the reduction is a coordinate change where isotropic radial
coordinate $r$ is replaced by a coordinate called $\vartheta$. And
the relation $\vartheta(r)$ is a curved space harmonic function
with the curvature contribution to the harmonic equation given by
$d\chi/dr$.

The resulting Toda-like system is integrable and there are several
known classes of the solutions which can be written explicitly
with the use of elementary functions. Fortunately even the
restricted class of analytically known solutions can be relevant
for realistic brane configurations and it is possible to test
properties of the quite wide class of brane configurations by
studying the exact solutions. In this paper several examples
of such solutions are discussed in much detail to properly interpret 
supersymmetric properties of the solutions. We examine the single 
brane solutions with and without a dilaton. The first when $D=11$ 
can be treated as the supergravitational description of branes 
in M-theory and the second (for $D=10$) as the analogous description 
of branes in superstring theories. The solutions are in fact 
families of solutions parametrized by $\lambda$ giving a charge 
of the brane and $R$ describing a localization of a horizon 
expressed in terms of the isotropic coordinate $r$. The dilatonic
solution additionally depends on a parameter $c$ giving the 
classical description of a tachyon condensate. Also the three 
magnetic brane configuration in $D=11$ supergravity possessing 
a three dimensional (spatial) intersection and one overall 
transverse direction is studied. 

A transformation $r \rightarrow 1/r$ together with 
$R \rightarrow 1/R$ and $c\rightarrow -c$ gives a 
duality in the family of the solutions. The parameter $R=0$ 
(together with $c=0$) corresponds to supersymmetric solution when 
the BPS inequality $\calE \geq Q$ is saturated, where $\calE$ and 
$Q$ are respectively energy and charge density of the brane. 
For $R \neq 0$ supersymmetry is broken. But the solution dual 
to the supersymmetric one i.e. given by $R=\infty$ although 
nonsupersymmetric again saturates the BPS bound. The apparent 
contradiction can be explained by checking that in this case 
complementary parts of supersymmetry are broken by gravitino
components with vector index respectively tangent and transversal
to the brane worldvolume. So considering the model as
dimensionally reduced only to the directions orthogonal to the
brane (or only to the directions parallel) it seems like one part
of supersymmetry is preserved. It proves that it is possible to
construct a nonsupersymmetric model characterized by properties
usually reserved only for supersymmetric ones.

A character of the duality $r \rightarrow 1/r$ is rather unclear
in full generality. While it preserves a metric tensor and the
dilaton field and reverses sign of antisymmetric tensor one can
see (from the form of supersymmetry transformations) that its
action on fermions is more complicated (at least flips their
chirality). Since we considered only bosonic truncated model we
cannot verify this fact - the model should be extended to
incorporate fermions with possibly nonzero vacuum values. Such an
extension is rather natural with nonsupersymmetric solutions
because preserving supersymmetry condition immediately leads to
$\Psi_M=0$ while there is no analogous condition in the
nonsupersymmetric case.

Other generalizations of the model given in this paper can be
obtained when one discusses intersections at angles instead of the
orthogonal ones or localized branes instead of the delocalized
ones. It would be very interesting to find exact solutions in this
cases and check if any of them can serve as backgrounds for
realistic compactification models.

The possibility of finding the exact nonsupersymmetric solutions
is very intersecting. The solutions can be used to verify various
mechanisms postulated to break supersymmetry in a way leading to
the Standard Model. With the general solution at hand one is able
to check if conditions under which the mechanisms work can be
consistently derived from the equations of motion of the
underlying theory. In any realistic case however, it is necessary
to go beyond the bosonic solutions of the equations of motion
discussed in this paper and include fermions if one wants to take
into account fermionic condensates, calculate masses, chiralities,
coupling constants etc. and compare it to the Standard Model or
its extensions with broken supersymmetry.

%%%%%%%%%%%%%%%%%%%%%%%%%%%%%%%%%%%%%%%%%%%%%%%%%%%%%%%%%%%%%%%%%
%%%%%%%%%%%%%%%%%%%%%%%%%%%%%%%%%%%%%%%%%%%%%%%%%%%%%%%%%%%%%%%%%
%%%%%%%%%%%%%%%%%%%%%%%%%%%%%%%%%%%%%%%%%%%%%%%%%%%%%%%%%%%%%%%%%

\end{document}